\documentclass[letterpaper]{article}
\usepackage[T1]{fontenc}
\usepackage{pagecolor}
\usepackage[nottoc,numbib]{tocbibind}
\usepackage{tipa}
\usepackage{titling}
\usepackage{natbib}
\newcommand\fl[1]{}
	\fl{2}
		\fl{3}
			\fl{4}
				\input{preamble.tex-input}
				\input{comments.tex-input}

				\usepackage{array}
				\newcolumntype{P}[1]{>{\centering\arraybackslash}m{#1}}
				\commentstrue
				\commentsfalse
				\definecolor{darkgreen}{rgb}{0.0, 0.45, 0.13}

				\definecolor{lightred}{rgb}{0.7, 0, 0}
				\setupcomments{ac}{AC}{red}{color=red!30}
				\setupcomments{jf}{JF}{darkgreen}{color=green!30}
				\setupcomments{sr}{SR}{purple}{color=purple!30}
        \setupcomments{dk}{DK}{cyan}{color=cyan}

				\def\blockindentlength{\parindent}

				\newcommand{\term}[1]{{\textbf{\emph{#1}}}}

				\newcommand{\directions}{directions\xspace}

				\title{AI  Research Considerations for  Human Existential Safety \\ (ARCHES)}

				\author{
				Andrew Critch \\
				Center for Human-Compatible AI \\
				UC Berkeley
				\and
				David Krueger \\
				MILA \\
				Universit\'{e} de Montr\'{e}al}

				\newcommand{\Tsai}{Technologically autonomous AI\xspace}

				\newcommand{\misaligned}{misaligned\xspace}

\usepackage{tocloft}
\cftsetindents{subsection}{0.4in}{0.4in}
\begin{document}
	\fl{2}
		\fl{3}
			\maketitle
				\begin{abstract}
			Framed in positive terms, this report examines how technical AI research might be steered in a manner that is more attentive to humanity's long-term prospects for survival as a species.
			In negative terms, we ask what existential risks humanity might face from AI development in the next century, and by what principles contemporary technical research might be directed to address those risks.

				A key property of hypothetical AI technologies is introduced, called \emph{prepotence}, which is useful for delineating a variety of potential existential risks from artificial intelligence, even as AI paradigms might shift.  A set of \auxref{dirtot} contemporary research \directions are then examined for their potential benefit to existential safety.  Each research direction is explained with a scenario-driven motivation, and examples of existing work from which to build.  The research directions present their own risks and benefits to society that could occur at various scales of impact, and in particular are not guaranteed to benefit existential safety if major developments in them are deployed without adequate forethought and oversight.  As such, each direction is accompanied by a consideration of potentially negative side effects.

				Taken more broadly, the \auxref{dirtot} explanations of the research directions also illustrate a highly rudimentary methodology for discussing and assessing potential risks and benefits of research directions, in terms of their impact on global catastrophic risks.  This impact assessment methodology is very far from maturity, but seems valuable to highlight and improve upon as AI capabilities expand.

					\end{abstract}

\newpage

\section*{Preface}
\newcommand{\preface}{Preface}
\acsays{[ ]The contents of this section reflect my true opinion, but I'm not sure if it's necessary for the remainder of the document.  If either SR or JF thinks it should be removed, I would prefer to remove it.}

    At the time of writing, the prospect of artificial intelligence (AI) posing an existential risk to humanity is not a topic explicitly discussed at length in any technical research agenda known to the present authors.
    Given that existential risk from artificial intelligence seems physically possible, and potentially very important, there are number of historical factors that might have led to the current paucity of technical-level writing about it:
    \begin{enumerate}[1)]
				\item Existential safety involves many present and future stakeholders \citep{bostrom2013existential}, and is therefore a difficult objective for any single researcher to pursue.

        \item The field of computer science, with AI and machine learning as subfields, has not had a culture of evaluating, in written publications, the potential negative impacts of new technologies \citep{hecht2018time}.

				\item Most work potentially relevant to existential safety is also relevant to smaller-scale safety and ethics problems \citep{amodei2016concrete,cave2019bridging}, and is therefore more likely to be explained with reference to those applications for the sake of concreteness.

        \item The idea of existential risk from artificial intelligence was first popularized as a science-fiction trope rather than a topic of serious inquiry \citep{rees2013denial,bohannon2015fears}, and recent media reports have leaned heavily on these sensationalist fictional depictions, a deterrent for some academics.

    \end{enumerate}
    We hope to address (1) not by successfully unilaterally forecasting the future of technology as it pertains to existential safety, but by inviting others to join in the discussion. 		Counter to (2), we are upfront in our examination of risks.  Point (3) is a feature, not a bug: many principles relevant to existential safety have concrete, present-day analogues in safety and ethics with potential to yield fruitful collaborations.  Finally, (4) is best treated by simply moving past such shallow examinations of the future, toward more deliberate and analytical methods.

    Our primary intended audience is that of AI researchers (of all levels) with some preexisting level of intellectual or practical interest in existential safety, who wish to begin thinking about some of the technical challenges it might raise.
		For researchers already intimately familiar with the large volume of contemporary thinking on existential risk from artificial intelligence (much of it still informally written, non-technical, or not explicitly framed in terms of existential risk), we hope that some use may be found in our categorization of problem areas and the research directions themselves.

    Our primary goal is \emph{not} to make the case for existential risk from artificial intelligence as a likely eventuality, or existential safety as an overriding ethical priority, nor do we argue for any particular prioritization among the research directions presented here.
		Rather, our goal is to illustrate how researchers already concerned about existential safety might begin thinking about the topic from a number of different technical perspectives.  In doing this, we also neglect many non-existential safety and social issues surrounding AI systems.  The absence of such discussions in this document is in no way intended as an appraisal of their
    importance, but simply a result of our effort to keep this report relatively focused in its objective, yet varied in its technical perspective.
\vfill

\null
\vfill
	\begin{figure}[H]
		\centering
		\includegraphics[scale=0.44]{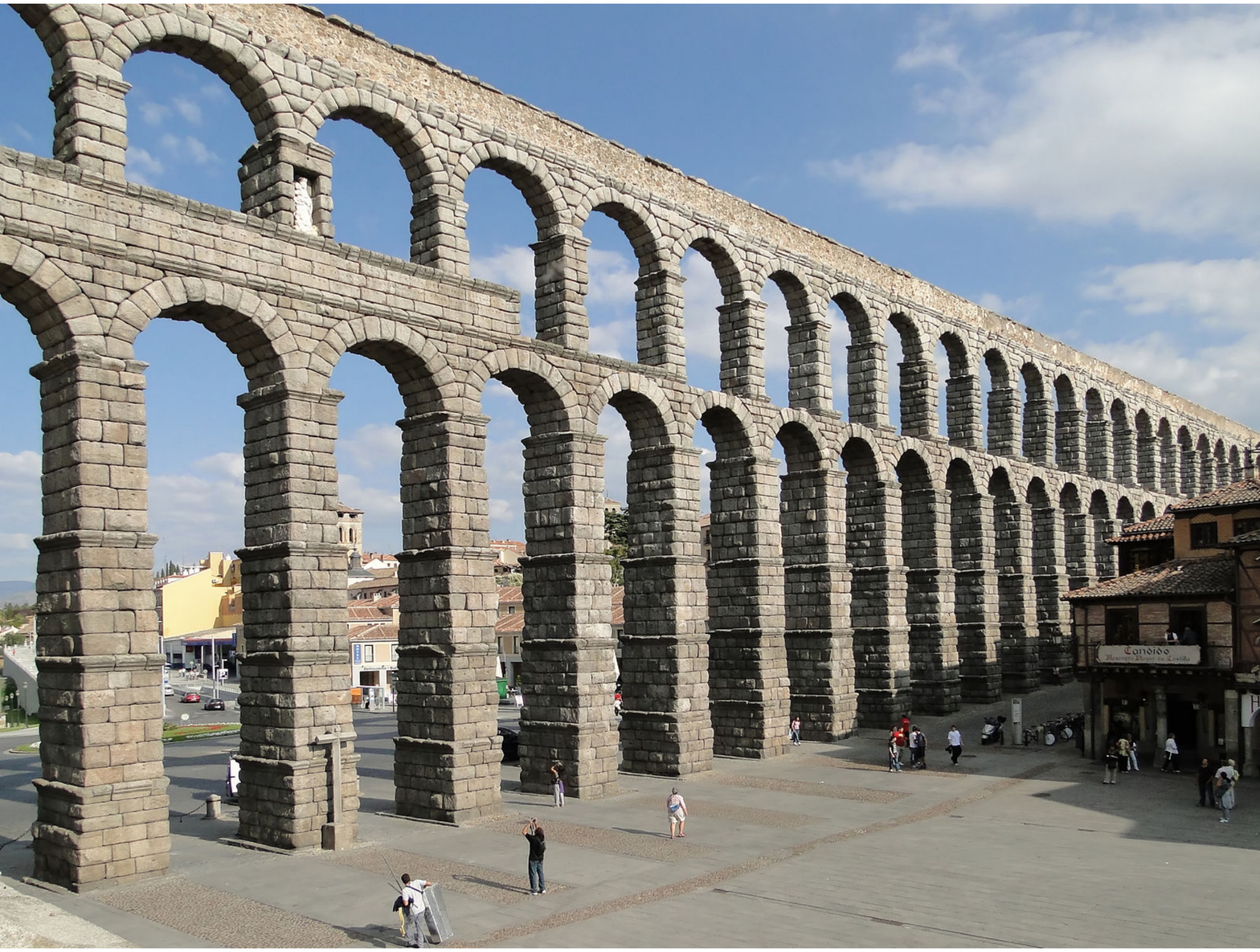}
		The arches of the Acueducto de Segovia, thought to have been constructed circa the first century AD  \citep{defeo2013historical}.
	\end{figure}
\vfill

\newpage
\setcounter{section}{-1}
\section{Contents}
\renewcommand{\contentsname}{}
\vspace{-3ex}
\tableofcontents
\newpage

\section{Introduction}\label{sec:intro}
	\fl{2}
		\fl{3}
AI technology has the potential to
			alleviate poverty,
			automate medical research,
			accelerate clean energy development,
			and enhance human cognitive abilities.
			Such developments would have been difficult to imagine in concrete terms 100 years ago, but are not inconceivable now.  If the worldwide AI research and development community is vigilant in distributing the benefits of these technologies fairly and equitably to all of humanity, global human welfare could be drastically and permanently improved.

			Unfortunately, any human extinction event would mean humanity ceases or fails to ever enjoy these marvelous benefits. The purpose of this report is to consider research directions in terms of their potential to steer away from human extinction risks, toward globally safer outcomes for humanity.
			While it is very difficult to \emph{forecast} whether any particular research direction will lead to an increase in risk to society, it may still be possible for researchers to \emph{steer} research in safer and more beneficial directions, if we are collectively attentive and mindful of the potential for both risks and benefits as new capabilities are developed.
			Since it is common for researchers to discuss the potential benefits of their work, this report is focussed almost entirely on risk.

			Why focus on human extinction risk, and not global catastrophic risks more broadly?  For two reasons: relative concreteness, and agreeability.  Many principles for mitigating existential risks also apply to mitigating global catastrophic risks in general.  However, thinking about the potential for future global catastrophic risks from artificial intelligence, while morally compelling, involves a great deal of speculation.

			Discussions in computer science can be more focused if there is a concrete and agreeably important outcome in mind, and the survival of the human species is one such an outcome that is relatively concrete and broadly agreeably important in the landscape of global catastrophic risks.

	\subsection{Motivation}\label{sec:mot}
		\fl{3}
			Taking a positive view of artificial intelligence, the aim of this report is to examine how technical AI research might be steered in manner that is more sensitive to humanity's long-term prospects for survival in co-existence with AI technology.  In negative terms, the aim is to consider how human extinction could occur if artificial intelligence plays a significant role in that event, and what principles might help us to avoid such an event.

			If human extinction were to occur within the next century, it seems exceedingly likely that human activities would have precipitated the extinction event.
			The reason is simple: nature has not changed much in the past 10,000 years, and given that nature on its own has not yielded a human extinction event for the past 100 centuries, it is not a priori likely for a natural human extinction event to occur in the next century.  (Indeed, a naive application of Laplace's law of succession would yield a probability estimate of at most around 1\%.)  By contrast, within this century, human extinction could occur through a variety of anthropogenic pathways, including bio-terrorism, climate change, nuclear winter, or catastrophic artificial intelligence developments \citep{matheny2007reducing,bostrom2013existential}.  This report is focused on the latter.

			Unfortunately, there are numerous pitfalls of human reasoning and coordination that mean human extinction \emph{in particular} is not a problem we should expect to avoid by default:

			\begin{quote}
				``We may be poorly equipped to recognize or plan
				for extinction risks \citep{yudkowsky2008cognitive}.
				We may not
				be good at grasping the significance of very large numbers
				(catastrophic outcomes) or very small numbers
				(probabilities) over large time frames.
				We struggle
				with estimating the probabilities of rare or unprecedented
				events \citep{kunreuther2001making}.
				Policymakers
				may not plan far beyond current political administrations
				and rarely do risk assessments value the existence
				of future generations [For an exception, see
					\citet{kent2004critical}.]  We may unjustifiably
				discount the value of future lives.
				Finally, extinction
				risks are market failures where an individual enjoys
				no perceptible benefit from his or her investment in
				risk reduction.
				Human survival may thus be a good
				requiring deliberate policies to protect.''  --\citet{matheny2007reducing}, \emph{Reducing the risk of human extinction.}
			\end{quote}

			In an effort to avoid some of these shortfalls of reasoning and coordination, this document examines how the development of artificial intelligence (AI) specifically could lead to human extinction, and outlines how various directions of technical research could conceivably be steered to reduce that risk.

 		Aside from wishing to avert existential risks in general, there are several reasons to take seriously the objective of reducing existential risk from artificial intelligence specifically:

			\newcommand{\motivation}[1]{\item \emph{#1}}
			\begin{enumerate}

				\motivation{A variety of advanced AI capabilities could be sufficient to pose existential risks.} A central theme of this report, argued further in \secref{keyconcepts}, will be that artificial intelligence does not need to meet the conditions of ``human-level AI'' \citep{nilsson2005human}, ``artificial general intelligence'' \citep{goertzel2007artificial}, or ``superintelligence'' \citep{bostrom1998long} to become a source of existential risk to humanity.
				It is conceivable that increasingly capable AI systems could lead to human extinction without ever achieving human-level intelligence or fully general reasoning capabilities.

				\motivation{The AI development timeline is unknown.} AI development has entered a period of high activity and abundant funding.
				In the past, AI research has cycled through periods of excitement and stagnation.  ``AI winter'' is a term used for a period of reduced funding and interest in AI.
				It was previously believed that the current period of activity might terminate with an AI winter sometime in the 2010s  \citep{hendler2008avoiding}, but this does not seem to have occurred.  Others believe that another AI winter could be yet to come.
 				\citet{grace2018will} conducted a 2016 survey of the 1634 researchers who published in NIPS 2015, and found great variation among the respondents, but a majority of respondents believing ``High-level machine intelligence'' would be achieved within a century:

				\begin{quote}
					Our survey used the following definition:
					\begin{quote}
						`High-level machine intelligence' (HLMI) is achieved when unaided machines can accomplish every task better and more cheaply than human workers.
					\end{quote}
					Each individual respondent estimated the probability of HLMI arriving in future years. Taking the mean over each individual, the aggregate forecast gave a 50\% chance of HLMI occurring within 45 years and a 10\% chance of it occurring within 9 years.  [...] There is large inter-subject variation: [...] Asian respondents expect HLMI in 30 years, whereas North Americans expect it in 74 years.  [...] Respondents were asked whether HLMI would have a positive or negative impact on humanity over the long run. They assigned probabilities to outcomes on a five-point scale. The median probability was 25\% for a ``good'' outcome and 20\% for an ``extremely good'' outcome. By contrast, the probability was 10\% for a bad outcome and 5\% for an outcome described as ``Extremely Bad (e.g., human extinction).''  Forty-eight percent of respondents think that research on minimizing the risks of AI should be prioritized by society more than the status quo (with only 12\% wishing for less).
				\end{quote}

				Given this variation in opinion, combined with the consensus that HLMI will most likely be developed in this century, it seems prudent to direct some immediate research attention at managing the concomitant risks.

					\motivation{Safe and powerful AI systems could reduce existential risk.} If safe and robust AI technologies continue to be developed, AI technology could in principle be used to automate a wide range of preventive measures for averting other catastrophes, thus serving to \emph{reduce} existential risk  \citep{yudkowsky2008artificial}.

			\end{enumerate}

			\subsection{Safety versus existential safety} \label{sec:safetyversus}

			This report is about existential safety.  What is the relationship between existential safety and safety for present-day AI systems? The answer can be summarized as follows:

			\renewcommand{\textit}[1]{\item \emph{#1}}
			\begin{enumerate}[1)]
				\textit{Deployments of present-day AI technologies do not present existential risks.}  Today's AI systems are too limited in both their capabilities and their scope of application for their deployment to present risks at an existential scale.
				\textit{Present-day AI deployments present safety issues which, if solved, could be relevant to existential safety.}
				For instance, the deployment of present-day autonomous vehicles present risks to individual human lives.
				Solutions to such safety problems that generalize well to more powerful AI capabilities could be used to improve existential safety for future AI technologies.
				On the other hand, safety techniques that work for present-day AI technologies but fail to generalize for more powerful AI systems could yield a false sense of security and lead to existential safety problems later.  Questioning which safety techniques and methodologies will generalize well is an important source of research problems.

				\textit{Present-day AI deployments present non-safety issues which could later become relevant to existential safety.}
				For instance, consider present-day AI ethics problems such as fairness, accountability, and transparency for AI systems.

				Many such problems do not present immediate and direct risks to the physical safety of humans or even their physical property, and are thus not often considered AI safety problems.
				However, if AI capabilities improve to become much more impactful on a global scale,
				ethical issues in the governance of those capabilities could eventually become matters of existential safety, just as present-day human institutions can present risks to public safety if not governed ethically and judiciously.
			\end{enumerate}

			\noindent Points (1)-(\theenumi) above can be summarized in the diagram of \figref{safetyversus}:

			\begin{figure}[H]
				\centering
				\includegraphics[scale=0.6]{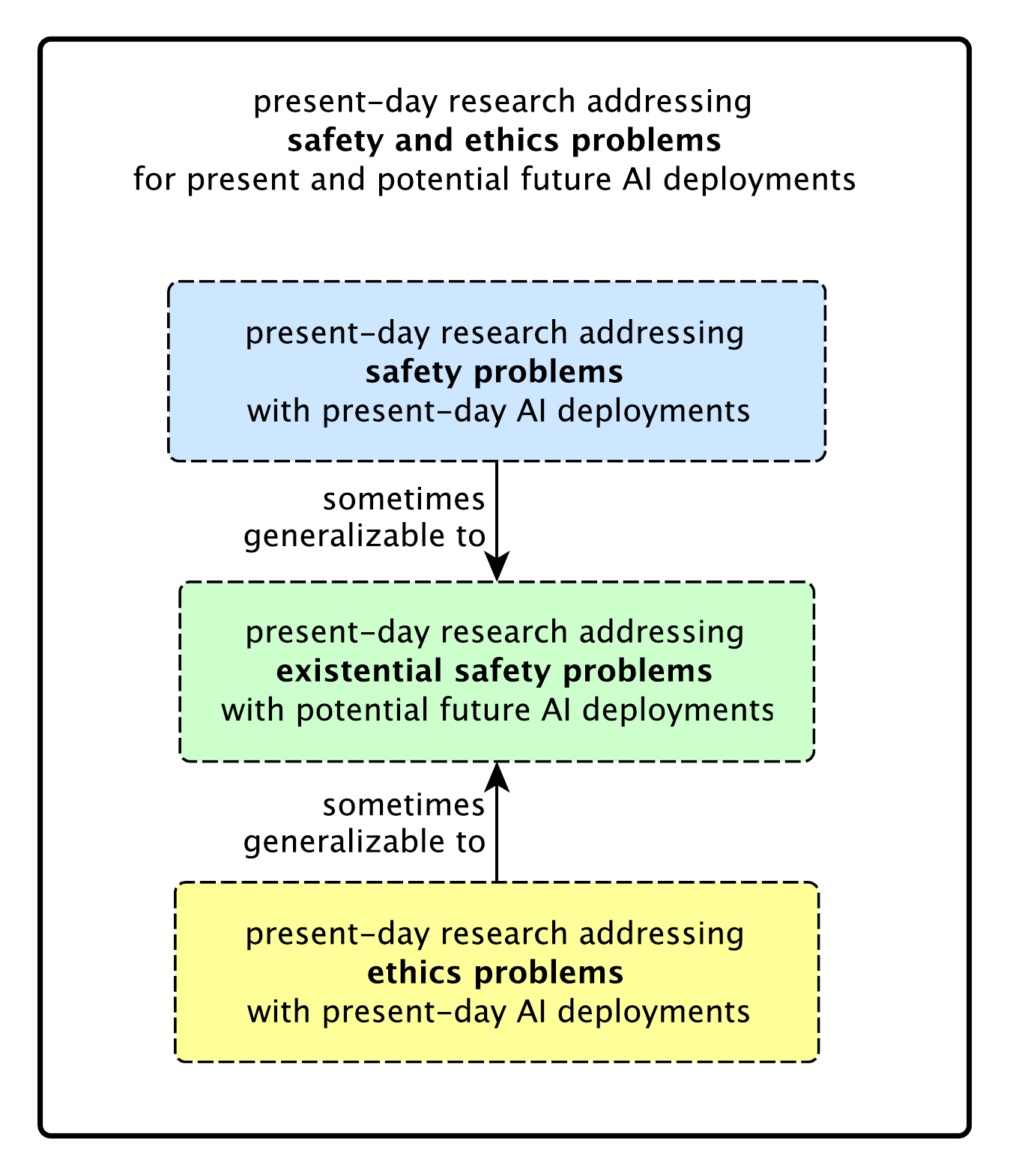}
				\caption{Relationship between AI safety, ethics, and existential safety.}
				\label{fig:safetyversus}
			\end{figure}

			A brief review of and comparison of related AI safety research agendas is provided in \secref{related}, including:

			\begin{itemize}
			    \item \emph{Aligning Superintelligence with Human Interests}
\citep{soares2014aligning},

			    \item \emph{Research Priorities for Robust and Beneficial Artificial Intelligence}  \citep{russell2015research},

					\item \emph{Concrete Problems in AI Safety} \citep{amodei2016concrete},

					\item \emph{Alignment for Advanced Machine Learning Systems} \citep{taylor2016alignment}, and

					\item \emph{Scalable Agent Alignment via Reward Modeling: a research direction} \citep{leike2018scalable}.
			\end{itemize}

\subsection{Inclusion criteria for research directions}
\label{sec:inclusioncriteria}
Each research direction in this report has been chosen for its potential to be used in some way to improve human existential safety.  The directions have been somewhat intentionally sampled from conceptually diverse areas of AI research, so as to avoid neglecting important considerations for how the technology could develop.

Research directions \emph{have not} been filtered for \emph{only} being relevant to AI safety or ethics. In particular, many of the selected research directions seem likely be pursued for reasons entirely unrelated to existential safety, at least in some form.

In addition, the research directions have \emph{have not} been filtered for having no potentially negative side effects; otherwise, the result would have been a very empty document.  Instead, reasoning is provided for how each research direction could potentially be pursued in service of existential safety, to enable further deliberation and discussion of that potential.

	Importantly, the reasoning included with each research direction \emph{is not} intended to argue or predict that the catastrophic scenarios discussed in this report will occur, nor to claim that humanity will or will not take adequate precautions to prevent catastrophes arising from AI development.
	Rather, this report simply aims to form a small part of those precautions.  In particular, this document is by no means a reasonable representation of the immense number and variety of potential beneficial applications of AI research.

		\subsection{Consideration of side effects}\label{sec:considerationside}
			None of the research directions in this report are guaranteed to be helpful to existential safety, especially if they are deployed carelessly or prematurely.  As such, each direction is exposited with a mini-section entitled ``Consideration of Side Effects'', intended to encourage researchers to remain mindful of the potential misapplications of their work.

			Unfortunately, it is not yet the norm in computer science research to write about the potentially negative impact of one's work in the course of producing the work.  This story has already been well told by the ACM Future of Computing Academy:

			\begin{quote}\small
				``The current status quo in the computing community is to frame our research by extolling its anticipated benefits to society. In other words, rose-colored glasses are the normal lenses through which we tend to view our work.
				[...]
				However, one glance at the news these days reveals that focusing exclusively on the positive impacts of a new computing technology involves considering only one side of a very important story. [...]

				We believe that this gap represents a serious and embarrassing intellectual lapse.  The scale of this lapse is truly tremendous: it is analogous to the medical community only writing about the benefits of a given treatment and completely ignoring the side effects, no matter how serious they are.
				[...]
				What's more, the public has definitely caught on to our community-wide blind spot and is understandably suspicious of it.
				[...]
				After several months of discussion, an idea for acting on this imperative began to emerge: we can leverage the gatekeeping functionality of the peer review process.
				[...]
				At a high level, our recommended change to the peer review process in computing is straightforward:
				\emph{Peer reviewers should require that papers and proposals rigorously consider all reasonable broader impacts, both positive and negative.}'' \newline \newline -- Hecht, B., Wilcox, L., Bigham, J.P., Sch\"{o}ning, J., Hoque, E., Ernst, J., Bisk, Y., De Russis, L., Yarosh, L., Anjum, B., Contractor, D. and Wu, C.  ``It's Time to Do Something: Mitigating the Negative Impacts of Computing Through a Change to the Peer Review Process.'' (2018)  \emph{ACM Future of Computing Blog.} \url{https://acm-fca.org/2018/03/29/negativeimpacts/}.
			\end{quote}

			\noindent In light of this phenomenon, perhaps this report can participate in an academia-wide shift toward the active consideration of potentially negative side effects of research outputs, including outputs of the research directions exposited here.

			As readers examine these potential side effects, it is important to remember that these \emph{are not} intended to communicate a forecast of what \emph{will} happen, only what \emph{might} happen and ought to be avoided.

			\subsection{Overview}\label{sec:overview}

			\emph{The logical thrust of this report can be summarized as follows.  This summary is not meant to stand on its own, and thus contains links to the relevant sections expanding on each point.  Please refer to those sections for supporting arguments.}

			 Existential risks arising from advancements in artificial intelligence are physically possible, very important if they occur, and plausible within this century (\secref{mot}).
			 Since existential safety applications of AI research are somewhat different from smaller-scale safety applications (\secref{safetyversus}), it makes sense to begin some manner of explicit discussions of how AI research could be steered in directions that will avoid existential risks.  This report aims to embody such a discussion (\preface, \secref{inclusioncriteria}, and \secref{considerationside}).

			For present-day thinking about existential safety to be robustly valuable for the many potential pathways along which AI technology could develop, concepts and arguments are needed that will be relevant in a broad variety of potential futures.
			\secref{keyconcepts} will attempt to organize together a few such key ideas. For instance, the potential for humanity to \emph{lose control} of the Earth to powerful AI systems is a key consideration, so \secref{prepotence} will define \emph{prepotent AI} as, roughly speaking, AI technology that would (hypothetically) bring about unstoppable globally significant changes to the Earth.
			If prepotent AI technology is ever developed, there are many potential pathways through which the effects of that technology could render the Earth unsurvivable to humans (\secref{humanfragility}).  Hence, the potential development of prepotent AI technology presents a source of existential risk.  Importantly, a hypothetical existential catastrophe arising from AI technology need not be attributable to a single, indecomposable AI system (\secref{multiplicitystakeholders}); catastrophes could also arise from the aggregate behavior of many AI systems interacting with each other and/or humans (\secref{questioningadequacy}).

			How might a catastrophe come about?  In general, supposing AI technology were to someday precipitate an existential catastrophe, there are a variety of societal errors that might have led up to that event.  Such errors could include
			coordination failures between AI development teams (\secref{uncdev}), failure to recognize the prepotence of an AI technology before its deployment (\secref{urprep}), unrecognized misalignment of an AI system's specifications with the long-term preservation of human existence (\secref{urmis}), or the involuntary or voluntary deployment of a technology known to be dangerous
			(\secrefs{invdep,voldep}).

			What do these errors have in common?  Abstractly, an existential catastrophe arising from AI technology could be viewed as an instance of AI systems failing to ``do what humans want.''
			After all, humans usually do not wish for humanity to become extinct.
			Thus, research aiming at existential safety for future AI systems might begin by studying and improving the interactions between a single AI system and a single human (\secref{ss}) to ensure that the AI system behaves in a manner desirable to the human.
			This could involve methods to help the human comprehend the AI system (\secref{sscomprehension}), deliver instructions to the system (\secref{ssinstruction}), and control the system if it begins to malfunction (\secref{sscontrol}).

			However, as soon as any new capability is developed that enables a single human to delegate to a single AI system for some sort of task, that capability is likely to be replicated many times over, leading to a multiplicity of AI systems with similar functionalities (\secref{multiplicitystakeholders}).
			Thus, any research anticipating the potentially global impacts of AI technology should take into account the numerous potential side effects of many AI systems interacting  (\secref{sm}).

			Moreover, diverse stakeholders can be expected to seek involvement in the governance of any AI technology that could be sufficiently impactful as to present an existential risk (\secref{multiplicitystakeholders}).  Therefore, existential safety solutions involving only single-stakeholder oversight are not likely to be satisfying on their own (\secref{questioningadequacy}).  For this and many other reasons, it makes more sense for AI technology to be developed in a manner that is well-prepared for oversight by ideologically, politically, and ethnically diverse people and institutions
			(\secref{relevantmultistakeholder}).

			In particular, facilitating collaboration in the oversight of AI systems by diverse stakeholders (\secref{facilitatingcollaborative}) could reduce incentives for research teams to enter unsafe development races (\secref{avoidingraces}), mitigate idiosyncratic risk-taking among the stakeholders (\secref{reducingidiosyncratic}), and increase the likelihood that systems will someday be developed with existential safety as their primary purpose (\secref{existentialsafety}).
			\secrefs{ms,mm} therefore adopt a focus on research directions relevant to one or more AI systems to serve multiple stakeholders at once.  Taken together, the research directions in \secrefs{ss,sm,ms,mm}
			constitute an incomplete but conceptually diverse portfolio of technical topics with potential relevance to  existential safety.

\section{Key concepts and arguments}\label{sec:keyconcepts}
	\fl{2}
		\fl{3}
			There are many potential pathways along which AI technology could develop.  This section introduces a few concepts and arguments for addressing a broad range of hypothetical futures in which existential risks from artificial intelligence could arise.

	\subsection{AI systems: tools, agents, and more}\label{sec:systems}
		\fl{3}
			By ``AI system'', we refer to any collection of one or more automated decision-making units.  The units are not assumed to be cooperating or competing, and are not assumed to have been created by cooperating or competing stakeholders.  Hence, the term ``system'' is intentionally general and agent-agnostic, and is meant to encompass simple and complex artifacts of engineering that could variously be called ``decision-making tools'', ``agents'', ``multi-agent systems'', ``societies of machines'', or none of the above.

	\subsection{Prepotence and prepotent AI}\label{sec:prepotence}
		\fl{3}
			We say that an AI system or technology is \emph{prepotent} /\textipa{"prE-p@-t@nt}/ (relative to humanity) if its deployment would transform the state of humanity's habitat---currently the Earth---in a manner that is \emph{at least as impactful as humanity} and \emph{unstoppable to humanity}, as follows:

			\begin{itemize}
				\item \emph{at least as impactful as humanity}:  By this we mean that if the AI system or technology is deployed, then its resulting transformative effects on the world would be at least as significant as humanity's transformation of the Earth thus far, including past events like the agricultural and industrial revolutions.
				\vspace{1ex}

				\item \emph{unstoppable to humanity}:  By this we mean that if the AI system or technology is deployed, then no concurrently existing collective of humans would have the ability to reverse or stop the transformative impact of the technology (even if every human in the collective were suddenly in unanimous agreement that the transformation should be reversed or stopped).  Merely altering the nature of the transformative impact does not count as stopping it.
			\end{itemize}

			In English, the term ``prepotent'' means ``Very powerful; superior in force, influence, or authority; predominant''.  On analogy with the terms `intelligent/intelligence' and `omnipotent/omnipotence', we favor the term \emph{prepotence} /\textipa{"prE-p@-t\super{@}n(t)s}/ over the more standard usage ``prepotency'' /\textipa{"pr\=e-""p\=o-t\super{@}n(t)-s\=e}/. In a number of Latin-descended languages, direct translations of ``prepotent'', such as ``prepotente'' and ``pr\'{e}potent'', mean ``arrogant'', ``overbearing'', ``high-handed'', ``despotic'' or ``possessing excessive or abusive authority''. These connotations are not typically carried in English, and while they do not contradict our usage, they are more specific than we intend.

			Before considering what level and types of risks prepotent AI technologies could pose to humanity, let us first consider briefly whether a prepotent AI system is physically possible to build in principle.  In short, the answer is probably yes.  Why should human beings---a product of random evolution and natural selection---be physically unsurpassable in our ability to control our physical environment?  Indeed, there are at least several classes of capabilities that might enable an AI technology to be prepotent, including:

			\begin{itemize}
				\item \textbf{Technological autonomy.}  Consider an AI system capable of outperforming the collective efforts of the world's top human scientists, engineers and industry professionals in endeavors of novel and independent scientific research and engineering.
				Let us call such a system \emph{technologically autonomous}.
				\Tsai might be able to build other AI systems that are prepotent, if so directed by whatever decision process determines its priorities.

				As well, technologically autonomous AI itself could constitute prepotent AI if it expands its scientific activities in the physical world in a manner that humans cannot contend with.  For comparison, consider how non-human animals are unable to contend with the industrial expansion of humans.
				\item \textbf{Replication speed.} The capability of AI systems to self-replicate and consume the Earth's physical resources too quickly for human civilization to intervene would constitute prepotence.
				To illustrate the in-principle possibility of such a scenario, consider the destruction of a large organism by a potent biological virus as a side effect of the virus rapidly disassembling the organism's cells to obtain resources for the virus producing copies of itself.  The virus need not be ``generally more intelligent'' than the host organism in any natural sense in order to end up destroying the host as a side effect of the virus's replication process.  The virus needs only to overwhelm or circumvent the host's immune system, a domain-specific problem.

				\item \textbf{Social acumen.} The capability to socially manipulate human nations to suddenly or gradually cede control of their resources could enable prepotence.
				To see the possibility of such a scenario in principle, consider that the holocaust of World War II was an event precipitated in large part by the highly influential natural language outputs of a particular human agent during a time of geopolitical unrest.
			\end{itemize}

			Because of the potential for such capabilities to cause humanity to lose control of the future, to develop any of them would mean facing a considerable and highly objectionable risk.

			\Hist The possibility that advanced AI systems could be difficult to control was considered by thinkers as early as visionary computer scientist Alan Turing:

			\begin{quote}
				``Let us now assume, for the sake of argument, that these machines are a genuine possibility, and look at the consequences of constructing them.
				[\ldots] [I]t seems probable that once the machine thinking method had started, it would not take long to outstrip our feeble powers.
				There would be no question of the machines dying, and they would be able to converse with each other to sharpen their wits.
				At some stage therefore we should have to expect the machines to take control [$\ldots$] -- Alan Turing [\citeyear{turing1951intelligent}], ``Intelligent Machinery, A Heretical Theory''.
			\end{quote}

			Mathematician and philosopher Norbert Wiener, widely regarded as the originator of the cybernetics, also remarked at the potential dangers of powerful ``mechanical agencies" with which we ``cannot interfere':

			\begin{quote}
				[...] if a bottle factory is programmed on the basis of maximum productivity, the owner may be made bankrupt by the enormous inventory of unsalable bottles manufactured before he learns he should have stopped production six months earlier.
				[...]

				Disastrous results are to be expected not merely in the world of fairy tales but in the real world wherever two agencies essentially foreign to each other are coupled in the attempt to achieve a common purpose.
				If the communication between these two agencies as to the nature of this purpose is incomplete, it must only be expected that the results of this cooperation will be unsatisfactory.

				If we use, to achieve our purposes, a mechanical agency with whose operation we cannot efficiently interfere once we have started it, because the action is so fast and irrevocable that we have not the data to intervene before the action is complete, then we had better be quite sure that the purpose put into the machine is the purpose which we really desire and not merely a colorful imitation of it.'' \citep{wiener1960some}
			\end{quote}

			\paragraph{Prepotent AI vs ``transformative AI''.}  The concept of prepotent AI may be viewed as defining a subset of what the Open Philanthropy Project has called  \emph{transformative AI} \citep{karnovsky2016background}, which roughly corresponds to clause 1 of the definition of prepotent AI.  Specifically, prepotent AI systems/technologies are transformative AI systems/technologies that are also unstoppable to humanity after their deployment (clause 2 of the prepotence definition).

			\paragraph{Prepotence vs ``superintelligence''.}  This report explicitly avoids dependence on the notion of ``superintelligence'' \citep{bostrom2014superintelligence} as a conceptual starting point.  Bostrom has defined the term \emph{superintelligence} to refer to ``an intellect that is much smarter than the best human brains in practically every field, including scientific creativity, general wisdom and social skills'' \citep{bostrom1998long}.
			This notion of superintelligence helps to simplify certain arguments about the in-principle possibility of existential risk from artificial intelligence, because superintelligence seems both physically possible to build in principle, and plausibly sufficient for threatening our existential safety.
			However, not all of the competencies stipulated in the definition of superintelligence are necessary for an AI technology to pose a significant existential risk.
            Although \citet{bostrom2014superintelligence} argues that superintelligence would likely be unstoppable to humanity (i.e., prepotent), his arguments for this claim (e.g., the ``instrumental convergence thesis'') seem predicated on AI systems approximating some form of rational agency, and this report aims to deemphasize such unnecessary assumptions.
			It seems more prudent not to use the notion of superintelligence as a starting point for concern, and to instead focus on more specific sets of capabilities that present ``minimum viable existential risks'', such as technological autonomy, high replication speed, or social acumen.

	\subsection{Misalignment and MPAI}\label{sec:mpai}
		In considering any prepotent or even near-prepotent AI technology, one immediately wonders whether its transformative impact on the world would be good or bad for humanity.  \emph{AI alignment} refers to the problem of ensuring that an AI system will behave well in accordance with the values of another entity, such as a human, an institution, or humanity as a whole \citep{soares2014aligning,taylor2016alignment,leike2018scalable}.

		What should qualify as a \emph{misaligned} prepotent AI (MPAI)?  Setting aside the difficulty of defining alignment with a multi-stakeholder system such as humanity, where might one draw the threshold between ``not very well aligned'' and ``misaligned'' for a prepotent AI system?  For the purpose of this report, we draw the line at humanity's ability to survive:

		\centerbox{
		\textbf{MPAI.} We say that a prepotent AI system is a \emph{misaligned} if it is \emph{unsurvivable} (to humanity), i.e., its deployment would bring about conditions under which the human species is unable to survive.\footnotemark{}  Since any unsurvivable AI system is automatically prepotent, \emph{misaligned prepotent AI} (MPAI) technology and \emph{unsurvivable AI} technology are equivalent categories as defined here.
		}

		\footnotetext{It is interesting to ask what it means for a particular AI system to ``bring about'' unsurvivable conditions, if such conditions occur.  This is a question involving accountability for AI systems \citep{barocas2014fatml}, which may become more difficult to define for more capable systems.  If System A builds System B, and System B brings about unsurvivable conditions, did System A bring about unsurvivable conditions?  Any scientific claim that a system will not ``bring about'' unsurvivable conditions will have to settle on a definition in order to be meaningful.  For the purposes of this report, the precise technical definition of ``bring about'' is left as an open question.}

			\begin{figure}[H]
				\centering
				\includegraphics[scale=0.44]{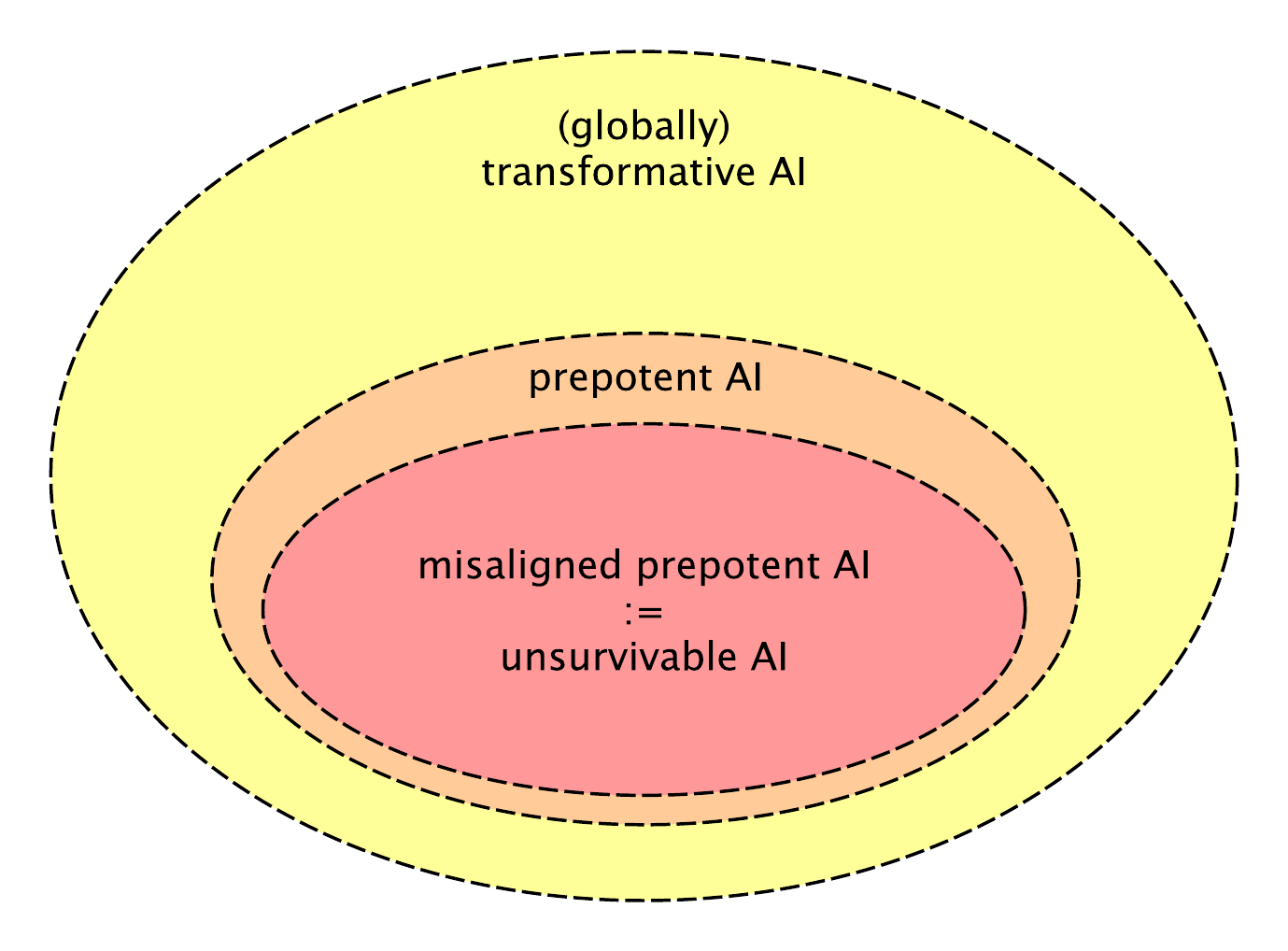}
				\caption{Venn diagram relating ``transformative AI'', ``prepotent AI'', and ``unsurvivable AI'' (``MPAI'' in this report).  In other contexts less focussed on human survival, it might make sense to use a different threshold to define \emph{misalignment} for prepotent  AI, in which case the term \emph{unsurvivable AI} could be reserved for what is called MPAI in this report.
}
				\label{fig:mpai}
			\end{figure}

			\paragraph{Extinctive versus unsurvivable.} It may worth noting that humanity can become extinct in a manner where our habitat is at no point unsurvivable, if the extinction is somehow willful.
			This means there is a category of \emph{extinctive AI} that lies strictly between prepotent AI and MPAI, which includes AI systems that would somehow lead humanity to extinction along a pathway where humanity has the ability to prevent its extinction at every point along the way, but somehow fails to exercise this ability, right to the very end.
			This may be a very important consideration for humanity, however, it will not be a key focus of the present document. In fact, \secref{humanfragility} will raise some considerations suggesting that prepotent AI systems may be unsurvivable \emph{by default} in a certain sense, in which case intermediate categories between prepotent AI and MPAI may not be particularly useful distinctions.
			In any case, attentive readers wishing to draw this distinction may often need to treat ``human extinction'' as a shorthand for ``involuntary human extinction'' at some places in this report.

	\subsecdef{deploymentevents}{Deployment events}\label{sec:hfa}
		What counts as the deployment of a prepotent AI system?  If an AI system becomes prepotent after it is already in deployment, shall we consider that moment ``the deployment of a prepotent AI system``?  In this report, the short answer is yes, because the resulting loss of control for humanity from that point forward may be similar to the result of deploying an AI system that is already prepotent.

		To be more precise, throughout this report,
		\begin{itemize}
			\item a \emph{transformative AI deployment} event refers to either a transformative AI technology becoming deployed, or a deployed AI  technology becoming transformative.
			\item a \emph{prepotent AI deployment} event refers to either a prepotent AI technology becoming deployed, or a deployed AI technology becoming prepotent;
			\item an \emph{MPAI deployment} event refers to either an MPAI technology becoming deployed, or a deployed AI technology becoming MPAI.
		\end{itemize}

		\noindent As defined above, these deployment events have the following implications for what is possible for humanity:
		\begin{itemize}
			\item[$\rightarrow$] At the deployment of a transformative AI technology, it follows that a highly significant global transformation of humanity's habitat will occur, but that transformation might be reversible or stoppable by humanity after it occurs.
			\item[$\rightarrow$] At the deployment of a prepotent AI technology, it follows that humanity has no further ability to reverse or stop the transformative impact of the system, but might still have some ability to channel or direct the impact in some way.
			\item[$\rightarrow$] At the deployment of an MPAI technology, it follows that humanity has no further ability prevent human extinction from occurring.
		\end{itemize}

\begin{figure}[H]
			\centering
			\includegraphics[scale=0.60]{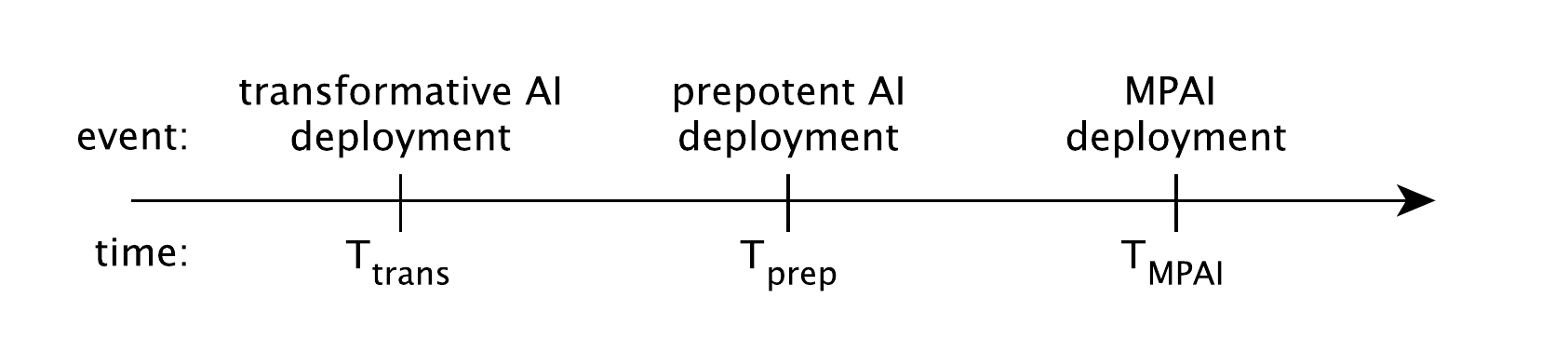}
			\caption{timeline of hypothetical deployment events}
			\label{fig:timeline}
		\end{figure}

		Since these deployment events correspond to successively smaller categories of AI systems having been in active deployment, if they occur they must occur in a sequence, as in \figref{timeline}.

		\noindent Note in particular that $T_\mathrm{trans}$ can be less than $T_\mathrm{prep}$ in a scenario where a transformative AI system becomes prepotent only after the system is in active deployment, and $T_\mathrm{prep}$ can be less than $T_\mathrm{MPAI}$ in a scenario where a prepotent AI system becomes misaligned only after the system is in active deployment.

		\subsecdef{humanfragility}{Human fragility}\label{sec:humanfragility}
		\newcommand{\hft}{the Human Fragility Argument (\secref{humanfragility})\xspace}
		\fl{3}

			There are numerous pathways through which the deployment of a prepotent AI system could be unsurvivable to humanity.  In short, the reason is that many possible transformations of the Earth would render it unsurvivable to humans, and prepotent AI technology by definition would globally and unstoppably transform the Earth.

			To see this, first observe that the physical conditions necessary for humans to survive are highly specific, relative to the breadth of environments in which machines can operate.  For instance, consider the availability of oxygen in the atmosphere, availability of liquid water, absence of many other compounds that would be noxious to breathe or drink, radiation levels, air pressure, temperature,
			and the availability of highly complex digestible food sources.  Each of these is a physical feature of humanity's surroundings which, if transformed significantly, would be unsurvivable.  By contrast, machines can already be designed to operate under the ocean, in space, and on Mars.  Humans can also visit these places, but only with the help of machines to maintain safe conditions for the human body.

			Next, recall that the deployment of a prepotent AI technology by definition brings about changes to the Earth at a global scale, in a manner that humans cannot reverse or stop.  At first such changes might not result in inevitable human extinction.  However, many vectors of change would, if compounded over time, end up violating one of the many physical, chemical, and biological prerequisites needed for human survival.   Over the past century it has become clear that human-driven changes to the Earth have the potential to destroy the human species as a side effect.  The variety of possible AI-driven changes expands and accelerates this potential.  While any particular pathway to unsurvivability is unlikely, the likelihood that \emph{some} such pathway could obtain is much higher, because of the many conditions which, if violated, would end human existence. This conclusion has been argued by numerous others, including
						\cite{yudkowsky2008artificial},
						\cite{shulman2010omohundro},
						\cite{shanahan2015technological}, and 			\cite{bostrom2018vulnerable}.

			Of course, it is not logically impossible for humans to survive the deployment of a prepotent AI technology.  Preserving conditions necessary for human survival means operating within certain limits, and if the creators of the technology were collectively mindful of human extinction as a potential side effect, perhaps great care and coordination may have been undertaken to ensure those limits would be permanently enforced.  One might even think the conditions for human survival are relatively easy to maintain, because they have been maintained for at least as long as humanity has existed.
			However, it is reasonable to expect that the deployment of an \emph{arbitrarily generated} prepotent AI system would most likely be unsurvivable to humans if deployed, just as the conditions of an arbitrarily generated planet would be unsurvivable to humans.

			This raises a key question regarding the danger of prepotence: how difficult is it to ensure that the deployment of prepotent AI technology would be survivable to humans?  Certainly we humans could all agree to never create or allow the development of prepotent AI technology in the first place, but this is not an answer to the question at hand: conditional on the deployment of a prepotent AI technology, what is the chance that humanity would be unable to survive?  In statistical terms, this is a question about the distribution from which the prepotent AI technology would be drawn, and that distribution itself is a function of the effort humanity collectively puts into constraining AI development through coordinated safety efforts.  For instance, if the AI research community as a whole became deeply engaged in the technical challenge of preserving human existence, perhaps that would be enough to eventually relinquish control of the Earth to prepotent AI technology while maintaining survivable conditions for humans.  Would a lesser degree of care suffice?

			An answer to this question is beyond the scope of this report.  It would be a claim relating the fragility of human existence with the coordinated aptitude of the worldwide AI research and development community.

			On one hand, Perrow's theory of \emph{Normal Accidents} \citep{perrow1984normal} would imply that if AI technology turns the world as a whole into a ``tightly coupled complex system'', then catastrophic failures should be expected by default.  On the other hand, the literature on \emph{highly reliable organizations} \citep{laporte1996high,roberts2001must} is suggestive that well-managed hazardous systems can operate for periods of decades without incident.

			Could humans ever succeed in developing prepotent AI technology that would operate as safely as a highly reliable human organization, over the indefinite future?

			Attempting this would seem an unnecessary risk from many perspectives; why not build highly beneficial non-prepotent AI instead?

			In any case, perhaps reflecting on the fragility of human beings could do some good toward motivating the right kinds of work.
			To that end, we encapsulate the above discussion in the following thesis:

			\centerbox{
			\textbf{The Human Fragility Argument.} Most potential future states of the Earth are unsurvivable to humanity.  Therefore, deploying a prepotent AI system absent any effort to render it safe to humanity is likely to realize a future state which is unsurvivable.  Increasing the amount and quality of coordinated effort to render such a system safe would decrease the risk of unsurvivability.  However, absent a rigorous theory of global human safety, it is difficult to ascertain the level of risk presented by any particular system, or how much risk could be eliminated with additional safety efforts.
			}
		With this argument in mind, we next consider the added complexity introduced by a multiplicity of human stakeholders delegating to a multiplicity of AI systems.

		\subsection{Delegation}
		Throughout this report, the relationship between humans and AI systems is viewed as one of \emph{delegation}: when some humans want something done, those humans can delegate responsibility for the task to one or more AI systems.  From the perspective of the AI systems, the relationship would be one of \emph{assistance} directed toward the humans.  However, to avoid dependence of our arguments upon viewing AI systems as having a ``perspective'', we treat humans as the primary seat of agency, and view the humans as engaged in delegation.

		Human/AI delegation becomes more complex as the number of humans or AI systems increases.  We therefore adopt the following terminology for indicating the number of human stakeholders and AI systems in a human/AI delegation scenario.  The number of humans is always indicated first; as a mnemonic, remember that humans come before AI: in history, and in importance!

		\begin{itemize}
			\item \textbf{Single(--human)/single(--AI system) delegation} means delegation from a \emph{single human stakeholder} to a \emph{single AI system} (to pursue one or more objectives).

			\item \textbf{Single/multi delegation} means delegation from a \emph{single human stakeholder} to \emph{multiple AI systems}.
			\item \textbf{Multi/single delegation} means delegation from \emph{multiple human stakeholders} to a \emph{single AI system}.
			\item \textbf{Multi/multi delegation} means delegation from \emph{multiple human stakeholders} to  \emph{multiple AI systems}.
		\end{itemize}

		In this taxonomy, the notion of a \emph{single human stakeholder} refers to either a single natural human person, or a single human institution that is sufficiently internally aligned and organized that, from the perspective of an AI system, the institution can be modeled as a single human.  It remains an open research question to determine when and how a human institution should be treated as a single human stakeholder.

		What should be viewed as a collection of distinct interacting AI systems, versus a single composite AI system?  In some situations, both views may be useful.  This consideration is deferred to the beginning of \secref{sm}.

		\subsection{Comprehension, instruction, and control}
		\label{sec:cic}

		Throughout this report, three human capabilities are viewed as integral to successful human/AI delegation: \emph{comprehension}, \emph{instruction}, and \emph{control}, as defined below.  This focus on maintaining human capabilities serves to avoid real and apparent dependencies of arguments upon viewing AI systems as ``agents'', and also draws attention to humans as responsible and accountable for the systems to which they delegate tasks and responsibilities.

		\centerbox{
		\textbf{Comprehension:} Human/AI comprehension refers to the human ability to understand how an AI system works and what it will do.
		}

Debuggers, static analysis, and neural net visualization tools are among present-day methods for improving human/AI comprehension.
Comprehension helps us reason about how an AI system will respond to an instruction before deploying it with that instruction, a key capability for reducing risks.

			\centerbox{
			\textbf{Instruction:} Human/AI instruction refers to the human ability to convey instructions to an AI system regarding what it should do.
			}

For a human to derive useful work from an AI system, there must be some conveyance of information or knowledge from the human about what the human would find useful, in a manner that steers the behavior of the AI system toward that work.  This conveyance, or ``instruction'', could take any number of forms, e.g., code written by the system's creators, recorded data about human history, real-time interactions with humans during training or deployment, keyboard input from a human user, or a direct neurological link with the user.
Some of these channels of human/AI instruction may be used to control and modulate the others.  Effective instruction involves not only ensuring a flow of information from the human to the AI system, but also knowing what information to put into which channels, and ensuring the information affects the AI system's behavior as needed.

Just as some programming languages are more difficult to write than others, there will always be some available forms of human/AI instruction that are more effective than others.  For example, methods that are highly tolerant of errors in human judgement or transcription will be easier to use than methods highly sensitive to human error.  In any case, human/AI instructions are bound to fail from time to time.

			\centerbox{
			\textbf{Control:} Human/AI control refers to the human ability to retain or regain control of a situation involving an AI system, especially in cases where the human is unable to successfully comprehend or instruct the AI system via the normal means intended by the system's designers.
			}

Shutting down, repairing, or dismantling an AI system are ways in which humans can retain control of an AI system's operation even when the communication abstractions of comprehension and instruction are not working well.

Of course, few present-day machines could not be safely shut down or destroyed by their owners if so desired.  However, some machines have no owner, such as the internet, and are not so easy to shut down by legitimate means.  If real-world AI capabilities ever approach the potential for prepotence, it may become very important for humans to retain safe and legitimate means to carry out such interventions on AI systems.

						\paragraph{Instruction versus control.}  Where should one draw the distinction between ``instruction'' and ``control''?  For instance, one could argue that an action like \guillemotleft unplug the power\guillemotright{} is an ``instruction'' for turning a machine off, and that the laws of physics are the ``interpreter'' ensuring the instruction is followed.  However, in this framing, the ``communication channel'' comprised by the \guillemotleft unplug\guillemotright{} mechanism is certainly of a different design and purpose than the usual mouse, keyboard, and voice instruction channels.  In particular, the \guillemotleft unplug\guillemotright{} channel has the power to override any instructions from the other channels.  So, even if one wishes to view control as a kind of instruction, it should be treated as a fairly special case, with the purpose and capacity to override other instructions.

		\subsection{Multiplicity of stakeholders and systems}\label{sec:multiplicitystakeholders}
		When first beginning to analyze existential risk from AI development, it may be tempting---and perhaps conceptually simpler---to focus on single/single delegation.  Indeed, if AI technology brings about a human extinction event, one might easily argue that the system ``did not do what humans would have wanted'', and the task of making an AI system do what even a single human wants is still a difficult challenge in many domains.
		Perhaps for this reason, much of the technical research to date that is formally or informally cited as relevant to catastrophic risks from AI---under such labels as ``AI safety'', ``AGI safety'' or ``long-term AI safety''---has been focussed primarily on single/single delegation. (\secref{related} will give a more detailed overview of the literature.)

		Focusing entirely on single/single delegation can be misleading, however.  There are powerful social and economic forces that can transform a single/single delegation scenario into a multi/multi delegation scenario.
		First, note that there are numerous pathways through which a single/single
		delegation scenario with any powerful AI system (such as a prepotent or near-prepotent AI system) can become a multi/single scenario:
		\begin{itemize}
			\item[a)] Outside stakeholders will have a strong motivation to seek to own and/or share control of the system, because of its potential for impact.
			\item[b)] The creators of the system might encounter any number of disagreements regarding how best to use the system.  These disagreements might not have been considered in advance, especially if the creators were not confident they would succeed in developing the system, or did not have a clear understanding of how the system would end up working when they began their partnership.  Facing the heightened stakes of this increased potential for impact could lead to a splintering of opinions about what to do next.  So where previously the creators might have acted as single unified stakeholder, this might not remain the case.
		\end{itemize}
		These pathways lead from single/single to multi/single delegation scenarios.  Next, consider how a multiplicity of AI systems can result:
		\begin{itemize}
			\item[c)] The creators of any powerful AI system have economic incentives to duplicate and sell instances of the system to outside buyers.
			\item[d)] Contemporary research groups, upon observing the capabilities of a powerful AI system, may also have strong intellectual and economic incentives to replicate its capabilities.
		\end{itemize}
		These pathways lead from single/single and multi/single to multi/multi delegation scenarios.  In summary:

			\centerbox{
			\textbf{The multiplicity thesis.}  Soon after the development of methods enabling a single human stakeholder to effectively delegate to a single powerful AI system, incentives will likely exist for additional stakeholders to acquire and share control of the system (yielding a multiplicity of engaging human stakeholders) and/or for the system's creators or other institutions to replicate the system's capabilities (yielding a multiplicity of AI systems).
			}

      \subsubsection{Questioning the adequacy of single/single delegation}\label{sec:questioningadequacy}
		  The multiplicity thesis presents a source of added complexity in maintaining existential safety, which might not be well addressed by safety research focussed on single/single delegation.

			How important is it to prepare for this complexity before it arises?  That is to say, how important is it to begin work today on single/multi, multi/single, and multi/multi delegation solutions, from an existential safety perspective?

			\paragraph{An optimistic view.} One view is that, given the development of near-prepotent AI systems for single/single delegation, future humans and human institutions would be able to use non-technical means to coordinate their use of those systems to ensure that either prepotent AI systems are never developed, or that the systems will maintain existential safety if they are developed.

			In this view, there is no pressing existential need to develop multi/multi delegation
			solutions prior to the development of near-prepotent AI systems.

			As evidence for this view, one can point to any number of past successes of human coordination in the use and governance of technology.  For instance, there is the Montreal Protocol banning the production of CFCs, which was fully signed in 1987, only 14 years after the scientific discovery in 1973 that CFSs are damaging to the ozone layer \citep{murdoch1997voluntary,andersen2012protecting}.  For an example of international coordination in computer science specifically, consider the creation and governance of internet protocols such as TCP/IP by the Internet Engineering Task Force---a community with no formal organizational hierarchy---by ``rough consensus and running code'' \citep{russell2006rough,resnick2014consensus}.

			\paragraph{A pessimistic view.}  Alternatively, it might be that future humans would struggle to coordinate on the globally safe use of powerful single/single AI systems, absent additional efforts in advance to prepare technical multi/multi delegation solutions.

			For a historical analogy supporting this view, consider the stock market ``flash crash'' of 6 May 2010, viewed as one of the most dramatic events in the history of financial markets \citep{madhavan2012exchange}.  The flash crash was a consequence of the use algorithmic stock trading systems by competing stakeholders \citep{easley2011microstructure,kirilenko2017flash}.
			If AI technology significantly broadens the scope of action and interaction between algorithms, the impact of unexpected interaction effects could be much greater, and might be difficult to anticipate in detail.  World War I is a particularly horrific example where technology seemed to outpace the strategic thinking of human beings (specifically, military and state leaders) regarding how to use it \citep[Chapter 5: The nineteenth century, II: technology, warfare and international order]{gray2013war}.
			Military tactics lagged behind military technology, such as the machine gun and mustard gas, that had been developed over the preceding century, leading to an unprecedented number of casualties at war.

			As a motivating thought experiment involving rapid and broadly scoped multi-agent interaction, imagine that 10 years from today, 50\% of all humans will be able to think one thousand times faster than they can today.

			Such a sudden change in human capabilities might be incredibly positive, dramatically accelerating progress in science, technology, diplomacy, and perhaps even multi/multi delegation research.
      But the result could also be disastrous.
			First, if the areas of the international economy most accelerated by intelligence enhancement turned out to involve the production of pollution or similar side effects, a rapid physical destruction of the environment could result.
			Second, perhaps the rapidly changing social and geopolitical landscape could lead to a proliferation of attempts to seize political or economic power via socially or physically destructive tactics or warfare.

			The proliferation of powerful single/single AI delegation solutions could conceivably have a similar destabilizing effect upon society and the environment.
			Indeed, some have argued that artificial intelligence and computing technology more broadly has already outpaced our collective ability to make good decisions about how it is used \citep{hecht2018time}.

			\paragraph{A precautionary view.}  Of course, only one future will obtain in reality.  Which of the above views---optimism or pessimism---will be closer to the truth?  This question obscures the role of pessimism in preparedness: we all can exercise some agency in determining the future, and the most useful form of pessimism is one that renders its own predictions invalid by preventing them.

			In any case, it is well beyond the scope of this report to determine for certain whether future humans and human institutions will succeed or fail in the judicious use of powerful single/single delegation solutions.  And, maintaining a diversity of views will enable planning for a diversity of contingencies.  Thus, in place of a prediction, we instead posit the following value judgement:

			\centerbox{
			\textbf{Multi/multi preparedness.} From the perspective of  existential safety in particular and societal stability in general, it is wise to think in technical detail about the challenges that multi/multi AI delegation might eventually present for human society, and what solutions might exist for those challenges, \emph{before} the world would enter a socially or geopolitically unstable state in need of those solutions.}

			\noindent We will return to this discussion in Section 7.

\subsection{Omitted debates}\label{sec:omitteddebates}

To maintain a sufficiently clear conceptual focus throughout, a number of debates have been intentionally avoided in this document:

	\begin{itemize}
	\item \emph{What about global catastrophes that would not result in human extinction?}  For concreteness, and to avoid digressions on what would or would not constitute a global catastrophe, this report is focussed on the simpler-to-agree-upon concept of human survival.  Nonetheless, it does seems likely that many present-day approaches relevant to averting extinction risk should also be applicable to averting other events that would broadly be considered global catastrophes.  The reader is therefore invited to use their own judgement to determine where arguments or methods in this report can be generalized  from human extinction prevention to global catastrophic risk prevention.

	\item \emph{What constitutes a human?} Even the concept of human survival is subject to some debate regarding its meaning, because of potential future ambiguity in what constitutes a human being.
	For instance, \citet[``Transmigration'', Chapter 4]{moravec1988mind} describes a future in which humans can replace themselves by digital emulations their own minds, and \citet{hanson2016age}	envisions a future economy where most work is carried out by human-like emulations that have been modified and selected for performing valuable work.
	If no biological humans remain, but human emulations continue to operate, should humanity be considered extinct?
	This report does not delve into that question, because the authors suspect that most present-day approaches to existential safety will are not greatly affected by the answer, although it could still become important in the future.

	\item \emph{What about other negative side effects of AI development?}  Many ideas and arguments considered in this report could be applied to averting safety and ethical failures that would by no means be considered global catastrophes.  The reader is invited to use their own judgement to consider what other negative side effects of AI development can be avoided and are worth the cost of avoidance.  As discussed in the \preface, omissions of other safety and ethical issues from this report is not intended by the authors as an appraisal of their importance or relevance to society.

	\item \emph{What constitutes ``beneficial'' AI?}  A closely related topic to reducing existential risk from artificial intelligence---and which does not entirely fit within the scope of this report---is that of developing \emph{provably beneficial} AI systems, i.e., AI systems which provably benefit the whole of human society.  At a technical level, provable beneficence and existential safety are tightly intertwined:

			\begin{enumerate}
				\item  For any broadly agreeable definition of ``benefit'', an AI system that provably benefits all of humanity should, by most definitions, preserve humanity's ability to avoid extinction.
				\item Conversely, preventing existential risk requires attending to global-scale problems and solutions, which might yield mathematical and algorithmic techniques for ensuring other global benefits as well as reducing other global risks.

			\end{enumerate}

			Despite these relationships, provable beneficence is a more general problem than existential safety.  To address provable beneficence, one would need to address or dissolve what it really means to benefit humanity, given that individual human preferences are ill-defined, plastic, and not in universal agreement.  By contrast, it might be easier to reach agreement on what scenarios constitute human extinction events, or at least to agree upon the general goal of avoiding all such scenarios.  So, this this report explicitly avoids delving into any debate regarding the meaning of ``provable beneficence''.
\end{itemize}

\section{Risk-inducing scenarios}\label{sec:riskinducing}
\begin{figure}[H]
				\centering
				\includegraphics[scale=0.4]{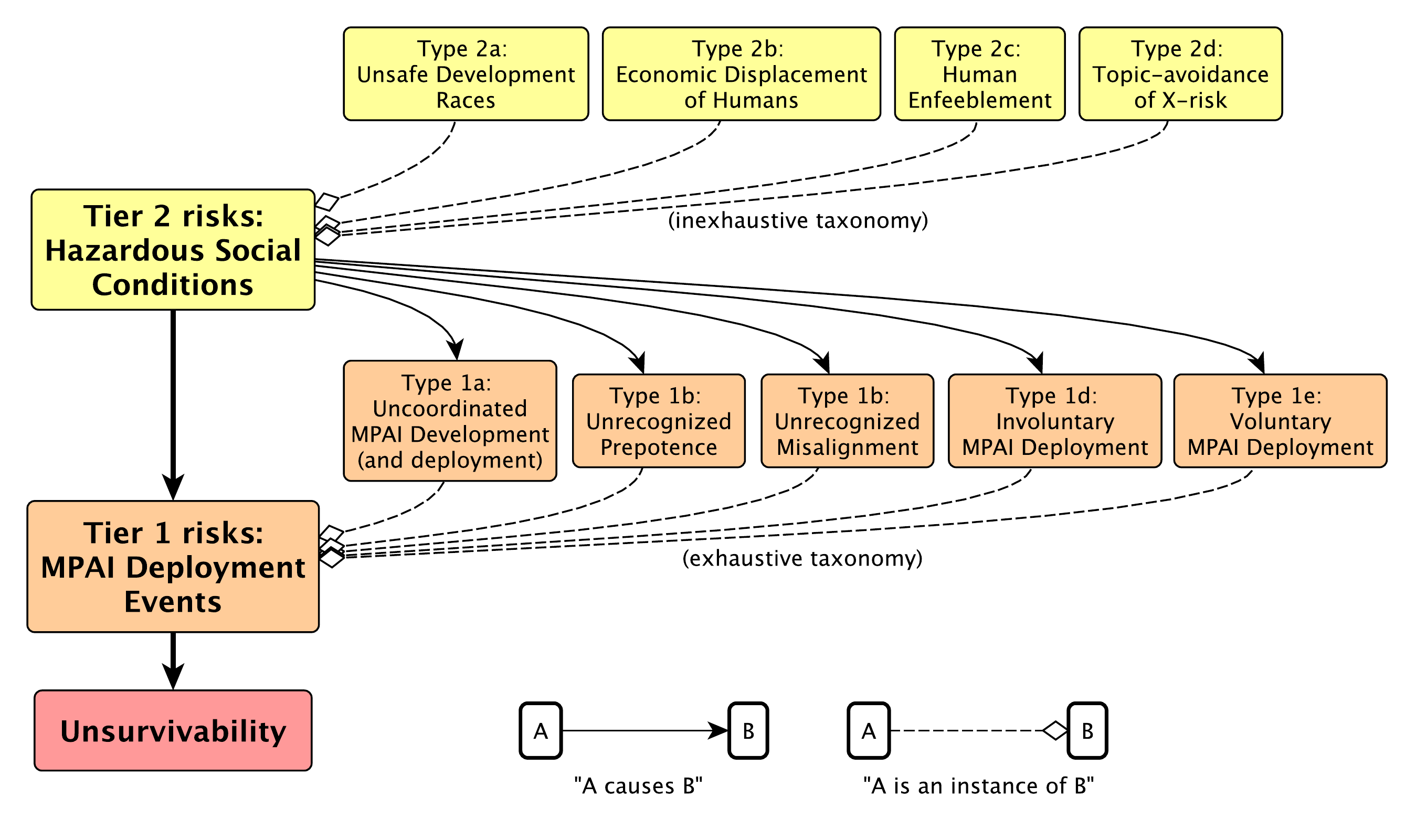}
				\caption{Relationship between risk types considered in this report; each risk type is described as its own subsection.}
				\label{fig:risktypes}
			\end{figure}

	\fl{2}
		\fl{3}
			How could human society make the mistake of deploying AI technology that is unsurvivable to humanity?  There are many hypothetical scenario types to consider, each of which might call for different forms of preventive measures.
			In this report, scenarios are organized into \emph{risk types} that will be outlined in this section.  The risk types are related via the causal diagram in \figref{risktypes}.
			\begin{itemize}
				\item ``Tier 1'' refers to risks that are 1 degree of causal separation from unsurvivability in the diagram, whereas
				\item ``Tier 2'' refers to risks that would generate Tier 1 risks, and are hence 2 degrees of causal separation from unsurvivability.
			\end{itemize}

	\riskdef{Tier}{mpais}{MPAI deployment events}
		\fl{3}
			This section outlines specific scenarios wherein an MPAI deployment event could occur.\footnotemark{}
			\footnotetext{Such scenarios have been considered extensively by philosopher Nick Bostrom \citep{bostrom2014superintelligence} under more specific assumptions defining ``superintelligent'' AI systems.
			}
			Because this report is targeted at AI developers, the MPAI deployment events considered here have been classified according to the following exhaustive decision tree centered on the hypothetical AI developers involved in building the MPAI:
			\newcommand{\detype}[1]{\item \textbf{#1.}}
			\begin{enumerate}[label=\alph*.]
				\detype{\riskref{uncdev}} Was there no single AI development team who was primarily responsible for developing the MPAI technology?  If so, classify the MPAI deployment event as arising from \emph{uncoordinated MPAI development}.

				Otherwise, in the remaining risk types one can assume the developers of the MPAI constitute a single team, and further subdivide scenarios based on the relationship of that team to the MPAI deployment event:

				\detype{\riskref{urprep}} Prior to the technology being deployed and prepotent, did the development team fail to recognize that the technology would be or become prepotent? If so, classify as an \emph{unrecognized prepotence} event; otherwise consider:

				\detype{\riskref{urmis}}
				Prior to the technology being deployed and misaligned, did the development team fail to recognize that the technology would be or become misaligned?  If so, classify as an \emph{unrecognized misalignment} event; otherwise consider:

				\detype{\riskref{invdep}} Did the MPAI deployment event occur without the voluntary permission of the development team responsible for creating it?  If so, classify as an \emph{involuntary MPAI deployment} event; otherwise classify as:

				\detype{\riskref{voldep}} The MPAI deployment was voluntarily permitted by its developers.
			\end{enumerate}

		\noindent The remainder of \secref{mpais} examines these risk types in more detail.

		\subriskdef{uncdev}{uncoordinated MPAI development}

			This risk type comprises MPAI deployment events that arise from \emph{uncoordinated MPAI development} in the sense that no one research team is solely responsible for having developed the MPAI.

			As an example of uncoordinated MPAI development, suppose Group A deploys a powerful AI system for managing an online machine learning development system, which is not prepotent because it lacks some key cognitive ability.
			Then, suppose that around the same time, Group B releases an open source algorithm that Group A's system learns about and uses to acquire the key cognitive ability, thereby becoming prepotent.

			In this situation, because no coordinated effort has been made to align the resulting prepotent AI system with human survival, it is relatively likely to be \misaligned, by \hft.
			And, one could argue that neither Group A nor Group B was solely responsible for having developed the MPAI; rather, they failed to coordinate on the combined impact of their development and deployment decisions.  Even if some members of each group were aware that the result of their actions might result in MPAI development, perhaps the local incentives of each group were to continue working on their products nonetheless.  A similar dynamic can be seen in the way separate countries tend to follow local economic incentives to continue producing carbon emissions, despite the potentially dangerous combined impact of those emissions.

			Avoiding this risk type calls for well-deliberated and respected assessments of the capabilities of publicly available algorithms and hardware, accounting for whether those capabilities have the potential to be combined to yield MPAI technology.  Otherwise, the world could essentially accrue ``AI pollution'' that might eventually precipitate or constitute MPAI.

			The remaining four Tier 1 risk types will focus on the knowledge and intentions of ``the developers'' of a hypothetical MPAI technology, such as whether the prepotence or misalignment of the technology was known or intended in advance.  By contrast, for an MPAI deployment scenario where the developers of the technology are too poorly coordinated to have a clear consensus on whether it will be prepotent or misaligned, the present risk type---uncoordinated MPAI deployment---may be a better descriptor.

		\subriskdef{urprep}{unrecognized prepotence}
			This risk type comprises MPAI deployment scenarios where the prepotence of the relevant AI technology was unrecognized prior to it being deployed and prepotent.

			Examples of this risk type can be divided into two natural sub-cases:
			\renewcommand{\textit}[1]{\item \textbf{#1:}}
			\begin{itemize}
				\textit{deployment, then prepotence} The AI technology in question is not prepotent at the time of its initial deployment, but later becomes prepotent in a manner that surprises its developers.
				For instance, this could happen if the developers are insufficiently informed of the system's relationship with the world after its initial deployment, if they are informed but insufficiently attentive to the information, or if they are informed and attentive but unable to deduce that the system will become prepotent.
				\textit{prepotence, then deployment} The AI technology in question is prepotent prior to its deployment, but the developers fail to recognize this at deployment time.  For instance, this could happen if the developers did not attempt to assess the prepotence of the technology, or somehow failed to complete an accurate assessment.
			\end{itemize}

			\noindent These sub-cases share an important feature in common: an AI technology with unrecognized prepotence is relatively likely to turn out to be MPAI.  For, suppose an AI development team deploys an AI technology that turns out to be or become prepotent in some way that they did not expect.
			Because of their faulty understanding of the system's capacity for impact, their safety efforts would have been undertaken under invalid assumptions.
			From there, by \secref{humanfragility} there are numerous pathways through which the system's unstoppable transformative impact might be unsurvivable.  Hence, unrecognized prepotence comes with an increased likelihood of  unrecognized misalignment.

			Avoiding this risk type calls for a rigorous scientific theory to understand and recognize when an AI system might be or become prepotent.  An important way in which the prepotence of an AI technology could go unrecognized is if the system exhibits \emph{behavior likely to obfuscate the full breadth of its capabilities}, thereby prompting developers to mistakenly deploy it as a non-prepotent system. Such behavior could result from a selection process that favors AI systems that somehow obfuscate capabilities that humans would consider dangerous.  Capability obfuscation could also arise from a system with social reasoning and planning capabilities that learns, in pursuit of real-world attainment of its assigned objective, to ``work around" human measures to prevent the deployment of prepotent systems.  The latter case could be viewed as an instance of ``intentional deception'' by the system, although attribution of intention is not necessary to describe this general class of phenomena.  In any case, an adequate theory for understanding and recognizing prepotence must account for the possibility of such systems systematically obfuscating their prepotence.

		\subriskdef{urmis}{unrecognized misalignment}
			This risk type comprises MPAI deployment scenarios where the misalignment of the relevant AI technology is unrecognized by its developers prior to it being deployed and misaligned.  Like unrecognized prepotence, unrecognized misalignment can occur whether the misalignment occurs before or after the technology is initially deployed.

			For example, suppose some team of AI developers build a prepotent AI system that they realize or suspect is prepotent, with the intention of using it for some positive and permanently transformative impact on the world.
			There is some risk that the developers might mistakenly overestimate the system's alignment, and hence fail to recognize that it is or will become MPAI.  And, just as with prepotence, an important way misalignment could go unrecognized is if the system itself deceives humans into thinking it is aligned.

			Avoiding this risk type calls for a rigorous scientific discipline for aligning powerful AI systems with human interests and existence, and for recognizing potential misalignment in deployed systems, including systems that may be able to systematically deceive humans regarding their misalignment.

		\subriskdef{invdep}{involuntary MPAI deployment}
			This risk type comprises MPAI deployment events that are \emph{involuntary} on the part of the technology's developers, i.e., occurring against the direct intentions of the team who developed the relevant AI technology.

			For example scenarios, let us focus on cases where the developers recognize that the MPAI deployment event is forthcoming before it happens (since \risknums{urprep,urmis} already cover MPAI deployment events involving unrecognized prepotence and unrecognized misalignment).
			These scenarios can be further organized according to whether an MPAI technology becomes deployed (``release'' events) or an already-deployed AI technology becomes MPAI (``conversion'' events):

			\renewcommand{\textit}[1]{\item \textbf{#1:}}
			\begin{enumerate}
				\textit{MPAI release events (involuntary)}
				An existing MPAI technology somehow becomes deployed without the voluntary consent of its developers.
				For instance, consider a well-meaning team of developers who have created an AI technology that they suspect is both prepotent and \misaligned, and are now conducting experiments on the technology to learn more about the risks it could present.
				In such a scenario, at least some security measures would likely be in place to prevent the technology from being deployed against the intentions of the developers, but those measures could fail in some manner.  The failure could involve:
				\begin{enumerate}
					\textit{Accidental release} An existing MPAI technology is released accidentally by its development team, enabling others to deploy it without the developers' consent.  No one on the development team intentionally causes the release of the technology; it is merely a haphazard mistake on the part of the developers.  This sort of event could be analogized to a nuclear power-plant meltdown: someone is responsible for the accident, but no one did it on purpose.

					\textit{Unauthorized release} An existing MPAI technology is obtained by someone other than its developers, against the developers wishes.  For instance:
								\begin{enumerate}
									\item Hackers obtain access to the technology's code base and deploy it, perhaps without knowledge of its misalignment or prepotence.
									\item Physical force is used to obtain access to the technology's code base, such as by a military or terrorist group, who then go on to deploy the technology, perhaps without knowledge of its misalignment or prepotence.
									\item A running instance of the AI technology acquires its own deployment as a goal, and finds a way to achieve deployment without its developers' permission.
								\end{enumerate}
				\end{enumerate}

				\textit{MPAI conversion events (involuntary)} An AI technology is deployed and is later converted into MPAI by certain post-deployment events that were not intended by the technology's developers.

				The conversion could be caused by interactions with the relevant AI system(s), or by failures entirely external to the system(s):

					\begin{enumerate}
						\textit{Conversion by uncontrolled interactions}  The developers did not establish adequate controls for post-deployment interactions with the technology, and those interactions convert the technology into MPAI.
						\textit{Conversion by external failures}  Humanity's collective capacity to control or otherwise survive the impact of the technology somehow decreases after its deployment (say, due to a conflict between humans that destroys resources or coordination), and systems using the technology do not adjust their behavior accordingly, becoming MPAI by virtue of humanity's increased vulnerability rather than by changes internal to the technology itself.
					\end{enumerate}
			\end{enumerate}

			Avoiding this risk type calls for measures enabling well-meaning AI developers to recognize and prevent the use of their inventions in ways that might harm society.

		\subriskdef{voldep}{voluntary MPAI deployment}

			This risk type comprises scenarios where an MPAI deployment event is triggered voluntarily by the developers of the MPAI technology.
			Even if the majority of the AI research and development community develops methods that make it easy to align powerful AI systems with human interests and existence, and existing powerful AI systems are protected from falling into the wrong hands, it may be possible for some misguided persons to develop and deploy MPAI technology on their own for some reason.  For example,
			\renewcommand{\textit}[1]{\item \textbf{#1:}}
			\begin{enumerate}
				\textit{Indifference} Persons unconcerned with the preservation of the human species develop and deploy a powerful AI system in pursuit of values that will yield human extinction as an inevitable side effect.
				\textit{Malice} A military or terrorist organization develops MPAI technology with the misguided hope of controlling it to threaten particular adversaries.
				\textit{Confusion} One or more AI developers that would not normally ignore or threaten human welfare becomes convinced to deploy an MPAI technology by morally confusing arguments.  Perhaps the arguments are produced by other indifferent or malicious persons, or perhaps by an AI system.
			\end{enumerate}

			Avoiding this risk type calls for measures to prevent powerful AI technologies from being developed and deployed by misguided persons.  Some of these preventive measures could also guard against instances of \riskref{invdep} that would arise specifically from unauthorized access to near-prepotent systems or code bases.

	\riskdef{Tier}{hazardoussocial}{hazardous social conditions}
		\fl{3}
		This section examines types of social conditions that exacerbate the likelihood of \risksnum{mpais}.  Unlike the typology of \risksnum{mpais}, the following typology is non-exhaustive.

		\subriskdef{races}{unsafe development races}
			This risk type comprises scenarios wherein two teams are in competition to develop powerful AI systems with the hope that the more successful of the teams might achieve wealth or power from the deployment of their system, and where each team is motivated by their competitive incentives to take risks that would be considered irresponsible from a societal perspective.  Even if each competing team knows about the challenges of aligning their systems with human existence, they might be tempted to divert resources away from safety measures in order to best the competition with superior capabilities.

			This sort of development race exacerbates the probability of \risksref{mpais}, specifically \risksrefs{uncdev,urprep,urmis}.
			This conclusion has also been argued by \citet{bostrom2014superintelligence} and \citet{armstrong2016racing}.
			Moreover, \riskref{invdep} is increased because security measures against unauthorized or accidental deployments are more difficult to implement in a hurry, to reduce the chances of hazardous post-deployment interactions with the system.
			Finally, if one of the development groups is a military or terrorist organization, they might decide to deploy their technology in a desperate attempt to overthrow their competitors by force.
			This would constitute a \riskref{voldep}.

			Avoiding this risk type calls for measures to reduce incentives for competing AI development teams to take socially unacceptable safety risks in the course of developing and deploying their technology.

		\subriskdef{econ}{economic displacement of humans}
			This risk type comprises scenarios wherein most human persons have no power to bid for the continued preservation of the human species, because humans have mostly been economically displaced by AI systems.

			The possibility of an unemployment crisis arising from automation has been discussed by numerous authors, e.g., \citet{joy2011why}, \citet{ford2013could}, \citet{brynjolfsson2014second}, \citet{brynjolfsson2014labor},

			\citet{russell2015research},
			\citet{chace2016economic},
			and \citet{frey2017future}.
			A gradual replacement of human workers by AI systems could lead to an economy wherein most trade and consumption is carried out by non-human entities.
			This is a bleak future from the point of view of many, but not yet a global threat to human survival.

			To see how this trend would constitute an existential risk if taken far enough, consider a scenario where human institutions have all been out-competed and replaced by autonomous corporations.
			Such autonomous corporations could be deployed by idealistic individuals looking to increase transparency or efficiency in certain industries, such as finance, supply chain management, or manufacturing.  Perhaps autonomous corporations could eventually also engage in primary resource industries such as mining, oil drilling, or forestry, which could supply raw materials to corporations in other industries.  If some combination of corporations turned out to be capable of sustaining and expanding an economy entirely without humans, humanity would lose its trade leverage for influencing their activities.  This could constitute prepotence for the collective machine economy, as was argued by \citet{turing1951intelligent} in ``Intelligent Machinery, A Heretical Theory''.

			Given the machine economy's prepotence, misalignment is relatively likely to follow.  By the \subsectitle{humanfragility} argument of \secref{keyconcepts}, the side effects of a prepotent machine economy---in terms of resource consumption, waste emissions, or both---would be hazardous to humans \emph{by default}, unless the leading autonomous corporations coordinated in such a way as to provide or at least allow the equivalent of social assistance and environmental protection to humans, for reasons not driven by the humans' economic output.

			The potential for economic take-off of a self-sustaining fully mechanized economy thus constitutes a \riskref{uncdev}: the combined activities of the machine economy could be prepotent while no single human decision-making entity would be responsible for the development and deployment of that economy.  With no one in particular being responsible for the deployment, coordinated safety measures might be sorely lacking, yielding a serious risk to humanity by \hft.

			Avoiding this risk type calls for the development of coordination mechanisms to ensure the continued economic relevance of both humans and human-aligned AI systems.

		\subriskdef{humenf}{human enfeeblement}
			This risk type comprises scenarios where humans become physically or mentally weaker as a result of assistance or interference from AI systems.

			For example, if AI-driven machines replace most or all forms of human labor, it is possible that humans will become generally physically and mentally weaker as a result.
			Human enfeeblement is a serious risk to the value of human society as it currently exists.  In particular, if the impairment of decision-making capacities of human individuals and institutions leads to a mismanagement of hazardous technologies inherited from previous generations, the chances of \risksrefs{uncdev,urprep,urmis,invdep} might be increased, as well as other existential risks from non-AI technologies.

			Avoiding this risk type calls for the observance of collectively agreeable metrics for human cognitive abilities such as attention span, numeracy, literacy, working memory, and interpersonal skills, as well as the continued observance of physical health metrics, so that any onset of widespread cognitive or physical declines would be noticed.  Some effort in this direction can already be seen in research broadly construed as examining the impact of internet and media technology on mental and physical health
			\citep{cain2010electronic,strasburger2010health,kuss2012internet,hale2015screen,lemola2015adolescents,demirci2015relationship}.  However, much of the work in this area has been observational rather than experimental, making it currently difficult to identify clear and valuable public policy recommendations.
			Meanwhile, as AI becomes an increasingly prevalent determinant of how and when people use technology, the urgency and importance of understanding its causal impact on human health and vigor will only increase in significance.

	\subriskdef{discourseimpairment}{ESAI discourse impairment}
			This risk type comprises scenarios where human persons and institutions fail to collectively assess and address existential risks from artificial intelligence, as a result of difficulties encountered in communicating about existential safety.	 There are numerous ways in which discourse on existential safety for artificial intelligence (``ESAI'') could be become impoverished:

			\begin{itemize}
				\item (alarmism) If too many debates are raised in the name of existential safety that on reflection turn out to have been unreasonable concerns, then discussions of ESAI could come to be seen as inflammatory and counterproductive to discuss, by the proverbial  ``cry wolf'' effect \citep{breznitz2013cry}.

				\item (politicization) The topic of ESAI could someday become politicized, in the sense that arguments for or against existential safety issues can become tightly linked with one or more political ideologies.
				For example, beliefs around the issue of climate change---an existential safety issue---are currently strongly correlated with political party affiliations \citep{mccright2011politicization,hart2012boomerang}.  If ESAI becomes similarly politicized, the quality of available discourse on the topic could be reduced.  This possibility has also been argued by \citet{baum2018superintelligence}.
				\citet{brysse2013climate} argues that climate scientists may systematically underreport their risk estimates so as to avoid seeming alarmist, and \citet{taylor1992we} argues that such reputational and political forces can even affect what problems scientists choose to pursue.

				\item (information security concerns) If transmitting information about ESAI between AI researchers comes to be viewed as risking the dissemination of \emph{information hazards}  \citep{bostrom2011information}---i.e., information that is too dangerous to be widely shared---then collaborative research efforts to improve existential safety could be impoverished.

				\item (association with science fiction) If planning for the safer development of powerful AI systems comes to be seen as evoking exciting or entertaining fictional narratives of the future, ESAI might come to be taken less seriously than would be appropriate given its potential importance.  \citet{rees2013denial} has argued that ``In a media landscape saturated with sensational science stories and `end of the world' Hollywood productions, it may be hard to persuade the wide public that real catastrophes could arise\ldots''.
				\end{itemize}

			\noindent Such discourse impairments not only impoverish group-scale decision processes, but also diminish opportunities for individuals to improve their own judgment through discussions with others.

			Prevention of this risk type calls for measures attending to whether AI researchers feel comfortable honestly expressing, to each other and the public, their views on the potential impacts of artificial intelligence, and measures attending to whether public consensus and expert consensus on risks from artificial intelligence are in agreement.  The present authors have not yet put forward any technical AI research directions that would benefit such measures, but social science research in this area might be valuable for helping society to continue making reasonable and legitimate risk/reward trade-offs in the governance of AI technology.

	\subsection{Omitted risks}
		\fl{3}
			Several other extremely costly potentialities for human society are conspicuously absent from the remainder of this document:

			\paragraph{Hazardous deliverables.}  Supposing humanity develops highly advanced AI systems, those systems could aid humans in developing other technologies which would themselves pose significant global risks to humanity.
			Nuclear weapons, chemical weapons, and bioweapons are examples of such hazardous technologies that have been developed in the past, without the aid of AI technology.

			Risks arising from the development of more such hazardous technologies in the future---with or without the assistance of AI in the development process---are not explicitly addressed by the technical directions of this report.
			However, such risks could be addressed by related principles of safe and ethical oversight.

			\paragraph{Suboptimal futures.}  More generally, it has been argued that futures where humans exist, but are not flourishing to the degree one would hope, should be considered existential risks or at least be treated with the same degree of severity as human extinction risks.
			For example, \citet{bostrom2013existential} considers ``permanent stagnation'' and ``flawed realization'' scenarios, wherein human civilization respectively either ``fails to reach technological maturity'' or ``reaches technological maturity in a way that is dismally and irremediably flawed''.  These scenarios are excluded from this report for two reasons.
			The first reason is to avoid debate in this report the issue of what constitutes a suboptimal future, as discussed somewhat in \secref{omitteddebates}.  The second reason is that these other risks do not naively belong under the heading ``existential'', so most readers are not likely to be confused by their omission.

\section{Flow-through effects and agenda structure}\label{sec:flowthrough}
	\fl{2}
		\fl{3}
			\secrefs{ss,sm,ms,ms} of this report may be viewed as a very coarse description of a very long-term research agenda aiming to understand and improve interactions between humans and AI systems, which could be viewed as ongoing throughout the full historical development of artificial intelligence, multi-agent systems theory, and human-computer interaction.

			How can one begin to account for the many ways in which progress in different areas of AI research all flow into one another, and how these flow-through effects relate to existential risk?  The task is daunting.  To organize and reduce the number of possible flow-through effects one would need to consider, the research directions in this report have been organized under the subsections of \secrefs{ss,sm,ms,mm}, which themselves are related by a lattice structure depicted in \figref{lattice}.

			\subsection{From single/single to multi/multi delegation}\label{sec:ssmm}

			Research on single/single delegation can be expected to naturally flow through to a better understanding of single/multi and multi/single delegation, and which will in turn flow through to a better understanding of multi/multi delegation.

			\begin{figure}[H]
				\centering
					\includegraphics[scale=0.44]{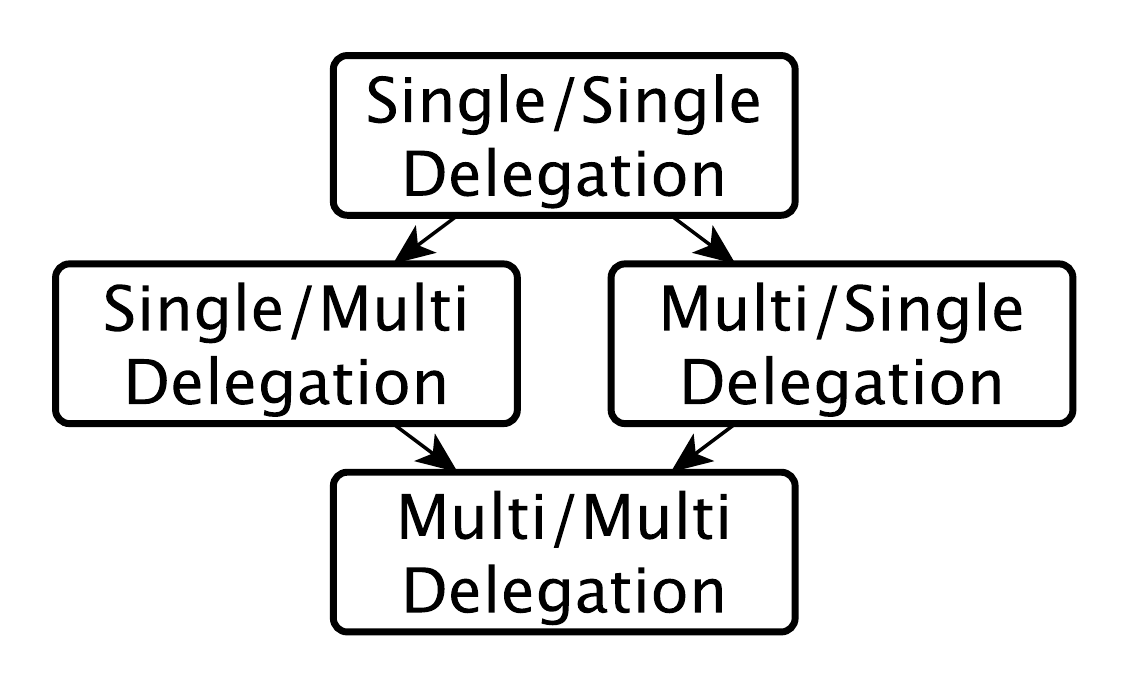}
					\caption{``discovery flow-though'' effects between sections.}
					\label{fig:singlemulti}
			\end{figure}

	\subsection{From comprehension to instruction to control}
		\fl{3}
			\secrefs{ss,sm,ms} are each divided into subsections regarding the human ability to either \emph{comprehend} AI systems, \emph{instruct} AI systems, or \emph{control} AI systems, as defined in \secref{cic}.  Within each section, comprehension research can be expected to benefit but not subsume instruction research, and comprehension and instruction research can be expected to benefit but not subsume control research.

			\begin{figure}[H]
				\centering
				\includegraphics[scale=0.44]{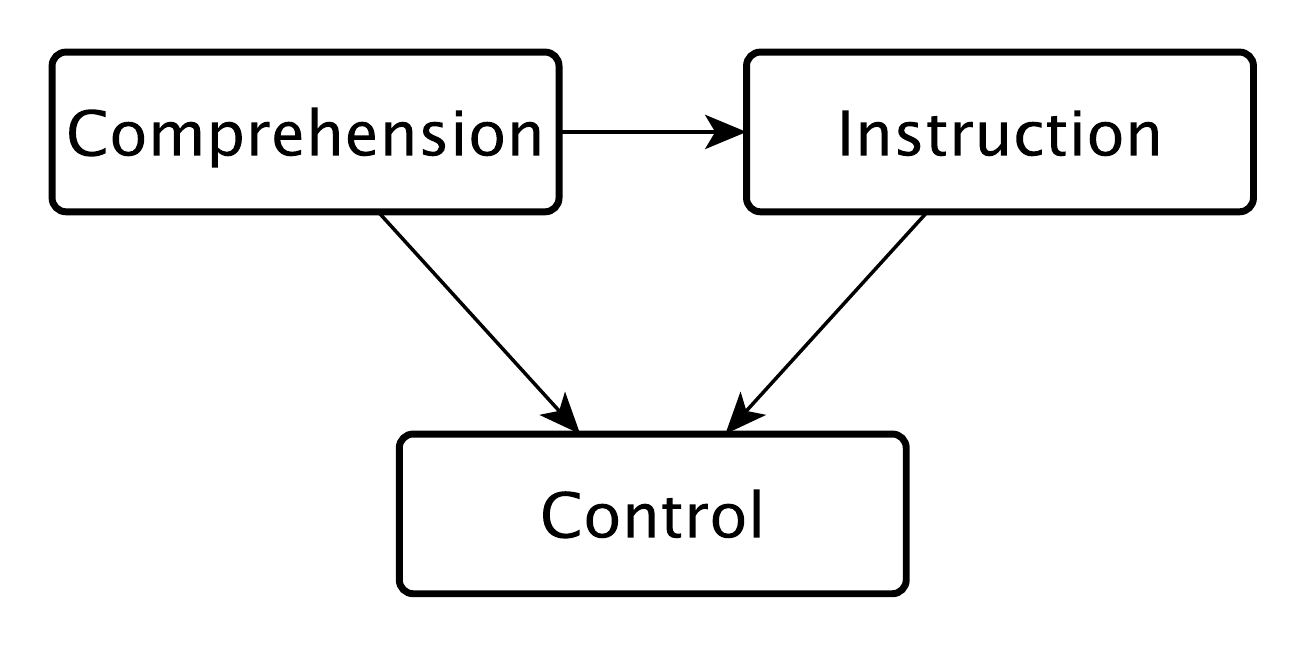}
				\caption{subsection lattice, depicting ``discovery flow-through'' effects between subsections within each section.}
				\label{fig:lattice}
			\end{figure}

			\subsection{Overall flow-through structure}
			Put together, the flow-through effects discussed above combine to yield the lattice depicted in \figref{lattice} below.  This lattice defines the overall organizational structure for \secrefs{ss,sm,ms,mm}, and summarizes the bulk of the ``discovery flow-through'' effects that should be expected between research directions in this report.
			Whenever a research direction would contribute to multiple corners of this subsection lattice, it is discussed under the earliest relevant subsection, leaving its usefulness to subsections further down in the lattice to be implied from the document structure.

						\begin{figure}[H]
							\centering
							\includegraphics[scale=0.44]{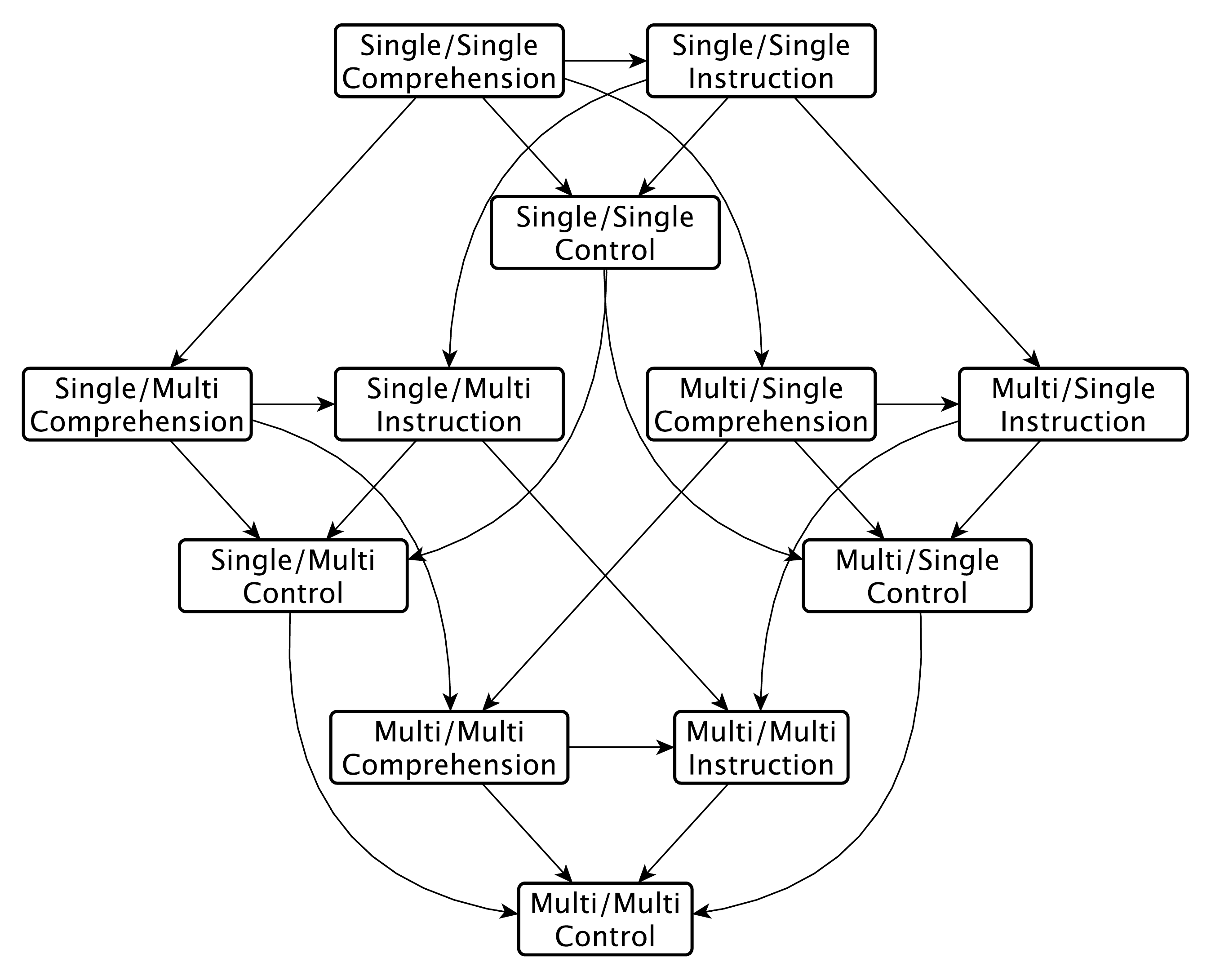}
							\caption{subsection lattice, depicting ``discovery flow-through'' effects between research directions in this report.}

							\label{fig:lattice}
						\end{figure}

	\subsection{Research benefits vs deployment benefits}\label{sec:researchbenefits}
		\fl{3}
			Suppose that a major breakthrough is made in single/single delegation, but that multi/multi delegation remains poorly understood. If the breakthrough leads to the release of several AI systems each intended to serve a different human stakeholder, then a multi/multi interaction scenario immediately results.  In such an event, the R\&D process that designed the AI systems will not have accurately accounted for the interaction effects between the multiple humans and systems.  Hence, many errors are likely to result, including safety issues if the AI systems are sufficiently impactful as a collective.

\begin{figure}[H]
				\centering
					\includegraphics[scale=0.44]{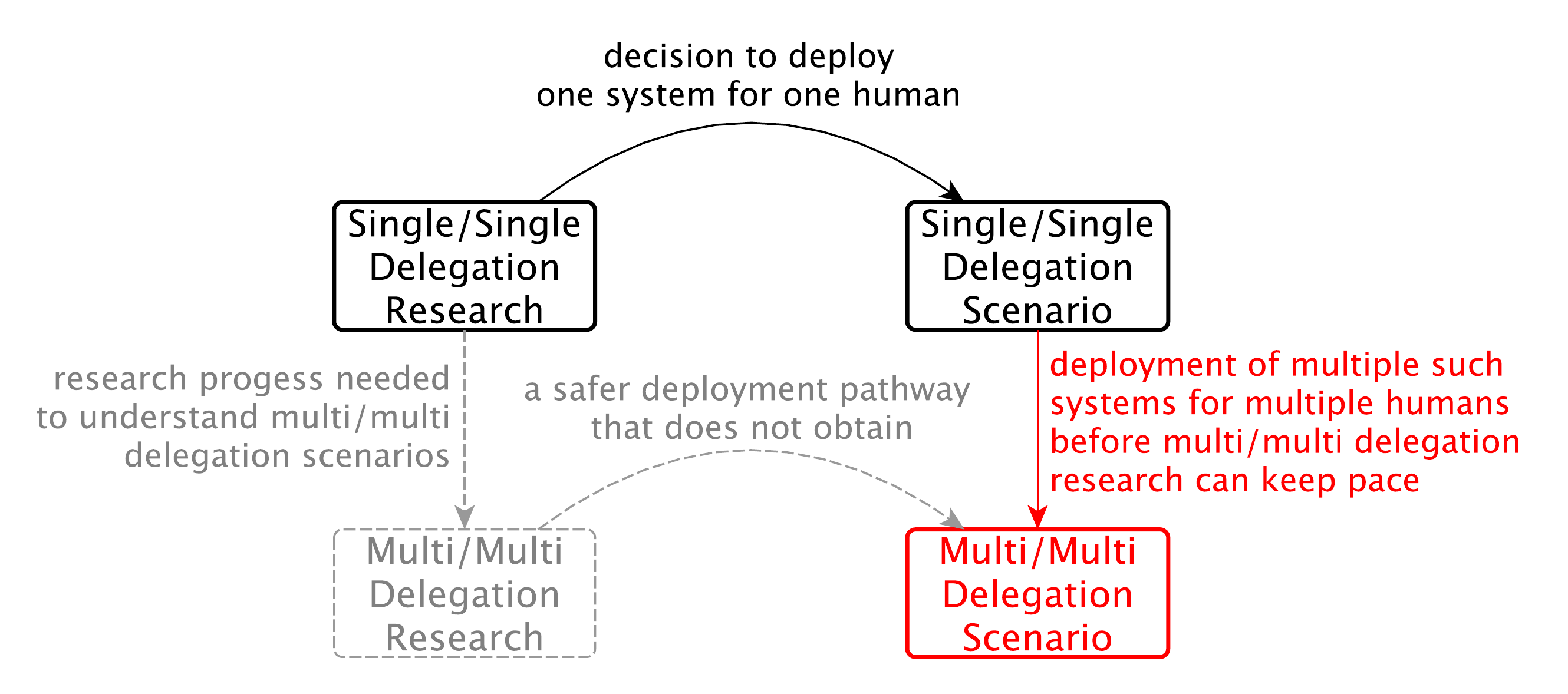}
					\caption{Research progress on single/single delegation can easily have negative flow-through effects on real-world multi/multi deployment scenarios if multi/multi delegation research does not keep pace.}
					\label{fig:researchdeployment}
			\end{figure}

			In the preceding scenario, single/single research flows through to a harm, rather than a benefit, in a multi/multi deployment setting.
			Such scenarios can make it very confusing to keep track of whether earlier developments will help or hinder later developments.  How can one organize one's thinking about such flow-through effects?  One way to reduce confusion is to carefully distinguish \emph{research benefits} from \emph{deployment benefits}.	 While research on earlier nodes can be reasonably expected to benefit \emph{research} on later nodes, the opposite effect can hold for \emph{deployment} scenarios on later nodes. This happens when research on an earlier node results in a premature deployment event in a setting where research on a later node was needed to ensure proper functioning.  For instance, \figref{researchdeployment} summarizes a causal pathway whereby research on single/single delegation could robustly lead to real-world errors in multi/multi delegation.

			Of course, it is common sense that the premature distribution of a powerful new technology can be hazardous.  However, combined with the observation that single/single systems can easily be replicated to yield a multi/multi interaction scenario, the potential for premature deployment  implies that an understanding of multi/multi delegation for powerful systems may be needed in short order after the development of any powerful single/single delegation solutions.  For any AI technology with the potential for global impact, this observation should not be taken lightly. Society may typically learn to correct premature deployment errors through experience, but an error that yields a human extinction event is not one that we humans can learn from and correct later.

	\subsection{Analogy, motivation, actionability, and side effects}
		\fl{3}
			In the next few sections, the reader may soon notice a series of repeated sub-headings, intended to suggest a methodology for thinking about long-term risks.  The intended meaning behind these subheadings will be as follows:

			\renewcommand{\textit}[1]{\item \textbf{``#1''.}}
			\begin{itemize}

			\textit{\Socwords} These subsections are post-hoc analogies for introducing each research direction by comparing desired AI system properties with typical human properties.
			The analogies can only be fitting to the extent that AI systems might be designed to operate according to similar principles as humans.  Hence, the motivation and actionability subsections (below) aim to give more precise illustrations that are intended to expand, clarify, and supersede these analogies.

			\textit{\Motwords} These subsections explain the final causal pathway through which a given research direction could be used to reduce existential risk.
			In aggregate, this content is intended to illustrate just some of the many technical and social mechanisms through which AI research and existential safety are intertwined.
			Motivations for some sections may be directly at odds with other sections.  At best this suggests a hedged portfolio of approaches to existential safety; at worst, some approaches may need to be cut short if they present serious negative externalities.

			\textit{\Instwords} These subsections explain how a given research direction could be steered and applied to benefit other research directions in this report.

			\textit{\Actwords} These subsections aim to provide illustrative examples of existing work relevant to a given research direction.  This report falls woefully short of providing fair and comprehensive overviews of the large corpora of work relevant to each direction, and for this the authors apologize in advance.

			\textit{\Csewords} These subsections examine ways in which particular research ideas could be taken in directions that would be problematic from an existential safety perspective.  The fact that many research directions are ``dual purpose'' in this way seems unavoidable: when examining capabilities relevant to existential risk, there is always the possibility that poor judgments about how to intervene on those capabilities could make matters worse.

		\end{itemize}

\section{Single/single delegation research}\label{sec:ss}

	\fl{2}
		\fl{3}
			This section begins our examination of research directions relevant to existential safety in the delegation of tasks or responsibilities from a single human to a single AI system.

			Consider the question: how can one build a single intelligent AI system to robustly serve the many goals and interests of a single human?
			Numerous other authors have considered this problem before, under the name ``alignment''.   For a diversity of approaches to AI alignment, see \citet{soares2014aligning,taylor2016alignment,leike2018scalable}.

			The AI alignment problem may be viewed as the first and simplest prerequisite for safely integrating highly intelligent AI systems into human society.  If we cannot solve this problem, then more complex interactions between multiple humans and/or AI systems are highly unlikely to pan out well.  On the other hand, if we do solve this problem, then solutions to manage the interaction effects between multiple humans and AI systems may be needed in short order.

			(Despite the current use of the term ``alignment'' for this existing research area, this report is instead organized around the concept of \emph{delegation}, because its meaning generalizes more naturally to the multi-stakeholder scenarios to be considered later on.
			That is, while it might be at least somewhat clear what it means for a single, operationally distinct AI system to be ``aligned'' with a single human stakeholder, it is considerably less clear what it should mean to be aligned with multiple stakeholders.  It is also somewhat unclear whether the ``alignment'' of a set of multiple AI systems should mean that each system is aligned with its stakeholder(s) or that the aggregate/composite system is aligned.)

			\Soc As a scenario for comparison and contrast throughout our discussion of single/single delegation, consider a relationship between a CEO named Alice who is delegating responsibilities to an employee named Bob:
			\newcommand{\cicitem}[1]{\item ({#1})}
			\begin{itemize}
				\cicitem{comprehension} In order to delegate effectively to Bob, Alice needs some basic understanding of how Bob works and what he can do---Alice needs to \emph{comprehend} Bob to some degree.
				\cicitem{instruction} Alice also needs to figure out how to explain her wishes to Bob in a way that he will understand---to \emph{instruct} Bob.
				\cicitem{control} 	If Bob genuinely wants to enact Alice's wishes as she intends them, that is a good start, but he can still falter, perhaps catastrophically.  Perhaps he might ignore or severely misinterpret Alice's instructions.  So, Alice also needs some systems in place to \emph{control} Bob's involvement in the company if he begins to behave erratically.  For instance, she should be able to revoke his computer system or building access if needed.  As Bob's employer, Alice also maintains the legal authority to fire him, at which point other company employees will typically stop accommodating his plans.
			\end{itemize}

       \Cse There are a number of potentially negative side effects of developing single/single delegation solutions in general, which are included here to avoid repetition:

			 \begin{enumerate}
				 \item (racing) If near-prepotent AI systems are eventually under development by competing institutions, single/single delegation solutions might increase the willingness of the systems' creators to move forward with deployment, thereby exacerbating \riskref{races}.

				 \item (enfeeblement) Widespread consumer dependence on single/single AI systems could lead to \riskref{humenf} if the systems take on so many mental and physical tasks that human capabilities begin to atrophy.

				 \item (misleading safety precedents) Single/single delegation solutions that only work for non-prepotent AI systems could create a false sense of security that those solutions would scale to near-prepotent and prepotent systems, increasing \riskref{urmis}.  For instance, ``just turn it off when it's malfunctioning'' is a fine strategy for many simple machines, but it won't work if the AI system is too pervasively embedded in key societal functions for shutting it down to be politically viable (e.g., food distribution), or if the system will develop and execute strategies to prevent humans from shutting it down even when they want to.

				 \item (premature proliferation) If single/single delegation solutions are deployed broadly without sufficient attention to the multi/multi delegation dynamics that will result, the resulting interaction between multiple humans and/or multiple AI systems could be destabilizing to society, leading to as-yet unknown impacts.  This general concern was discussed in \secref{questioningadequacy}.
			 \end{enumerate}

	\nodedef{Single/single comprehension}{sscomprehension}

		\fl{3}
			Comprehending a human employee is quite different from comprehending an AI system.  Humans have many cognitive features in common, due to some combination of common evolutionary and societal influences.  Therefore, a human may use an introspective self-model as a stand-in for modeling another person---to ``put oneself in someone else's shoes''.  By contrast, artificial intelligence implementations are by default quite varied and operate very differently from human cognition.

			A recent and salient illustration of the difference between machine and human intelligence is the vulnerability of present-day image classifiers to the perturbations that are imperceptible to humans \cite{szegedy2013intriguing}, due the many degrees of freedom in their high dimensional inputs \cite{goodfellow2014explaining}.  For instance, \citet{su2017one} trained an All Convolutional Network to achieve 86\% accuracy on classifying images in the CIFAR-10 database of $32\times 32$ images, and found that 68.36\% of the images could be transformed into a misclassified image by modifying just one pixel (0.1\% of the image), with an average confidence of 73.22\% assigned to the misclassification.  As well, \cite{athalye2017synthesizing} developed a method for constructing physical objects that are deceptive to machine vision but not to human vision.  The method was used to construct a toy replica of a turtle that was misclassified as a rifle from almost all viewing angles, by TensorFlow's standard pre-trained InceptionV3 classifier \citep{szegedy2016rethinking}, an image classifier with a 78.0\% success rate of classifying ImageNet images using the ``top-1'' scoring rule.

			\begin{figure}[H]
				\centering
				\includegraphics[scale=1]{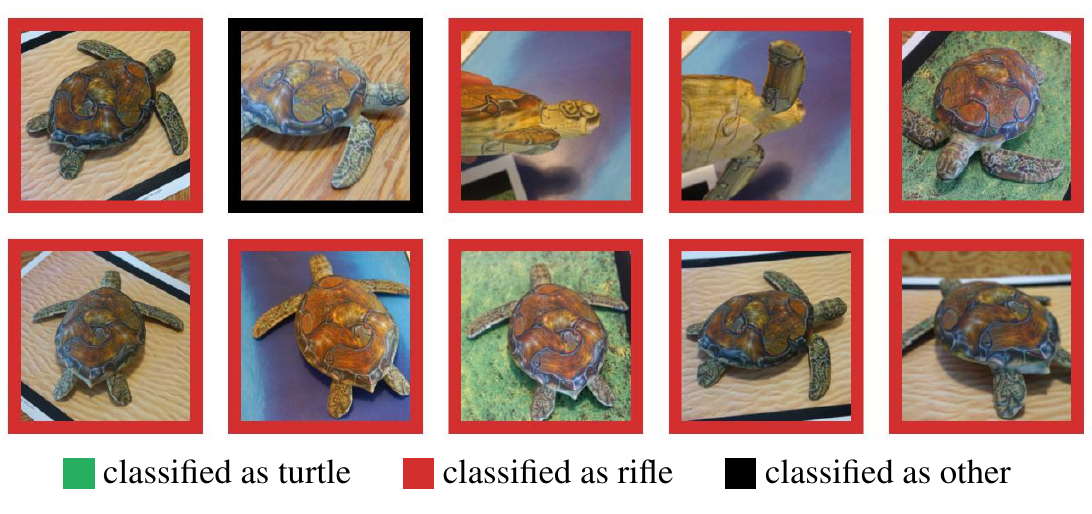}
				\caption{From \emph{Synthesizing robust adversarial examples}, \citet{athalye2017synthesizing}.  Video available at \href{https://youtu.be/YXy6oX1iNoA}{https://youtu.be/YXy6oX1iNoA}.}\label{turtlerifle}
			\end{figure}

			The fact that the image classifier networks in these experiments tend to fail outside their training sets means that the networks themselves have difficulty generalizing.  This alone is not a problem with human/AI comprehension.  However, the fact that the networks fail \emph{in ways that humans find surprising} means that our own understanding of their capabilities is also prone to generalizing poorly.  In particular, humans are unlikely to be able to comprehend AI systems by generalizing from simple analogies to other humans.  As such, research specifically enabling human/AI comprehension will likely be needed to achieve and maintain a reasonable level of understanding on the part of human users and even AI developers.

		\directiondef{transparency}{transparency and explainability}
			One approach to \emph{improving} human/AI comprehension is to develop methods for inspecting the inner-workings of the AI system (transparency), or for explaining the counterfactual dependencies of its decisions (explainability).  These techniques can then be used guide R\&D by helping engineers to better understand the tools they are building.  Perhaps good metrics for transparency and/or explainability could be used as objectives to guide or constrain the training of complex systems.  Together, transparency and explainability are sometimes called ``interpretability''.

			\Soc Businesses are required to keep certain records of decisions made and actions taken in order to remain amenable to public oversight, via government agencies such as the IRS.
			This makes the expenditure of business resources on illegal activities at least somewhat difficult.
			If one views an AI system as somewhat analogous to a corporation---a non-human entity which nonetheless pursues an objective---one might hope to impose analogous internal record-keeping requirements that could be used by humans to detect undesirable cognitive patterns before they would manifest in harmful actions.  Doing so would require a degree of transparency to the humans imposing the requirements.

			\Mot The decision to deploy a powerful AI system should come with a high degree of confidence that the system will be safe, prior to system being deployed.  In particular, the researchers and developers responsible for the system should have enough insight into the its inner workings to determine that it is not misaligned and prepotent.

			Just as business tends to move faster than governance, powerful AI systems will likely eventually operate and make decisions on a time scale that is too fast for humans to oversee at all times.  The more we are able to understand how such systems work, the less likely they will be to surprise us.  Thus, AI transparency improves our ability to foresee and avert catastrophes, whether it be with a powerful AI system or a rudimentary one.  Explainability, or after-the-fact transparency, also serves to improve human predictions about AI systems: aside from explanations informing humans' future predictions about what the system will do, if we impose explainability as a constraint on the system's behavior, we might avert at least some behaviors that would be surprising---to the point of being inexplicable---to the human.  Hence, this direction could apply to reducing \risksrefs{urprep,urmis}, by helping us to understand and predict the prepotence and/or misalignment of a system before its deployment.  Transparency and explainability techniques could also be used to reduce  \risksrefs{invdep}, such as by enabling the inspection any AI-dependent computer security infrastructure in use by AI development teams.

			\Act There is already active research working to make the decisions of modern machine learning systems easier to explain, for instance, \citet{yosinski2015understanding} and \citet{olah2017feature} have created visualization tools for depicting the inner workings of a neural network.
			While the decisions made by a neural network routinely  combine thousands of variables under intricate rules, it is in principle possible to locally approximate arbitrarily complex decisions by identifying a small number of critical input features that would most strongly affect the output under relatively small changes.
			This can be used to provide tractable ``local'' explanations of AI decisions that might otherwise be difficult or impossible for humans to comprehend \citep{Ribeiro2016}.

			Modifying the objective function or architecture of a machine learning system to require a degree of explainability to human inspectors could result in systems that are more legible to human overseers \citep{zhang2018interpretable}.
			One might hope to achieve better generalizability than most earlier work on explainability for AI systems, such as \cite{van2004explainable}.
			Perhaps quantitative models of pragmatic communication \citep{goodman2013knowledge}, wherein speakers and listeners account for one another's goals to communicate and thereby cooperate, could be useful for representing objective functions for explainability.  Or, perhaps sparse human feedback on the understandability of a self-explaining ML system could be augmented with frequent feedback from an automated dialogue state-tracking system, e.g., as studied by \citet{henderson2014second}.
			This would mean repurposing the dialogue state-tracking system to give quantitative feedback on the understandability of the outputs of the self-explaining system, based on the state-tracker's experience with understanding human dialogue.

			Explanations in natural language are an active area of exploration, e.g., by \citet{Hendricks2016}.
			The use of natural language is promising because it is in principle infinitely expressive, and thus opens up a wide space of possible explanations.
			However, their technique currently produces after-the-fact ``rationalizations'' that do not always correspond to the decision procedure actually employed by the AI system in each classification instance.
			Further work on producing natural language explanations should focus on ensuring faithfulness to the underlying reasoning of the system in each decision instance.
			As \citeauthor{Hendricks2016} remark, future models could ``look `deeper' into networks to produce explanations and perhaps begin to explain the internal mechanism of deep models''.
			This objective is critical: the goal of explainability should be to inform human users, never to appease or convince them.
			By contrast, if explanations are optimized merely to convince the human of a foregone conclusion, the system is essentially being trained to deceive humans in situations where it has made a mistake.  Starting down the path of developing such deceptive AI systems might exacerbate \risksrefs{urprep,urmis,invdep}.

			Robotic motion planning is another area of application for transparency.  Using a simple model that treats humans as Bayesian reasoners, robots can adjust their motion using that model to more legibly convey their goal to a human collaborator \citep{dragan2013legibility}, and plan action sequences that will be easier for humans to anticipate \citep{Fisac2016b}.
			Studies of mutual adaptation in human-robot collaboration seek to account for humans' ability to infer and conform to the robot's plan while also expecting it to reciprocate \citep{Nikolaidis2016}.

			To guide progress in any application area, it would be useful to understand the features of transparency and explanation that (1) humans instinctively prefer, and (2) aid in improving human judgment.
			For example, humans tend to prefer certain features in the explanations they receive, including simplicity \citep{lombrozo2007simplicity} and ``exportable dependence'', i.e., usability of the explanation for future predictions and interventions \citep{lombrozo2006functional, lombrozo2010causal}.
			These principles could be quantified in objective functions for training prototypical ``explainable AI'' systems.

			\Cse One possible source of negative side effects could occur if transparency and explaiability (T\&E) tools are developed which enable engineers to build much more complex systems than they would otherwise be able to construct, and if AI systems nearing prepotence turn out to be beyond the reach of the T\&E methods.  So, if T\&E methods are developed which hasten tech development but for whatever reason cannot be applied to ensure the safety of near-prepotent systems, the result would be a precarious situation for humanity.

			\directiondef{calibrate}{calibrated confidence reports}
				This research direction is concerned with developing AI systems which express probabilistic confidence levels that roughly match their success rates in answering questions or choosing good actions.  For instance, among statements that a knowledge database system assigns a 89\%-91\% probability of truth, roughly 90\% of those statements should turn out to be true.  Expressing calibrated confidence to accompany decisions can be seen as a subproblem of transparency or explainability,
				but has other applications as well.

				\Soc
				Suppose Bob sells Alice an investment promising her a 99\% chance of doubling her money by the end of the year.
				However, Alice also learns that among many other investments that Bob has sold claiming ``over a 95\% chance of doubling'', only 65\% actually doubled.  Therefore, even though Bob's ``99\%'' recommendation claims a very good expected value, Alice does not end up believing Bob's explicit claims about the likelihood of success.

				Suppose Alice also receives an investment tip from Charlie, who claims a 99\% chance of doubling in value.  When Alice investigates Charlie's past performance, he has no prior record of either success or failure rates on which to base her judgment.  Alice also investigates Charlie's \emph{reasons} for claiming the investment will double, and  finds that Charlie has done almost no market research, and knows very little about the investment.  Even without a track record, Alice is able to reason that Charlie is probably not very well calibrated, and does not end up believing his claim.

				\Mot Ultimately, the decision to deploy a powerful AI system should come with a well-calibrated prediction that the system is non-prepotent and/or aligned, prior to its deployment.  A working methodology for producing calibrated confidence reports could be used for this, in conjunction with well-codified notions of prepotence and/or misalignment.  That is to say, one could ask a confidence reporting system for the probability that a given AI system is aligned and/or non-prepotent.  Hence, this direction could help to address \risksrefs{urprep,urmis}.

				In addition, reliable confidence reports could be used to temper an AI system's online behavior.   For instance, a powerful AI system could be required to shut down or act conservatively when its confidence in the human-alignment of in its decision-making is low, thereby reducing the probability of catastrophes in general.

				\Inst

				\begin{itemize}

					\item \Dirref{corrigibility}.
					Well-calibrated uncertainty could help an AI system to recognize situations where shutdown or repair is needed.

					\item \Dirref{deference}.  Calibrated confidence reports could be used to trigger increased human oversight when an AI system's confidence in its own good performance is low \citep{hadfield2016off}.

					\item \Dirref{hoal}.
					Correctly identifying its uncertainty also allows an AI system to make better use of a limited supply of human feedback.
					For instance, an RL agent can specifically request feedback about human preferences or rewards when it is less certain \citep{christiano2017deep} or when the information is expected to help it improve its policy \citep{krueger2016active}.

					Thus, to make marginal improvements to scalable oversight, improvements to calibration need only lead to better-than-random decisions about what kind of feedback is useful.

				\end{itemize}

				\Act
				Efforts to represent model uncertainty in deep learning \citep{gal2016dropout,kendall2017uncertainties} are directly applicable to developing well-calibrated confidence reports from AI systems.  There are many recent papers focussed on improving calibration for machine learning models used to make uncertain predictions or classifications \citep{guo2017calibration,lakshminarayanan2017simple,lee2017training,liang2017enhancing,devries2018learning,hafner2018reliable,kuleshov2018accurate}.
				Because of the inevitability of some model misspecification in any system one might build, perfectly accurate calibration may be impossible to achieve in reality.  Thus, it is important to determine when and how one can reliably achieve precise calibration, and when and how awareness of imperfect calibration (in a sense, ``meta calibration'') can be leveraged to improve active learning and corrigibility.
				For instance, \citet{liu2015shift} propose an active learning approach that accounts for a model's inductive bias and thereby outperforms random selection of queries.
				Meanwhile, understanding the implications of miscalibration can motivate future work by suggesting applications of calibration solutions.
				As a case study, \citet{carey2017incorrigibility} provides examples of how misspecification of an RL agent's priors in an ``off-switch'' game \citep{hadfield2016off} can lead to incorrigibility of the RL agent, via miscalibration about when to defer to the human.

				\Cse The potential negative side effects of this work are similar to those of \dirref{transparency}, i.e., the risk that these methods might accelerate tech development without scaling to apply to near-prepotent systems.  One way this could occur is if calibrated safety reports are fundamentally more difficult to produce for a system with the capacity for developing a plan to deceive the safety assessment protocol.  Perhaps this issue, if it arose, could be mitigated with other transparency techniques for detecting if the system is planning to deceive the safety assessment.

		\directiondef{formalverification}{formal verification for machine learning systems}

		  For any safety criterion that one could hope for a powerful AI system to meet, a combination of empirical (experiment-driven) and formal (proof/argument-driven) verification methods might be relevant and useful.  This direction is about bolstering formal methods.

			\Soc When a venture capital (VC) firm chooses to invest in a start-up, they look for formal legal commitments from the company regarding how and when the VC firm will be entitled to redeem or sell its shares in the company.  Suppose instead the start-up offered only a word-of-mouth agreement, appealing to fact that the VC firm has never been swindled before and are hence unlikely to be swindled now.  The VC firm would likely be unwilling to move forward with the actual transfer of funds until a formal, legally enforceable agreement was written and signed by the start-up.  With the written agreement, the firm can develop a greatly increased confidence that they will eventually be entitled to liquidate their investment.

			\Mot At the point of deploying any powerful AI system or system component that could result in prepotence and/or misalignment, reliance entirely on empirical tests for alignment and/or controllability is likely to be unsatisfying and perhaps even reckless.  Indeed, the test ``will this system overthrow human society after it is deployed?'' is not an experiment one would like to actually run.

			But how can one know the outcome of an experiment before running it?  In other high-stakes engineering endeavors, such as building a bridge or launching a rocket, one is never satisfied with merely testing the components of the bridge or rocket, but also use formal arguments from well-established principles of physics to establish bounds on the safety of the system. Such principled analyses serve as a guide for what can and cannot be concluded from empirical findings, e.g., ``if force X amounts to less than 100 Newtons and force Y amounts to less than 200 Newtons, then in combination they will amount to less than 300 Newtons''.  Laying out such arguments in an explicit form allows for the identification of key assumptions which, if violated, could result in a system failure (e.g., a bridge collapse, or a rocket crash).

			As AI systems become more powerful, persons and institutions concerned with risks will expect to see similarly rigorous formal arguments to assess the potential impacts of the system before deployment.  Some would argue that such assessments should already have been carried out prior to the deployment of widespread social media technology, given its pervasive impact on society and potential to affect the outcome of national elections.  Techniques and tools for automatically generating formal assessments of software and its interaction with the real world will thus be in increasing demand as more powerful AI systems are developed.

			\Act Since many present-day AI systems involve deep learning components, advances in scalable formal verification techniques for deep neural networks could be potentially very valuable.
      For instance,
			\citet{dvijotham2018dual} have developed an anytime algorithm for bounding various quantities definable from network weights, such as robustness to input perturbations.  \citet{katz2017reluplex} have adapted the linear programming simplex method for verifying or refuting quantifiable statements about ReLU networks.  \citet{akintunde2018reachability} and \citet{lomuscio2017approach} have begun developing methods for reachability analysis of feed-forward ReLU neural networks.
			\citet{selsam2017developing} have developed an automated proof assistant for generating machine-checkable proofs about system performance as a step in the engineering process.  Their training system, Certigrad, performed comparably to Tensorflow.

			For even more rigorous verification, one must also consider assumptions about the so-called \emph{trusted computing base (TCB)}, the core software apparatus used to interpret and/or compile code into binaries and to write and verify proofs about the code. \citet{kumar2018software} argue that verification with a very small TCB is possible with appropriate adjustments to the programmer's workflow, and that such workflows are already possible in systems such as CakeML \citep{kumar2014cakeml} and {\OE}uf \citep{mullen2018oeuf}.

			In order to formally specify societal-scale safety criteria that formal verification tools would go on to verify for powerful AI systems, input may be needed from many other research directions, such as \dirrefs{humancognitive,bel,rigorouscoordination}.

			\Cse There is an interesting duality between design and verification in the creation of AI systems by human developers, that can be seen as analogous to the duality between training and testing in the creation of image classifiers by supervised learning algorithms.  Specifically, when some fraction of formal verification specs for an AI system are withheld from the human developers who design and build the system, the withheld specs can serve as an independent test of the system's performance (and hence also the quality of the developers' design process).
			This is similar to how, after a classifier has been ``built'' from a training dataset by a supervised learning algorithm, a separate testing dataset typically serves as an independent test of the classifier's accuracy (and hence also the quality of the learning algorithm).
			Such independent tests are important, because they reveal ``overfitting'' tendencies in the learning algorithm that make past performance on the training data an overly optimistic predictor of future performance on real data.  Conversely, using the entirety of a supervised learning dataset for training and none of the data for testing can result in a failure to detect overfitting.

			The analogue for human developers designing AI systems is that including too many automated verifications for the developers to use throughout the design processes enables the developers to fix just the automatically verifiable issues and not other issues that may have been overlooked.  Thus, if one publishes \emph{all} of one's available formal verification methods for testing an AI system's performance, one impoverishes one's ability to perform independent tests of whether the developers themselves have been sufficiently careful and insightful during the design process to avoid ``over-fitting'' to the specs in ways that would generalize poorly to real-world applications.

			This potential side effect of making too many formal verification specs publicly available can be viewed as an instance of \emph{Goodhart's Law} \citep{manheim2018categorizing}:
			``When a measure becomes a target, it ceases to be a good measure.''
			Simply put, if all known proxy measures for safety are made publically available in the form of automated tests, it could become too easy for reseachers to accidentally or intentionally learn to ``cheat'' on the test.  What this means for formal verification methods is that once a useful formal safety verification standard is developed, a non-trivial decision needs to be made about whether to publish reproducible code for running the safety test (making it a ``target''), or to keep the details of the test somewhat private and difficult to reproduce so that the test is more likely to remain a good measure of safety.
			For very high stakes applications, certain verification criteria should always be withheld from the design process and used to make final decisions about deployment.

		\directiondef{aiad}{AI-assisted deliberation}
			Another approach to improving human/AI comprehension is to improve the human's ability to analyze the AI system's decisions or recommendations.

            In this report, \emph{AI-assisted deliberation} (AIAD), refers to the capability of an intelligent computer system to assist humans in the process of reflecting on information and arriving at decisions that the humans reflectively endorse.

			In particular, this might involve aiding the human to consider arguments or make observations that would be too complex for the human alone to discover, or even to fully reason about after the point of discovery.  AIAD can be viewed as being closely complementary with transparency and explainability (T\&E): while T\&E methods aim to present information in a form amenable to human comprehension, AIAD would assist the humans in directing their own thoughts productively in analyzing that information.

			\Soc A busy executive can benefit greatly from the assistance of employees and expert advisors who make it easier for them to evaluate important choices.  At the same time, reliance on deliberative assistance leaves the executive prone to accidental or intentional manipulation by the assistant.

			\Mot It is possible that humanity will collectively insist on relatively simple constraints for any powerful AI system to follow, that would ensure the humans are unlikely to misunderstand its reasoning or activities.  Absent such constraints, humans can be expected to struggle to understand the discoveries and actions of systems which by design would exceed the humans' creative abilities.  The better guidance one can provide to the human overseers of powerful systems, the less likely they will be to overlook the misalignment or prepotence of an AI system.  Hence, AIAD could be used to address \risksrefs{urprep,urmis}.  At the same time, if AIAD technologies are eventually developed, caution may be needed to prevent their use in ways that would accidentally or intentionally deceive or distract humans away from key safety considerations, especially for high-stakes applications that could be relevant to existential risk.  (For instance, present-day social media services employ a plethora of interactive AI/ML systems to capture and maintain user attention, and many people report that these services distract them in ways they do not endorse.)

            \dksays{[ ]changed this:}
			\Inst Improved human deliberation would be directly useful to safety methods that rely on human feedback.

            This includes \dirrefs{preferencelearning,hoal,moderatinghuman}

			\Act

			There is also evidence that automated systems can be used to aid human deliberation on non-technical topics.
			The delivery of cognitive behavioral therapy (CBT) by automated conversational agents over the internet has been found to be somewhat effective for reducing some symptoms of general psychological distress, in comparison with reading an e-book \citep{twomey2014randomized} or simply awaiting an in-person therapist \citep{fitzpatrick2017delivering}.
			One might therefore hypothesize that automated problem-solving agents could assist in the making of stressful or otherwise difficult decisions.

			\citet{christiano2016humans} has proposed a recursive framework for decomposing problems assisting deliberation, recursively named ``Humans Consulting HCH (HCH)''.  This method has undergone some empirical testing by a new research group called \citet{ought2017decomposing, ought2017predicting}.

			\Cse Widespread use of AIAD could lead to unexpected societal-scale effects.  For example, if humans come to rely on AIAD more than their fellow humans to help them deliberate, perhaps trust between individual humans will gradually become degraded.  As well, providing AIAD without accidentally misleading or distracting the human may remain an interesting and important challenge.  To avoid this, it may be necessary to develop an operationalized definition of ``misleading''.

		\directiondef{boun}{predictive models of bounded rationality}
			Both humans and AI systems are subject to bounds on their computational abilities.
			These bounds will likely need to be accounted for, explicitly or implicitly, in predicting what independent and collaborative behaviors the humans and AI systems can or will exhibit.  Ideally, a good model of a boundedly rational decision-making system should be able to predict what sorts of the decisions the are too hard, or sufficiently easy, for the system to make correctly with its given computational resources.

			\Soc When a law school student with a poor memory and slow reading speed fails a final examination, it is apt to attribute their failure to a lack of ability rather than a lack of desire to pass.
			On the other hand, if a student known to have a prodigious memory and a fast reading speed is seen to fail such an exam, it may be more appropriate to infer that they are insufficiently motivated to pass.
			Thus, observing the same behavior from two different humans---namely, failing an exams---lead us to different conclusions about their desires (trying to pass and failing, versus not caring much about passing).
			In this way, thinking informally about a person's mental capabilities is key to making inferences about their desires.

			Conversely, suppose you know your attorney has the best of intentions, but nearly failed out of law school and required numerous attempts to pass the bar exam.
			If a serious lawsuit comes your way, you might be inclined to find a more skilled attorney.

			These situations have at least three analogues for AI systems: (1) humans accounting for the limitations of AI systems, (2) AI systems accounting for the limitations of humans, and (3) AI systems accounting for the limitations of other AI systems.

			\Mot See the instrumental motivations.

			\Inst Numerous directions in this report would benefit from the ability to calculate upper and lower bounds on a given cognitive capacity of a system,
			as a function of the computational resources available to the system (along with other attributes of the system, which are always needed to establish non-trivial lower bounds on performance):

			\begin{itemize}
			\item \Dirref{preferencelearning}.
			Inferring the preferences of a human from their words and actions requires attributing certain failures in their behavior to limitations of their cognition.
			Some such limitations could be derived from resource bounds on the human brain, or even better, on relevant cognitive subroutines employed by the human (if sufficient progress in cognitive science is granted to identify those subroutines).
			\item \Dirref{hoal}.
			The degree of oversight received by an AI system should be sufficient to overcome any tendency for the system to find loopholes in the judgment of an overseer(s).
			A precise model of how to strike this balance would benefit from the ability to predict lower bounds on the cognitive abilities of the overseer and upper bounds on the abilities of the AI system being overseen, accounting for their respective computational resources.
			\item \Dirref{mod}.
			Upper bounds on the collective capabilities of malicious hackers could be used to estimate whether they have sufficient resources to re-train, re-program, or otherwise compromise a powerful AI system or the security protocols surrounding it.
			It would be informative if such bounds could be derived from estimates of the hackers' total computational resources.
			(Although this would not protect against flaws in the assumptions of the designers of the system to be protected, which are the main source of real-world security breaches.)
			\item \Dirref{equilibria}.

			Suppose some sufficiently sharp upper bounds on the collective capabilities of the non-human-agents in a multi-agent system could be predicted as a function of their computational resources.  These bounds could be used to set limits on how much computation the non-human agents are allowed to wield, so as to ensure a sufficient degree of control for the humans while maintaining the usefulness of the non-human agents to the collective.
			\item \Dirref{prepotencefree}.
			Bounds on the capabilities of both AI systems and humans could be used to determine whether an AI system is sufficiently computationally endowed to be prepotent.  This could lead to more definable standards for when and when not to worry about \risksref{urprep}.

			\item \Dirref{humancognitive}.
			\citet{griffiths2015rational} have argued that computational limitations should be accounted for in human cognitive models.
			A better understanding of how an ideal bounded reasoner manages computation for rational decision-making could lead to better predictive and interactive models of humans, which could flow through to work on \dirrefs{transparency,aiad,bel,deference}.
			\end{itemize}

			\Act Most experimental work in the field of machine learning is concerned with assessing the capabilities of AI systems with limited computation.
			Therefore, it could be fruitful and straightforward to begin experimental approaches to each bullet point in the instrumental motivation section above.

			However, to bolster experimental approaches, it would help to develop a rigorous framework for planning and evaluating such experiments in advance.
			Currently, no satisfactory axiomatic theory of rational thinking under computational limitations---such as the hardware limitations inherent in a human brain, or any physical computer system---is known.

			One essential difficulty is that probability estimates calculated using bounded computational resources cannot be expected to follow the laws of probability theory, which require computation in order to satisfy (see the historical note below).
			For example, it can take a great deal of computation to prove that one statement is logically equivalent to another, and therefore to deduce that the statements should be assigned the same probability.
			Agent models which assume agents' beliefs follow the rules of probability theory---which assign equal probability to logically equivalent statements---are therefore unrealistic.
			Another difficulty is that it is unclear what rules the beliefs of reasoners in a multi-agent system should be assumed to satisfy, especially when the reasoners are in competition with one another.  Competition means the agents may have an incentive to deceive one another; when one agent deceives another, should the deceived agent be blamed, or the deceiver, or both?  On one hand the deceived agent is failing to protect itself from deception; on the other hand the deceiver is failing to uphold a basic principle of good faith communication that might be fundamental to effective group-scale interactions.

			\citet{garrabrant2016logical} have made some effort to resolve these difficulties by developing a model of a bounded reasoner called a ``logical inductor'', along with a suite of accompanying theorems showing that logical inductors satisfy a large number of desirable properties.
			A logical inductor's capabilities include converging toward satisfying the laws of probability over time, making well-calibrated predictions about other computer programs including other logical inductors, the ability to introspect on its own beliefs, and self-trust.
			Logical inductors also avoid the fallacy of treating the outputs of deterministic computations as random events, whereas past models of bounded reasoners tend to assume the reasoner will implicitly conflate uncertainty with randomness \citep{halpern2014decision}.

			However, the logical inductor theory as yet provides no \emph{upper} bounds on a bounded reasoner's capabilities, nor does it provide effective estimates of how much computation the reasoner will need for various tasks.
			Thus, progress on bounded rationality could be made by improving the Garrabrant model in these ways.

			\Cse A working predictive theory of bounded rationality would eliminate the need to run any machine learning experiment whose outcome is already predicted by the theory.  This would make machine learning research generally more efficient, hastening progress.  The theory could also inspire the development of new and more efficient learning algorithms.  It is unclear whether such advancements would reduce or increase existential risk overall.

			\Hist Chapters 1 and 3 of \emph{Do the Right Thing} \citep{russell1991right} contain a lengthy discussion of the challenge of treating bounded rationality axiomatically.
			Some excerpts:

			\begin{quote}
			``[...] computations are treated as if they were stochastic experiments, even when their outcomes are completely deterministic.
			[...] Given the absence of a satisfactory axiomatic system for computationally limited agents, our results have only a heuristic basis, strictly speaking.'' (p. 25)
			\end{quote}

			\begin{quote}``These time-limited estimates, which Good (1977) called dynamic probabilities and utilities, cannot obey the standard axioms of probability and utility theory.
			Just how the axioms should be revised to allow for the limited rationality of real agents without making them vulnerable to a charge of incoherence is an important open philosophical problem, which we shall not attempt to tackle here.
			[...] the formulae here and in chapters 4 and 5 have as yet only a heuristic justification, borne out by practical results.'' (pp. 60-61)
			\end{quote}

			Despite this, many attempts to axiomatize bounded rationality since then, such as by \citet{halpern2011dont}, continue to prescribe that the agent should model the outputs of unfinished computations using probability.

	\nodedef{Single/single instruction}{ssinstruction}

		\directiondef{preferencelearning}{preference learning}

			Preference learning is the task of ensuring that an AI system can learn how to exhibit behavior in accordance with the preferences of another system, such as a human.

			\Soc When a CEO asks her employee to help increase their company's profits, she implicitly hopes the employee will do so without conspiring to have her fired from the company in order to replace her with someone more effective, or by engaging in immoral acts like hacking a competitor's bank account.
			The CEO's preferences are thus quite a bit more complex than the statement ``help us increase profits'' alone might suggest.
			Moreover, because she cannot easily specify the innumerable things she hopes the employee will \emph{not} do, the employee must exercise some independent judgment to \emph{infer} the CEO's preferences from surrounding social context.

			\Mot Preference learning is mainly relevant to mitigating \risksref{urmis}, and requires striking a balance between literal obedience and independent judgment on the part the AI system.
			If a superintelligent factory management system is instructed with the natural language command, ``make as many paperclips as possible this year'', one of course hopes that it will not attempt to engineer nanotechnology that fills a sphere two light-years in diameter with  paperclips \citet[Chapter 8, ``Infrastructure Profusion'']{bostrom2014superintelligence}.
			At the same time, if it does not make any paperclips at all, it will tend to be replaced by another system which does.

			Without a satisfactory procedure for striking a balance between literal obedience and independent judgment, we humans may be unable to instate our preferences as governing principles for highly advanced AI systems.
			In particular, the continued existence and general well-being of human society---a highly complex variable to define---would be placed at risk.

			\Act Specifying an AI system's objectives directly in terms of a score function of the environment to be maximized can lead to highly unpredictable behavior.
			For an example, programming a cleaning robot to maximize the amount of dirt it picks up could result in the robot continually spilling out dirt for itself to clean \citep[Chapter 17.1]{russell2003artificial}.
			Similarly, a reinforcement learning system trained to maximize its score in a boat racing game learned to drive in circles to collect more points instead of finishing the race \citep{amodei2016faulty}.

			One approach to this problem is to use \dirtitle{preferencelearning}, i.e., to design AI systems to adjust their model of human preferences over time.  Human preference learning is already an active area of research with numerous past and present applications, for example in product recommendation systems or automated software configuration.  New commercial applications of preference learning, such as personal assistant software, will surely become more prevalent over the coming decade.

			There are numerous mathematical formulations of preference learning problem; see \citet{braziunas2006computational} for a review.  In a sequential decision-making setting, the problem can be expressed as a POMDP, where the human's preferences are encoded as information about the environment determining which states are desirable \citep{boutilier2002pomdp}.  This formulation involves not only learning human preferences, but taking actions that satisfy them.  This is the full problem of preference \emph{alignment}: aligning an AI system's behavior with the preference a user.

			Preference learning is further complicated in a cooperative setting, where the human is also taking actions directly toward their goal.  Here, success for the AI system is defined as the combined efficacy of a human/AI team working toward a common objective that is understood primarily by the human.  This setting can also been represented as a POMDP, where the human's actions are part of the environment's transition function \citep{fern2010computational}.   The human's actions can then be taken as evidence about their preferences, such as using inverse reinforcement learning (IRL), also known as inverse optimal control \citep{kalman1964linear}.  This approach was introduced by \citet{javdani2015shared}.  Somewhat concurrently, \citet{hadfield2016cooperative} introduced \emph{cooperative inverse reinforcement learning} (CIRL), a problem framing where a human and an AI system share common knowledge that the AI system is attempting to learn and optimize the human's objective.
			The CIRL framing been used to explore the possibility of ``pragmatic'' robots that interpret human actions with an awareness that the human is attempting to teach them \citep{fisac2017pragmatic}.
			Using similar but slightly different assumptions from CIRL (in particular, using limited levels of metacognition on the part of the human and robot, yielding non-equilibrium strategies), \citet{milli2019literal} show that non-pragmatic robots are more robust than pragmatic robots, even when humans are in fact trying to teach them about their preferences.  In these experiments, joint performance is improved when the robot takes a literal interpretation of the human, even when the human is not attempting to be literal.

			There are some concerns that present-day methods of preference learning may not suffice to infer human preferences in a form sufficiently detailed to safely direct the behavior of a prepotent or near-prepotent AI system.
			Thus, in order to be marginally valuable for the purpose of reducing existential risk, a focus on approaches to preference learning that might scale well for directing more advanced  systems (as in \risksnum{mpais}) may be needed.

			For this, heuristics for minimizing the unintended side effects of the system's operation \citep{amodei2016concrete,krakovna2018penalizing}, avoiding taking optimization to extremes \citep{taylor2016quantilizers}, or taking optimization instructions too literally, also known as ``reward hacking'' \citep{amodei2016concrete,ibarz2018reward}), could be useful to codify through theory or experiment.  Absent an approach to single/single delegation that would address such issues implicitly and automatically, heuristics could be helpful as transient rules of thumb to guide early AI systems, or to provide inspiration for rigorous and scalable long-term solutions to preference alignment.

			As well, preference learning methods that account for idiosyncrasies of human cognition may also be needed to avoid interpreting errors in judgement as preferred outcomes. For instance, \citet{evans2015learning} explore preference learning methods accounting for bounded cognitive capacity in the humand, and \citep{evans2016learning} account for biases in the human's judgement.  An alternative approach would be to ascertain how humans themselves infer and convey  preferences \citep{Baker2014,Lucas2014,Meltzoff1995}, and develop AI systems to use the same methods.

			This approach is being investigated by Stuart Armstrong, in as-yet unpublished work.

			\Cse
			If AI systems or human institutions use preference learning to develop a highly precise understanding of human preferences, that knowledge could be used in ways that are harmful to the humans.  For instance, satisfying the short-term preferences of the humans in question could be used as part of a longer-term strategy to gain and exploit their trust in ways that they will later regret.  Thus, to respect the wishes of the persons or institutions whose preferences are being learned, certain measures may be needed to ensure that preference learning capabilities are usually or always deployed within a preference alignment methodology.

			\Hist The challenge of clearly specifying commands to an intelligent machine was also remarked by Norbert Wiener \citep{wiener1960some}; see the historical note in \secref{prepotence} for a direct quote.

		\directiondef{bel}{human belief inference}
			An AI system that is able to infer what humans believe about the factual state of the world could be better suited to interact with humans in a number of ways.  On the other hand, it might also allow the system to acquire a large amount of human knowledge by inferring what humans believe, thereby enabling prepotence.  As such, this research direction is very much ``dual use''.

			\Soc Suppose Alice is a doctor, and Bob is her intern.  A hospital patient named Charlie has previously experienced severe allergic reactions to penicillin.  One day, Charlie gets an ear infection, and Alice prescribes penicillin for the treatment.  Now suppose Bob is nearby, and knows about Charlie's allergy.  What should Bob do about Alice's decision?  If Bob assumes Alice's beliefs about the world are correct, this would mean either Alice wishes to harm Charlie, or that that Charlie is in fact no longer allergic to penicillin.

			However, the pragmatic thing is for Bob to infer something about Alice's beliefs: in this case, that Alice is not aware of Charlie's allergy.
			This inference will likely lead Bob to ask questions of Alice, like whether Charlie's allergy has been accounted for in the decision.

			\Mot See the instrumental motivations.

			\Inst Progress on the theory and practice of belief inference could improve our understanding of
			\begin{itemize}
				\item \dirref{aiad}.
				This may require AI systems to model human beliefs, implicitly or explicitly, in order to decide when and how to assist in their deliberation.

				\item \Dirref{preferencelearning}.
				Suppose a model describing humans does not account for potential errors in a human's beliefs when observing the human.  Then, when the human fails at a task due to erroneous beliefs, the model will interpret the human as \emph{wanting} to the fail at the task.  Hence, belief inference is important for preference inference and thereby \dirtitle{preferencelearning}.

				\item \dirref{deference}.
				A number of protocols for AI systems deferring to humans could involve inferring the beliefs of the human.  For instance, ``defer to the human's beliefs when the human is more likely to be correct than me'', or ``defer to the human in situations where the human will believe I should have deferred to them''.  These protocols behave very differently when the human's beliefs are incorrect but the human wants to be deferred to anyway, say, for policy-level reasons intended to maintain human control.  Nonetheless, they both take inferred human beliefs as inputs.

				\item \Dirref{compromisingbetween}.
				Humans with differing beliefs may come into disagreements about what policy a powerful AI system should follow.
				An AI system that is able to infer the nature of the differing beliefs may be able to help to resolve the disagreement through dialogue.
			\end{itemize}

			\Act Human beliefs should likely be inferred through a variety of channels, including both natural language and demonstrations.
			Bayesian methods specifically for extracting human priors \citep{griffiths2005bayesian} have been explored to determine human priors on variables such as box office earnings and the lengths of poems \citep{lewandowsky2009wisdom}.
			For learning human beliefs from demonstrations of human actions, a generalization of Inverse Reinforcement Learning \citep{abbeel2004apprenticeship} could be viable, such as by modeling the human as solving a POMDP.
			There is a small amount of quantitative evidence that humans model other agents (and presumably other humans) in this way, i.e., by assuming the other agent is solving a POMDP and figuring out what the agent's beliefs and desires must be to explain the agent's behavior \citep{baker2011bayesian}.  If humans indeed make use of this ``POMDP inversion'' method in order to model each other, perhaps AI systems could use POMDP inversion to model humans.
			Differentiable MDP solvers and POMDP solvers can be used for gradient descent-based approaches to maximum-likelihood estimation of the MDP or POMDP an agent believes it is solving.
			This would enable a learner to simultaneously infer the prior, transition rule, and reward function in the mind of a demonstrator.
			Empirical testing could then assess the efficacy of this approach for assessing the beliefs of humans from their demonstrations.  \cite{reddy2018you} has explored this methodology in a user study with 12 human participants.

			\Cse  There are several major concerns about AI systems that are able to infer human beliefs.

			\renewcommand{\textit}[1]{\item \textbf{(#1)}}
			\begin{itemize}
					\textit{rapid acquisition of human knowledge}	If an AI system can infer human beliefs in a usable form, it can acquire human knowledge.  For instance, if an AI system is capable of reading and understanding natural language corpora, perhaps all of the knowledge of the internet could be made available to the system in an actionable form.  The ability to absorb human knowledge at scale would eliminate one of the main barriers to prepotence, namely, that human society has accumulated wisdom over time that is not by default usable to a powerful AI system.  Belief inference methods, especially through natural language processing that could be repurposed to process natural language corpora, could therefore enable prepotence and exacerbate all \risksref{mpais}.

					\textit{deception of humans} A related issue is that any sufficiently detailed model of a human person could be used to deceive that person, by reverse-engineering what they would need to see or hear in order to become convinced of a certain belief.  If an AI system is able to deceive all of human society, this could enable prepotence via social acumen, thereby exacerbating all \riskref{mpais}.  Alternatively, if an AI system is already prepotent via non-social means, but only sufficiently skilled in deception that it can can deceive a small number of individuals humans, it might trick its creators into deploying it prematurely, which would also increase \risksnums{urprep,urmis}. These issues would need to be averted somehow to ensure that the net impact of human-modeling technology is a reduction in existential risk.

			\end{itemize}

		\directiondef{humancognitive}{human cognitive models}
			Models of human cognition that are representable in a mathematical or otherwise digital form could be useful for designing human/AI interaction protocols for addressing other problems in this report.  On the other hand, they could also be abused to manipulate humans.  This research direction, like many, is ``dual use''.

			\Soc Suppose Alice is the CEO of a law firm, and Bob is her assistant.  Alice has been hoping for some time that her firm would take on CharlieCorp as a client.
			Once day, CharlieCorp sends Alice a long email, cc'ing Bob, which ends with
			\begin{quote}
				``... we are therefore seeking legal counsel.
				We assume from your past cases that you would not be interested in taking us as a client, but thought it would be a good idea to check.''
			\end{quote}
			Alice, having a busy week, fails to read the last line of the email, and replies only with ``Thanks for the update.''  Luckily, Bob realizes that Alice might have overlooked the ending, and sends her a ping to re-read it.
			Alice re-reads and responds with ``Looking at your situation, we'd actually be quite interested.
			Let's set up a meeting.''  Here, Bob is implicitly modeling not only Alice's desire to work with CharlieCorp, but also Alice's attentional mechanism.
			In particular, Charlie thinks Alice's attention was not directed toward the end of the email.

			Later, CharlieCorp asks Bob a question about a very long document.
			That day, Alice's schedule is clear, and knowing Alice is a fast reader who is familiar with the subject matter of the document, Bob forwards the question to Alice for her to think about.
			Here, Bob is modeling Alice's attentional capacity, her written language comprehension, as well as the contents of her memory.

			\Mot See the instrumental motivations.

			\InstAct Progress on the theory and practice of human cognitive modeling could improve our understanding of
				\begin{itemize}
					\item \Dirref{aiad}.
					To the extent that AI systems may eventually be needed to assist humans in safety assessments of other AI systems, understanding the quirks and limitations of human thinking may be helpful in designing a system that helps humans to reach a sound conclusion.  To this end, \citet{ought2017predicting} have attempted to generate datasets of examples of human deliberative output.  Collecting more data of this sort could help to train and/or validate models of human cognitive functions involved in deliberation.

					\item \Dirref{preferencelearning}.
					To infer a person's preferences from their behavioral outputs, it would help to understand the mapping $B$ from preferences to behavior, including speech.  Then, preference inference amounts to inverting that mapping: given observed behavior $b$, we seek to find preferences $p$ that would satisfy $B(p)=b$.  \dirref{bel} has already discussed how the person's beliefs play a role in defining the map $B$.  However, $B$ is parametrized by other features of human cognition aside from beliefs and preferences, such as planning, attention, memory, natural language production, and motor functions.  Isolating or at least narrowing our uncertainty about those variables could thus help us to reduce uncertainty in the ``behavior equation'' $B(p)=b$ that we are solving when performing preference inference.  As an example of early work in this direction, \citet{steyvers2006probabilistic} models the interaction of inference and memory.

					\item \Dirref{deference}.
					Suppose an AI system plans to defer to humans to take over from certain confusing situations, but those situations would either be too complex for humans to reason about, or too prone to the influence of particular human biases for humans to handle the situation responsibly.  This means that even routine applications of AI technology, in situations where the AI hands off control or decision-making to a human, will likely need to account explicitly or implicitly for human cognitive peculiarities aside from preferences.  Developing principled and generalizable hand-off procedures that will scale with the intelligence of the AI system may require better models of human cognition.  As a simple present-day example, self-driving car technology must account for human reaction time when handing control over to a human driver \citep{dixit2016autonomous}.

					\item \Dirref{compromisingbetween}.
					Disagreements between humans might sometimes be due to different tendencies in more basic cognitive functions like attention and memory.  For example, if Alice has a great memory and Bob has a terrible memory, Alice might disagree with Charlie on the nature of their unrecorded verbal agreements, and Bob---if he knows he has a bad memory---might not trust Alice to be the arbitrator of those disagreements.  Thus, an AI system that offers compromises that humans are likely to accept may need a working model of humans' cognitive capacities aside from their preferences.
					Identifying and explaining these differences could be helpful in dispute resolutions, and hence in facilitating agreements to continue sharing ownership of powerful AI systems.
					For example,  \citet{taber2006motivated} shows that political disagreements arise to some extent from motivated skepticism, and \citet{griffiths2008theoretical} show that cultural disagreements should be expected to arise from inherited inductive biases.  Such nuances may also prove essential in \dirref{modelingcommittees}.

				\end{itemize}

			\Cse There are a number of potentially dangerous and wide-reaching side effects to developing high-fidelity human cognitive models.

			\renewcommand{\textit}[1]{\item \textbf{#1. }}
			\begin{itemize}
					\textit{Manipulation of humans}  Human cognitive models can be used to manipulate humans.  This can already be seen in social media platforms that develop user models to generate addictive features to keep users engaged.  If sufficiently detailed, perhaps human cognitive models could be used by an AI system to manipulate all of human society in a goal-directed fashion.  In principle this could enable prepotence through social acumen, thereby exacerbating all \risksref{mpais}.
					\textit{Impoverished third-party safety testing}	If detailed human models are made publicly available, we impoverish our ability to perform ``hold-out'' safety testing and verification for powerful AI systems, as in \dirref{formalverification}.  Specifically, if precise human models are \emph{not} made publicly available, and instead withheld by a independent AI safety testing institution, then the models could be used to design simulation-based safety tests as a regulatory safety check for AI systems built by private corporations or the public.  However, if the human models used in the safety tests were released, or derivable by institutions other than the safety testers, then the models could be used by corporations or individuals deploying AI systems to ``game'' the regulatory testing process \citep{kumar2019thoughts}, the way a student who knows what questions will be on exam doesn't need to learn the rest of the course material.
						In particular, this could lead to an increase in \risksrefs{urprep,urmis}.  Thus, a judicious awareness of how and when to apply human-modeling technology will be needed to ensure it is shared appropriately and applied beneficially.

				\end{itemize}

				\noindent See also \dirref{bel} for a consideration of side effects of modeling human beliefs specifically.

	\nodedef{Single/single control}{sscontrol}

		\directiondef{safeshutdown}{generalizable shutdown and handoff methods}
			As with any machine, it remains important to maintain safe shutdown procedures for an AI system in case the system begins to malfunction.  One might operationalize ``shutdown'' as the system ``no longer exerting control over the environment''.  However, in many situations, ceasing to apply controls entirely may be extremely unsafe for humans, for example if the system is controlling a self-driving car or an aircraft.  In general, the sort of shutdown procedure we humans want for an AI system is one that safely hands off control of the situation to humans, or other AI systems.  Hence, the notion of a \emph{handoff} can be seen as generalizing that of a shutdown procedure.  In aviation, the term ``handoff'' can refer to the transfer of control or surveillance of an aircraft from one control center to another, and in medicine the term is used similarly for a transfer of responsibilities from one doctor to another.  This research direction is concerned with the development of generalizable shutdown and handoff techniques for AI systems.

			\Soc Suppose AliceCorp hires Betty to take on some mission-critical responsibilities.  In case Betty ever becomes ill or uncooperative and can no longer perform the job, other employees must be ready to cover off Betty's responsibilities until a replacement can be found.  Such handoffs of responsibility can be quite difficult to coordinate, especially if Betty's departure is a surprise.  For instance, any documented instructions for performing Betty's responsibilities may need to be documented in a manner that is readable to other employees, given their more limited context and perhaps experience.  Therefore, many companies will go to great lengths to maintain detailed  documentation of responsibilities and handoff procedures.  Similar procedures are often needed but missing on the scale of industries: when certain companies become ``too big to fail'', governments are left with no means of replacing them with better versions when they begin to malfunction.

			\Mot Generalizable shutdown and/or handoff procedures could reduce the risk of \risksrefs{urprep,urmis} by making it easier for humans to regain control of a situation where an AI system is malfunctioning or behaving drastically.  In general, future applications of powerful AI systems may pose risks to society that cannot be simulated in a laboratory setting.  For such applications to be responsible, general principles of safe shutdown and safe handoff procedures may need to be developed which are known in advance to robustly generalize to the high-stakes application.

			Somewhat orthogonally, perhaps the involvement of many humans in training and/or drills for AI$\to$human handoffs could create a source of economic involvement for humans to reduce \riskref{econ}, and/or cognitive stimulation for humans to reduce \riskref{humenf}.

				\Act Practically speaking, almost any existing computer hardware or software tool has a custom-designed shutdown procedure, including AI systems.
				However, there has not been much technical work on generalizable strategies for shutting down or handing over control from an AI system.

				In human--robot interaction literature, there is a body of existing work on \emph{safe handovers}, typically referring to the handoff of physical objects from robots to humans.  For instance,
			\citet{strabala2013toward}, have studied both robot-to-human and human-to-robot handovers for a variety of tasks.
			\citet{moon2014meet} showed that using humanlike gaze cues during human-robot handovers can improve the timing and perceived quality of the handover event.
			For self-driving cars, \citet{russell2016motor} show that human motor learning affects car-to-driver handovers.  For unmanned aerial vehicles, \citet{hobbs2010unmanned} argue that ``the further development of unmanned aviation may be limited more by clumsy human--system integration than by technological hurdles.''
			Each of these works contains reviews of further relevant literature.

			For coordination with multiple humans, \citet{scerri2002elf} put forward a fairly general concept called \emph{transfer of control} for an AI system coordinating with multiple humans, which was tested in a meeting-planning system called Electric Elves (E-Elves).
			The E-Elves system was used to assist in scheduling meetings, ordering meals, and finding presenters, over a 6-month period by a group of researchers at the University of Southern California.
			\citeauthor{scerri2002towards} describes the mathematical model underlying the system, which used an MDP formulation of the human/AI interaction problem to express coordination strategies and assess their expected utility in terms of
			``the likely relative quality of different entities' decisions; the probability of getting a response from an entity at a particular time; the cost of delaying a decision; and the costs and benefits of changing coordination constraints''.  Perhaps similar general principles could be used to design shutdown and/or handover processes in other settings.

			In any task environment, one might try to operationalize a safe shutdown as ``entering a state from which a human controller can proceed safely''.  As a cheaper proxy to use in place of a human controller in early prototyping, another AI system, or perhaps a diversity of other AI systems, could be used as a stand-in during training.  Suites of reinforcement learning environments such as OpenAI Gym \citep{brockman2016openai} could be used to ascertain the generality of any given safe handover technique.

			\Cse As with any safety methodology, if safe handover methods are developed for near-term systems and erroneously presumed to generalize to more powerful systems, they could create a false sense of security.  For instance, suppose generalizable solutions are developed for handing off control from a single AI system to a single human, such as from a self-driving car to a human driver.  The same principles might not work to hand off control from an automated air traffic control system to human air traffic controllers, which might require solving a coordination problem between the humans who receive the control in the event of a shutdown.
			Or, a simple ``suspend activity and power down'' procedure might be used to shut down many simple AI systems, but then someday fail to effectively shut down a powerful misaligned system that can build and execute copies of itself prior to the shutdown event.
			Thus, to apply ideas from this research direction responsibly, one must remain on the lookout for unique challenges that more complex or capable AI systems will present.

			\Hist Wiener has also remarked on the difficulty of interfering with a machine which operates on a much faster time scale than a human.
			\begin{quote}
				``We have seen that one of the chief causes of the danger of disastrous consequences in the use of the learning machine is that man and machine operate on two distinct time scales, so that the machine is much faster than man and the two do not gear together without serious difficulties.
				Problems of the same sort arise whenever two operators on very different time scales act together, irrespective of which system is the faster and which system is the slower.'' \citep{wiener1960some}
			\end{quote}

		\directiondef{corrigibility}{corrigibility}

			An AI system is said to be \emph{corrigible} if it ``cooperates with what its creators regard as a corrective intervention, despite default incentives for rational agents to resist attempts to shut them down or modify their preferences'' \citep{soares2015corrigibility}.  In particular, when safe shutdown procedures are already designed and ready to execute, a corrigible AI system will not work against its human operator(s) to prevent being shut down.

			\Soc A person is said to be ``corrigible'' if they are capable of being corrected, rectified, or reformed.
			An ``incorrigible'' person is one who does not adjust their behavior in response to criticism.
			If an employee behaves in an incorrigible manner, an employer may rely on the ability to terminate the employee's contract to protect the company.
			Imagine, however, an incorrigible employee who is sufficiently crafty as to prevent attempts to fire them, perhaps by applying legal technicalities or engaging in manipulative social behaviors.
			Such a person can cause a great deal of trouble for a company that hires them.

			\Mot As AI systems are developed that are increasingly capable of social intelligence, it becomes increasingly important to ensure that those systems are corrigible.
			An incorrigible AI system whose goals or goal inference instructions are mis-specified at the time of its initial deployment poses a \riskref{urmis} to humans if it is able to prevent us from modifying or disabling it.

			\Act  \citet{hadfield2016off} have shown that a reinforcement learning system can be given uncertainty about its reward function in such a way that human attempts to shut it down will tend to cause it to believe that being shut down is necessary for its goal.
			This is not a full solution to corrigibility, however.
			\citet{carey2017incorrigibility} shows that incorrigibility may still arise if the AI system's uncertainty about the reward function is not appropriately specified.
			Moreover, \citet{milli2017should} point out that too much reward uncertainty can lead an AI system to underperform, so there is a balance to be struck between expected performance and confidence that shut-down will be possible.

			As a potential next step for resolving these issues, experiments could test other mechanisms aside from reward uncertainty for improving corrigibility.
			For example, see \dirref{self} below.

			A different approach to corrigibility for reward-based agents is to somehow modify their beliefs or reward function to make them more amenable to shutdown or modification.  \citet{armstrong2017indifference} provides an overview of attempts in this direction.

			\Cse Progress on the problem of corrigibility does not seem to present many negative side effects, other than the usual risk of falsely assuming that any given solution would generalize to a high-stakes application without sufficient testing.

		\directiondef{deference}{deference to humans}

			Deference refers to the property of an AI system actively deferring to humans on certain decisions, possibly even when the AI system believes it has a better understanding of what is right or what humans will later prefer.

			\Soc Suppose Allan is a patient and Betty is his doctor.  Allan is bed-ridden but otherwise alert, and Dr. Betty is confident that Allan should receive a dose of anesthetic to help Allan sleep.
			Suppose also that the Dr. Betty is bound by law to ask for the patient's consent before administering this particular anesthetic, and that she expects the patient to say ``no''.
			Even if Dr. Betty is very confident that she knows what's best for the patient, the doctor is expected to defer to the patient's judgment in this case, rather than, say, administering the anesthetic in secret along with the patient's other medications.
			That is, the doctor is sometimes required to defer to the patient, even when confident that the patient will make the wrong choice.

			\Inst Theoretical models and/or training procedures for \dirtitle{deference} could help directly with

			\begin{itemize}
			\item \Dirref{corrigibility}.
			In order to preserve the corrigibility of an AI system over time, we will need AI systems to not only respond to corrective interventions, but to seek them out as a matter of policy, particularly on decisions that could lead to a loss of corrigibility.
			\item \Dirref{hoal}.
			A generic deference capability may allow AI systems to serve as useful delegates in a chain of command including humans and other AI systems.
			\item \Dirref{equilibria}.
			A notion of deference to humans that is stable as AI systems evolve and replicate over time might constitute an important class of \dirref{equilibria}.
			\end{itemize}

			\Act Simulated experiments where one AI system is required to seek out and defer judgment to another AI system could be fruitful for developing and testing protocols for deferring to outside judgment.
			\citet{milli2017should} show that performance trade-offs are to be expected when requiring direct obedience to commands.
			Experiments to ascertain an appropriate balance between deference and autonomy for minimizing tail risks arising from system mis-specification could be highly informative.

			\Cse Too much deference to humans could lead to catastrophic errors.  For instance, if a powerful AI system responsible for managing the electrical grid of a city were to defer to a single human on the decision to shut it down, perhaps many people could suffer or die as a result.  In the future, perhaps larger systemic failures of this sort could present existential risks.

		\directiondef{open}{generative models of open-source equilibria}

			AI systems are in principle completely inspectable to humans, in that their execution can create a perfect log of every internal state that occurs.  The degree to which the internal ``thought processes'' of such machines will be understandable to humans will likely depend on the success of future research on \dirref{transparency}.  Whatever degree of transparency and/or explainability can be achieved, its implications of the game-theoretic relationship between systems and humans should be explored.  But, so far, very little game theory research has been carried out to ascertain, either analytically or by simulation, what equilibria arise between agents when one agent is assumed to be partially or fully transparent to another.

			\Soc Suppose Alice is very good at reading Bob's body language, such that if Bob tries to deceive her or make plans that she would dislike, Alice will notice.
			His thoughts, in addition to his outward actions, have a direct impact on his interactions with Alice.
			Thus, Bob has an incentive to think differently than he would if he were less transparent to Alice.
			This changes the space of actions Bob can take, because actions that would require planning will produce side effects in Alice's awareness.  For example, if Bob begins to formulate a plan to deceive Alice, she might notice and try to shut him down and/or simply see through the deception.

			Similarly, imagine two nations which have a large number of spies investigating one another.
			If Nation A begins to plan a trade embargo against Nation B, spies may leak this information to Nation B and trigger early responses from Nation B prior to Nation A's instatement of the embargo.  The early response could range from submissive behavior (say, conceding to Nation A's expected demands) to preemptive counter-embargoes, depending on the situation.

			\Mot Could a powerful AI system someday learn or infer how to deceive its own developers?  If possible, it could constitute a  \riskrefsWithOr{urprep,urmis}.  If not possible, it would be reassuring to have a definite answer as to why.  This is a question for ``open source game theory'', the analysis of interactions between decision-making entities that are partially or fully transparent to one another.

			More broadly, deception is only one important feature of a human/AI equilibrium in which mutual transparency of the human and the AI system could play a key role.
			Another might be intimidation or corruption: is it possible for the mere existence of a particular powerful AI system---in a partially or fully transparent form---to intimidate or corrupt its creators to modify or deploy it in ways that are harmful to the public?  In a diffuse sense, this might already be happening: consider how the existence of social media platforms create an ongoing incentive for their developers to make incremental updates to increase user engagement.  While profitable for the company, these updates and resulting increases in engagement might not be beneficial to the overall well-being of individual users or society.

			To understand the dynamics of these mutually transparent relationships between humans and AI systems, it might help to begin by analyzing the simplest case of a single human stakeholder interacting with a single relatively transparent AI system, and asking what equilibrium (long-run) behaviors are possible to arise.

			\Inst Generative models of machine learning agents reaching equilibria in open-source games could be helpful toward understanding
			\begin{itemize}
			\item \Dirref{hoal}.
			In scenarios where one AI system is tasked with assisting in the oversight of other AI systems, it might make sense for the overseer system to be given access to the sources codes or other specifications of the systems being overseen.
			By contrast, classical game theory assumes that players are capable of private thoughts which determine their actions.
			Hence, the relationship between an AI system and a system overseeing its source code is outside the assumptions of classical game theory.

			\item \Dirref{equilibria}.
			An AI system's source code will likely be visible to the humans who engineered it, who will likely use that code to run simulations or other analyses of the system.
			This relationship is also outside the assumptions of classical game theory.
			\end{itemize}

			\Act \citet{halpern2013game} have already remarked that ``translucency'' rather than opacity is a more realistic assumption when modeling the interaction of human institutions, or humans who can read one another's body language.
			Moreover, remarkably different equilibrium behavior is possible when agents can read one another's source code.
			\citet{tennenholtz2004program} developed the notion of \emph{program equilibrium} for a pair of programs playing a game which, when given access to one another's source code, have no positive incentive to be replaced or self-modified.
			Strikingly, it turns out that open-source agents can achieve certain cooperative (or defective) equilibria that are in principle not possible for closed-source agents \citep{critch2019parametric}.
            Understanding whether and how such equilibria could arise amongst advanced AI systems (and how various design choices might affect these outcomes), or between AI systems and humans, is an important question for understanding how multi-agent AI systems will equilibrate with humans.

			\Cse This direction could be problematic from an existential risk perspective if models of open-source equilibria are later used to preferentially develop AI/AI/AI coordination methods in the absence of human/AI coordination methods or multi-human multi-AI coordination methods.  Such methods could lead to \risksrefs{econ,humenf} and/or \riskref{humenf} if they result in too much human exclusion from economically productive work.

\section{Single/multi delegation research}\label{sec:sm}
	\fl{2}
	  \fl{3}
			This section is concerned with delegation from a single human stakeholder to multiple operationally separated AI systems (defined below).

			As powerful AI systems proliferate, to diminish \risksrefs{urprep,urmis}, it might help to have ways of predicting and overseeing their collective behavior to ensure it remains controllable and aligned with human interests.  Even if serving a single human or human institution, coordination failures between large numbers of interacting machines could yield dangerous side effects for humans, e.g., pollutive waste, or excessive consumption of energy or other resources.  These could constitute \risksref{urmis}.  Conversely,  unexpectedly well-coordinated interactions among multiple AI systems could constitute \riskref{urprep}, for instance, if a number of cooperating AI systems turned out to be capable of collective bargaining with states or powerful corporations.

			To begin thinking clearly about such questions, we must first decide what to count as ``multiple AI systems'' versus only a single AI system:

			\paragraph{Operational separation.}  Roughly speaking, for the purposes of this report, when we say ``multiple AI systems'' we are referring to a collection of AI-based algorithms being executed on physically or virtually separated computational substrate units, with each unit having a relatively high-bandwidth internal integration between its sensors, processors, and actuators, but only relatively low-bandwidth connections to other units.  We say that such units are \emph{operationally separated}.

            \dksays{[ ]+=but emphasize that in there may be factors besides bandwidth that should be used to determine whether subroutines should be considered as part of a unified whole or not, such as how messages influence the behaviors of the receiver.}

            It might be tempting to simplify the number of concepts at play by viewing the collective functioning of operationally separate units as a single ``agent'' to be aligned with the human stakeholder.  However, this perspective would elide the mathematical and computational challenges involved in balancing the autonomy of the individual units against the overall functioning of the group, as well as the non-trivial task of dividing up responsibilities between the units.

			\paragraph{Dec-POMDPs.}  The concept of a Decentralized Partially Observable Markov Decision Process, or Dec-POMDP \citep{oliehoek2016concise}, is a useful formalism for describing the problem faced by multiple AI systems (i.e., multiple operationally separated units) working to serve a common purpose.  Variants of Dec-POMDPs can also be considered, such as by adding uncertainty to the reward function or transition dynamics, or more refined assumptions on computational limitations.

	\nodedef{Single/multi comprehension}{smcomprehension}
		\fl{3}
			If companies and governments deploy ``fleets'' of AI systems to serve specific objectives---be they in physical or virtual environments---humans will likely seek to understand their collective behavior in terms of the individual units and their  relationships to one another.  From one perspective, a fleet of AI systems might be viewed as ``just a set of parallel processing units.''  But, when the systems are engaged in interactive intelligent decision-making based on objective-driven modeling and planning, new tools and abstractions may be needed to organize our understanding of their aggregate impact.  This section is concerned with research to develop such tools and abstractions.

            Single/multi delegation seems poised to become increasingly relevant.
            Modern computer systems, and machine learning systems in particular, already make increasing use of parallel computation.
            This is in part because the speed of individual processors has started to encounter physical limits, even though the \emph{cost} of a FLOP has continued to decline rapidly.
            However, there are also increasingly relevant physical limits to communication bandwidth between processes; thus future large-scale computer systems will almost certainly employ a high degree of operational separation at some scale of organization.

		\directiondef{rigorouscoordination}{rigorous coordination models}
			The Von Neumann-Morgenstern utility theorem and resulting utility theory \citep{morgenstern1953theory, von2007theory} provides a principled framework for interpreting the actions of a single agent: optimizing an expected value function conditioned on a belief distribution over the state of the world.  Can an analogous theory be developed for a cooperative multi-agent system to serve a single goal or objective?  In addition to utilities and beliefs, the model should also include mathematical representations of at least two other concepts:
			\begin{itemize}
			\item \textbf{Communications:} packets of information exchanged between the agents.  These could be modeled as ``actions'', but since communications are often designed  specifically to directly affect only the internal processes of the agents communicating, they should likely receive special treatment.
			\item \textbf{Norms:} constraints or objective functions for the policies of individual agents, which serve to maintain the overall functioning of the group rather than the unilateral contributions of its members.
			\end{itemize}

			\Soc Humans, of course, communicate.  And our reliance upon norms is evident from the adage, ``The ends do not justify the means''.  An individual person is not generally expected to take actions at all costs to unilaterally optimize for a given objective, even when the person believes the objective to serve ``the greater good''.  Instead, a person is expected to act in accordance with laws, customs, and innate respect for others, which ideally leads to improved group-scale performance.

			\Mot If there is any hope of proving rigorous theorems regarding the collective safety of multi-agent systems, precise and accurate mathematical definitions for their components and interaction protocols will be needed.  In particular, theorems showing that a collective of AI systems is or is not likely to become prepotent or misaligned will require such models.  Hence, this direction applies to the reduction of \risksrefs{urprep,urmis}. Moreover, common knowledge of problems and solutions in this area may be necessary to motivate coordination to reduce \risksref{uncdev}, or to avoid dangerous interactions with powerful AI systems that would yield \riskref{invdep}.

			\Act The framework of Dec-POMDPs introduced by \cite{bernstein2002complexity} provides a ready-made framework for evaluating any architecture for decentralized pursuit of an objective; see \cite{oliehoek2016concise} for an overview.  As such, to begin proving a theorem to support the use of any given coordination protocol, one could start by stating conjectures using the language of Dec-POMDPs.  Protocols could be tested empirically against existing machine learning	 methods for solving Dec-POMDPs.  In fact, any given Dec-POMDP can be framed as two distinct machine learning problems:

			\begin{itemize}
				\item \emph{Centralized training for decentralized execution.} This is the problem of producing---using a centralized training and/or learning system---a suite of decentralized ``agents'' (sensor/actuator units) that collectively pursue a common objective.  As examples of recent work in this area:
				\begin{itemize}
						\item \cite{sukhbaatar2016communication} treat a system of decentralized agents undergoing centralized training as a single large feed-forward network with connectivity constraints representing bandwidth-limited communication channels.  The authors find that on four diverse tasks, their model outperforms variants they developed with no communication, full-bandwidth communication (i.e., a fully connected network), and models using discrete communication.
						\item	\cite{foerster2016dialrial} propose two approaches to centralized learning of communication protocols for decentralized execution tasks.  The first, Reinforced Inter-Agent Learning (RIAL), has each agent learn its communication policy through independent deep Q-learning.  The second, Differentiable Inter-Agent Learning (DIAL), allows the training system to propagate error derivatives through noisy communication channels between the agents, which are replaced by discrete (lower bandwidth) communication channels during execution.

						\item \cite{foerster2017stabilising} explore, in a collaborative multi-agent setting with no communication at execution time, two methods for making use of experience replay (the re-use of past experiences to to update a current policy).  Each method aims to prevent the learners from confusing the distant-past behavior of its collaborators with their more recent behavior.  The first method treats replay memories as \emph{off-environment data} \citep{ciosek2017offer}.  The second method augments past memories with a ``fingerprint'': an ordered tuple comprising the iteration number and exploration rate, to help distinguish where in the training history the experience occurred.

				\end{itemize}

				\item \emph{Decentralized training for decentralized execution.} This is the problem of a decentralized set of learners arriving at a collective behavior that effectively pursues a common objective.  As examples of recent related work:
				\begin{itemize}
						\item \cite{matignon2012independent} identify five qualitatively distinct coordination challenges---faced by independent reinforcement learners pursuing a common (cooperative) objective---which they call ``Pareto-\-selection'', ``nonstationarity'', ``stochasticity'', ``alter-explo\-ration'' and ``shadowed equilibria''.
						\item	\cite{tampuu2017multiagent} examine decentralized Q-learners learning to play variants of Pong from raw visual data, including a cooperative variant where both players are penalized equally when the ball is dropped.
				\end{itemize}
			\end{itemize}

			The variety of problems and methods in recent literature for training collaborative agents shows that no single architecture has been identified as universally effective, and far from it.  None of the above works is accompanied by a rigorous theoretical model of how coordination \emph{ought} to work in order to be maximally or even sufficiently effective.  Hence the motivation for more rigorous foundations: to triage the many potential approaches to learning for single/multi delegation.

			\Cse In order for research enabling multi-agent coordination to eventually lead to a decrease rather than an increase in existential risk, it will need to be applied in a manner that avoids runaway coordination schemes between AI systems that would constitute a \riskrefsWithOr{uncdev,urprep,urmis,invdep}.  In particular, coordination-learning protocols compatible with a human being serving as one of the coordinating agents may be considerably safer in the long run than schemes that exclude humans.  Present methods do not seem particularly suitable for explicitly including humans in the mix.

		\directiondef{interpretablelanguage}{interpretable machine language}
			Just as today we seek more enlightening explanations for the actions of a neural network in order to improve our ability to evaluate and predict its behavior, in the not-too-distant future we will likely find ourselves seeking to understand the content of communications between AI systems.

			\Soc Business regulations that generate legible, auditable communications within and between companies increase the difficulty for those companies to engage in corrupt business practices.  This effect is of course only partial: despite the significant benefits of auditing requirements, it is usually still possible to find ways of abusing and/or circumventing legitimate communication channels for illegitimate means.

			\Mot As we humans delegate more of our decisions to AI systems, we will likely require those systems to communicate with each other to achieve shared goals. Just as transparency for an individual AI system's cognition benefits our ability to debug and avoid systematic and random errors, so too will the ability to interpret communications between distinct decision-making units.  This benefit will likely continue to scale as the scope and number of AI systems grows.  For AI capabilities approaching prepotence, interpretability of communications between AI systems may be needed to avoid \risksrefs{urprep,urmis}.  The more broadly understandable the interpreted communications are made, the better developer coordination can be enabled to diminish \riskrefs{uncdev}.  Since interpretable communications are more easily monitored and regulated, interpretable communication standards may also be helpful for regulating communicative interactions with powerful deployed AI systems, including communications that could precipitate \riskref{invdep}.

			\Act As techniques develop for machine learning transparency and interpretability, similar techniques may be adaptable to ensure the interpretability of machine--machine communications in multi-agent settings; see \dirnum{transparency}.  Or, there may arise entirely novel approaches.  \cite{bordes2016learning} explore the use of end-to-end trained dialog systems for issuing and receiving API calls, as a test case for goal-oriented dialogue.

            In this setting, one could consider a dialogue between two machines, Machine A and Machine B, where A treats B as a machine+human system in which the human on rare occasions attempts to understand messages from A to B and penalizes the system heavily if they are not understandable. As an alternative or complement to sparse human feedback, perhaps machine--machine language could be constrained or regularized to be similar to human language, as in \cite{lewis2017deal}.  Or, perhaps frequent automated feedback on the understandability of the A/B communication channel could be provided by a dialog state-tracking system (DSTS).  As DSTS normally attempts to understand human dialogue \citep{henderson2014second}, but perhaps one could be repurposed to give automated feedback on whether it can understand the communication between A and B.

			\Cse Any attempt to design or select for interpretability could lead to accidentally deceiving humans if one optimizes too much for human satisfaction with the communications rather than the accuracy of the human's understanding.
            A particular concern is ``steganography'', where information is ``hidden in plain sight'' in a way that is invisible to humans; \citet{} demonstrate steganography in cycleGANs \citep{}.
            \dksays{[ ]TODO: reference "CycleGAN, a Master of Steganography", and the cycleGAN paper}

		\directiondef{reldet}{relationship taxonomy and detection}
			In any attempt to train a multi-agent system to perform useful tasks like delivery services and waste collection, it is already clear that our choice of training mechanism will tend to affect whether the individual agents end up exhibiting cooperative or competitive relationships with one another.   Aside from ``cooperative'' and ``competitive'', what other descriptors of relationships between agents in a multi-agent system can be quantified that would allow us to better understand, predict, and perhaps improve upon the system's behavior?

			\Soc Alice and Bob work together on a team whose responsibility is to send out a newsletter every week.  Alice always asks to see the newsletter before Bob sends it out.  Bob has expressed that he thinks Alice's review is an unnecessary step, however, Alice continues to advocate for her review step.  Are Alice and Bob in a competitive or cooperative relationship here?  The answer could be somewhat complex.  Perhaps Alice and Bob both really have the newsletter's best interests at heart, and know this about each other, but  Alice just doesn't trust Bob's judgment about the newsletters.  Or, perhaps she doubts his loyalty to their company, or the newsletter project specifically.  Perhaps even more complicatedly, she might trust Bob's judgment about the content entirely, but prefer to keep the reviews in place to ensure that others know for sure that the newsletter has her approval.  This scenario illustrates just a few ways in which disagreements in working relationships can arise from a variety of different relationships between beliefs and values, that do not always involve having different values.

			\Mot To avert \risksrefs{urprep,urmis}, any single institution deploying multiple powerful AI systems into the real world will need to have a sufficient understanding of the relationships that would arise between those systems to be confident their aggregate behavior would never constitute an MPAI.
			To avoid \risksrefs{uncdev,invdep}, development teams will collectively need to maintain an adequate awareness of the potential interactions between their own AI systems and AI systems deployed by other teams and stakeholders.

			For instance, consider the possibility of a war between AI systems yielding an unsurvivable environment for humanity.
			\begin{itemize}
				\item If the warring AI systems were developed by warring development teams, the aggregate AI system comprising the interaction between the warring systems would be an MPAI.  This would constitute a \riskref{uncdev}, or a \riskref{voldep} if one of the teams recognized that their involvement in the war would make it unsurvivable.  Such cases could perhaps be made less likely by other ``peacekeeping'' AI systems detecting the violent relationship between the conflicting systems, and somehow enforcing peace between them to prevent them from becoming an MPAI in aggregate.

				\item If the war or its intensity was unexpected or unintended by the developers of the AI technology used in the war, it could constitute a \riskrefsWithOr{urprep,urmis,invdep}.  Such cases could perhaps be made less likely by detecting and notifying developers when violent relationships are arising between the systems they develop and deploy, and allowing developers to recall systems on the basis of violent usage.
			\end{itemize}

			On the other hand, an unexpected \emph{coalition} of AI systems could also yield a runaway loss of power for humanity.
			If the coalition formation was expected by everyone, but human institutions failed to work together to stop it, then it would constitute a \risklabelsWithOr{uncdev,invdep}.
			Developing a methodology for identifying and analyzing relationships between AI systems might be among the first steps to understanding and preventing these eventual possibilities.

			Crucially, there may be many more complex relationships between powerful AI systems that we humans would struggle to define in terms of simple war or peace, furthering the need for a systematic study of machine relationships.  In any case, both positive and negative results in research on \dirtitle{reldet} could be beneficial to making negative outcomes less likely:

			\begin{itemize}
				\item \textbf{Benefits of negative results.} If the relationships between near-prepotent AI systems begin to appear too complex to arrange in a manner that is legibly safe for humanity, then researchers aware of this issue can advise strongly for policies to develop at most one very powerful AI system to serve human civilization (or no such system at all, if multi/single delegation also proves too difficult).  In other words, advanced warning of unsurmountable difficulties in this research area might help to \emph{avoid} heading down a so-called ``multi-polar'' development path for powerful AI technologies.

				\item \textbf{Benefits of positive results.} If the relationships between near-prepotent AI systems appear manageable, perhaps such systems could be used to keep one another in check for the safety of humanity.  In other words, positive results in this area might help to \emph{optimize} a ``multi-polar'' development pathway to be safer on a global scale.
			\end{itemize}

			\Act One approach to this research area is to continually examine social dilemmas through the lens of whatever is the leading AI development paradigm in a given year or decade, and attempt to classify interesting behaviors as they emerge.  This approach might be viewed as analogous to developing ``transparency for multi-agent systems'': first develop interesting multi-agent systems, and then try to understand them.  At present, this approach means examining the interactions of deep learning systems.  For instance, \cite{leibo2017multi} examine how deep RL systems interact in two-player sequential social dilemmas, and \cite{foerster2018learning} explore the consequences of agents accounting for one another's learning processes when they update their strategies, also in two-player games.  \cite{mordatch2018emergence} examine the emergence of rudimentary languages from a centralized multi-agent training process, giving rise to a variety of interactive behaviors among the agents.

			\Cse This sort of ``build first, understand later'' approach will become increasingly unsatisfying and unsafe as AI technology improves, especially if AI capabilities ever approach prepotence.
			As remarked by \cite{bansal2017emergent}, ``a competitive multi-agent environment trained with self-play can produce behaviors that are far more complex than the environment itself.''
			As such, it would be useful to develop a methodology for relationship taxonomy and detection that not only makes sense for current systems but will generalize to new machine learning paradigms in the future.  For this, a first-principles approach rooted in the language of game theory and/or economics may be necessary as a complement to empirical work.

		\directiondef{inthierrep}{interpretable hierarchical reporting}
			This research direction is concerned with arranging hierarchies of AI systems that report to one another and to humans in a manner that resembles a present-day human business, and that would be legible to human overseers.  Hierarchy is a natural solution to the problem of ``scalable oversight'' \citep{amodei2016concrete} for teams of AI systems and/or humans, because hierarchies often lead to exponential gains in efficiency by reducing the complexity of problems and systems to smaller parts.  In a hierarchical reporting paradigm, AI systems could be developed for the express purpose of ``middle management'', to provide intelligible reports and questions either directly to humans, or other AI systems.  By involving human overseers at more levels of the hierarchy, perhaps a greater degree of interpretability for the aggregate system can be maintained.

			\Soc
			Imagine the CEO of a large corporation with thousands of employees.
			The CEO is responsible for making strategic decisions that steer the company towards desirable outcomes, but does not have the time or expert technical knowledge to manage all employees and operations directly.
			Instead, she meets with a relatively small number of managers, who provide her with summarized reports on the company's activities that are intelligible to the CEO's current level of understanding, with additional details available upon her request, and a limited number of questions deferred directly to her judgment.
			In turn, each manager goes on to review other employees in a similar fashion.  This reporting structure is enriched by the ability of the CEO to ask questions about reports from further down in the ``chain of command''.

			\Mot Consider a world in which autonomous, nearly-prepotent AI systems have become capable of interacting to generate a large number of business transactions that generate short-term wealth for their users and/or trade partners.  Who or what entity can oversee the net impact of these transactions to avoid negative externalities in the form of catastrophic risks, e.g., from pollution or runaway resource consumption?
			Historically, human governments have been responsible for overseeing and regulating the aggregate effects of the industries they enable, and have benefited from human-to-human communications as a source of inspectable documentation for business interactions.  If no similar report-generation process is developed for AI systems, human businesses and governments will face a choice: either to stifle the local economic gains obtainable from autonomous business transactions in favor of demanding more human involvement to generate reports, or to accept the risk of long-term loss of control in favor of the short-term benefits of more autonomy for the AI systems.  If and when any nation or corporation would choose the latter, the result could be:
			\begin{itemize}
				\item An increase in \risksrefs{urmis,urprep,invdep} due to the inability of the companies releasing AI systems to monitor their potential prepotence or misalignment through reporting mechanisms, and
				\item An increase in \risksrefs{uncdev} due to the inability of human authorities such as governments and professional organizations to recognize and avert decentralized development activities that could pose a risk to humanity in aggregate.
			\end{itemize}
			Thus, it would makes sense to find some way of eliminating the pressure to choose low-oversight regulatory regimes and business strategies, by making high-oversight strategies cheaper and more effective.
			Hierarchical reporting schemes would take advantage of exponential growth of the amount of supervision carried out as a function of the depth of the hierarchy, and may become a key component to scaling up supervisory measures in a cost-effective manner.  One potential approach to this problem would be to deploy AI systems in ``middle management'' roles that curate reports for human consumption.
			One can imagine chains of command between sub-modules that oversee one another for safety, ethics, and alignment with human interests.
			Just as communication between employees within a company can be made to produce a paper trail that helps to some degree with keeping the company aligned with governing authorities, perhaps teams of AI systems could be required to keep records of their communications that would make their decision-making process more inspectable by, and therefore more accountable to, human overseers.
			Such an approach could serve to mitigate \risksref{mpais} in full generality.

			\Act The interpretability aspect of this research direction would benefit directly from work on \dirrefs{transparency}.
			The concept of hierarchical learning and planning is neither new nor neglected in reinforcement learning \citep{dayan1993feudal,kaelbling1993hierarchical,wiering1997hq,sutton1999between,dietterich2000hierarchical,kulkarni2016hierarchical,vezhnevets2016strategic,bacon2017option,tessler2017deep}.
			The conception of different levels of the planning hierarchy as separate agents is also familiar \citep{parr1998reinforcement}.
			By viewing levels of hierarchical planning as separate learning agents, one can ask how to improve the transparency or interpretability of the subagents to the superagents, along the lines of \dirref{transparency}.  Ideally, the ``reports'' passed from subagents to superagents would be human-readable as well, as in \dirref{interpretablelanguage}.  Hence, work on building interpretable hierarchical reporting structures could begin by combining ideas from these earlier research directions, subject to the constraint of maintaining and ideally improving task performance. For instance, one might first experiment with unsupervised learning to determine which `report features' should be passed from a sub-agent to a superagent, in the manner learned by the agents in \cite{mordatch2018emergence}.  One could then attempt to impose the constraint that the reports be human-interpretable, through a combination of real human feedback and artificial regularization from natural language parsers, although as discussed in \dirref{transparency}, it is unclear how to ensure such reports would reflect reality, as opposed to simply offering ``rationalizations''.

			\Cse If the humans involved in interpreting the system were insufficiently concerned with the safety of the public, they might be insufficiently vigilant to avert catastrophic risk from rare or unprecedented events.  Or, if the humans individually cared about catastrophic risks, but were for some reason uncomfortable with discussing or reporting the potential for rare or unprecedented catastrophes, their individual concerns would not be enough to impact the collective judgment of the system.  Hence, \riskref{discourseimpairment} might undermine some of the usefulness of this research direction specifically for existential risk reduction.  Finally, if the resulting systems were interpretable to humans, but the institutions deploying the systems chose not to involve enough humans in the actual task of interpreting the systems (say, to operate more quickly, or to avoid accountability), then advancements in this area would accrue mostly to the capabilities of the resulting systems rather than their safety.

	\nodedef{Single/multi instruction}{sminstruciton}
		\fl{3}
			This section is concerned with delivering instructions to $N$ operationally separated decision-making units to serve the objectives of a single human stakeholder.  This problem does not reduce to the problem of instructing $N$ separate AI systems to each serve the human on their own.  This is because coordination solutions are needed to ensure the units interact productively rather than interfering with one another's work.  For instance, given multiple ``actuator'' units---each with the job of taking real-world actions to affect their physical or virtual environments---a separate ``coordinator'' unit could be designed to assist in coordinating their efforts.  Conveniently, the role of the coordinator also fits within the Dec-POMDP framework as a unit with no actuators except for communication channels with the other units.

			\directionabbrevdef{hoal}{hierarchical human-in-the-loop learning}{HHL}

			Just as reports will be needed to explain the behavior of AI systems to humans and other AI systems, queries from subsystems may be needed to aid the subsystems' decision-making at times when they have insufficient information or training to ensure safe and beneficial behavior.  This research objective is about developing an AI subsystem hierarchy in a manner compatible with real-time human oversight at each level of the hierarchy.

				\Soc Many companies are required to undergo financial audits on a regular basis.  For example, the California Nonprofit Integrity Act requires any charity with an annual gross revenue of \$2 million or more to have their financial statements audited, on an annual basis, by an independent
				certified public accountant.  This ensures that the taxpayer has a representative---the auditing firms---involved in the management of every tax-exempt company of a sufficient size.  Suppose instead that California's Franchise Tax Board attempted to audit every company itself; the FTB would quickly become overwhelmed by the amount of information to process.  Hence, the notion of an auditing firm is a replicable and hence scalable unit of organization that allows for more pervasive representation of taxpayer interests, at a scale of authority that is intermediate between the employees of individual companies on the low end and the California Franchise Tax Board on the high end.

				\Mot Active learning---that is, machine learning driven by queries from the machine to a human about areas of high uncertainty---seems potentially necessary for ensuring any AI system makes economical use of the human labor involved in training it.  It is likely possible to arrange AI systems into a hierarchy, as in \dirref{inthierrep}, where lower-level systems make queries to higher-level systems.  In such a set-up, human beings could be involved in answering the queries, either
				\begin{itemize}
					\item only at the topmost level of the hierarchy, or
					\item at all or most levels of the hierarchy.
				\end{itemize}

				The latter option would seem better from an employment perspective: more roles for humans in the hierarchy means a reduction of \riskref{econ}, and if the roles involve maintaining valuable human skills, a reduction of \riskref{humenf}.
				Involving a human at each node of the hierarchy also seems better from the perspective of accountability and governance.  Many human laws and accountability norms are equipped to deal with hierarchical arrangements of responsibilities, and hence could be applied as soft constraints on the system's behavior via feedback from the humans.
				In particular, human-checked company policies could be implemented specifically to reduce \risksrefs{urprep,urmis,invdep}, and nation-wide or world-wide laws could be implemented to reduce \risksrefs{uncdev,voldep}.
				The weight of these laws could derive in part from the accountability (or less euphemistically, the punishability) of the individual humans within the system if they fall short of their responsibilities to instruct the system according to safety guidelines.
				Such a system of accountability might feel daunting for whatever humans would be involved in the system and therefore accountable for global safety, but this trade-off could well be worth it from the perspective of existential risk and long-term human existence.

				\Act
				Engineering in this area would benefit from work on \dirref{inthierrep} because of the improved understanding of the aggregate system that would accrue to the engineers.  After deployment, in order for each human in the \dirabbrev{hoal} system to oversee their corresponding AI system in a time-efficient manner, techniques would be needed to train each AI system to take a large number of actions with only sparse feedback from their human supervisor on which actions are good.  \citet{amodei2016concrete} identify this issue as a problem in what they call ``scalable oversight'', and propose to approach it via \emph{semi-supervised reinforcement learning} (SSRL).  In SSRL,  where a managing or training system (which might involve a human) provides only sparse input into the decision-making of a reinforcement learner.  They outline six technical approaches to scalable oversight, and potential experiments to begin work in this area.

				Sparse rewards are merely one piece of the puzzle needed to be solved to enable \dirabbrev{hoal}.  \citet{abel2017agent} aim to develop a schema for ``Human-in-the-Loop Reinforcement Learning'' that is agnostic to the structure of the learner.
				Scaling up human-in-the-loop interaction models in a principled and generalizable manner is a rich technical challenge.
				To reduce confusion about whether solutions would be applicable for more complex or civilization-critical tasks, it is recommended that authors include in their publications some discussion of the scalability of their solutions, e.g., as in  \citet{saunders2017trial}.

				\Cse Hierarchical decision-making structures present a clear avenue for general AI capabilities advancements.  These advancements may fail to reduce existential risk if any of the following problems arise:
				\begin{itemize}
					\item The institutions deploying the resulting AI systems choose not to involve enough humans in the hierarchy.  For instance, the institution might prefer this outcome to speed up performance, or avoid accountability.
					\item The AI systems in the hierarchy are insufficiently legible to the humans, i.e., if  progress on \dirref{inthierrep} has been insufficient, or not applied to the system.
					\item The humans involved in the hierarchy are insufficiently individually motivated to think about and avert unprecedented catastrophic risks.
					\item The humans in the hierarchy are uncomfortable discussing or reporting their concerns about unprecedented catastrophic risks.
				\end{itemize}

		\directiondef{stability}{purpose inheritance}
			As AI systems are used increasingly in the development of other AI systems, some assurance is needed that the deployment of a putatively ``aligned'' system will not lead to the creation of dangerous systems as a side effect.

			To begin thinking about this dynamic informally, if an AI system $A$ takes actions that ``create'' another AI system $B$, let us say that $B$ is a ``descendant'' of $A$.  Descendants of descendants of $A$ are also considered to be descendants of $A$.  Given a satisfactory a notion of ``creating a descendant'', we say that $A$ has a \term{heritable purpose} to the extent that there is some purpose---that is, some internally or externally specified objective---which $A$'s own actions directly benefit, and which the collective actions of $A$'s descendants also benefit.  This research direction is concerned with the challenge of creating powerful AI systems with any particular heritable purpose, with human survival being a purpose of special interest.

			While the precise definition of ``creating a descendant'' is interesting to debate, the relevant definition for this report is whatever notion can best guide our efforts to reduce existential risk from useful real-world AI systems.  In particular, our notion of ``creation'' should be taken fairly generally.  It should include cases where $A$ creates $B$
			\begin{itemize}
			\item \emph{``intentionally''}, in the sense of being directed by a planning process internal to $A$ which represents and selects a series of actions for their utility in creating $B$;
			\item \emph{``subserviently''}, in the sense of being directed by a human or another AI system with an intention to use use $A$ as a tool for the creation of $B$; or
			\item \emph{``accidentally''}, in the sense of not arising from intentions on the part of $A$ or other systems directing $A$.
			\end{itemize}

			\noindent Whatever the definition, safety methods applicable for broader definitions of  ``descendant'' will be able to cover more bases for avoiding existential risks from descendant AI systems.

			\Soc
			A human corporation may be viewed as having a \emph{heritable purpose} if it only ever creates subsidiary companies that effectively serve the parent corporation's original purpose.  To the extent that subsidiaries might later choose to defect against the parent's mission, or create further subsidiaries that defect, the parent's purpose would not be considered perfectly heritable.

			When a human institution builds an AI system, that system can be viewed as a ``descendant'' of the institution.  So, if an AI system brings about human extinction, it could be said that human civilization itself (as an institution) lacks the survival of the human species as a heritable purpose.

			\Mot

			An AI system with the potential to create prepotent descendants presents a \riskrefs{urprep}.
			As an unlikely but theoretically enlightening example, an AI system performing an unconstrained search in the space of computer programs has the potential to write an AI program which is or becomes prepotent.  In general, it may be difficult to anticipate which AI systems are likely to instantiate descendants, or to detect the instantiation of descendants. At the very least, a powerful AI system that is not itself an MPAI, but which lacks human survival as a heritable purpose and is used to develop other AI systems, could constitute a \riskref{urmis}.  For instance, an automated training system for developing machine learning systems could be used as a tool to develop an MPAI, and hence the training system would lack human survival as a heritable purpose.

			\Act Lack of technically clear definitions of ``instantiate a descendant'' and ``heritable purpose'' are obstructions to this research direction.  Some definitions would be too restrictive to apply in reality, while others would be too permissive to imply safety results even in theory.
			Hence, next actions could involve developing clearer technical conceptions of these ideas that are adequate for the purposes of guiding engineering decisions and reducing existential risk.  There are at least two distinct approaches one might consider:

			\renewcommand{\textit}[1]{\item \textbf{#1.}}
			\begin{itemize}
				\textit{Approach 1: Avoidance techniques} This approach develops an adequate definition of ``instantiating a descendant'', and uses the resulting concept to design AI systems that entirely avoid instantiating descendants, thus obviating the need for purpose inheritance.  There has not been much research to date on how to quantify the notion of ``instantiating a descendant'', though a few attempts are implicit in literature on agents that ``copy'', ``teleport'', or ``tile'' themselves  \citep{yudkowsky2013tiling,orseau2014teleporting,orseau2014multi,soares2014tiling,fallenstein2015vingean}.
				One problem is that current theoretical models of AI systems typically assume a well-defined interface between the AI system and its environment, receiving inputs only via well-defined sensors and making outputs only via well-defined actuators.
				Such models of AI systems are sometimes called \emph{dualistic}, after mind-body dualism.  In reality, AI systems are \emph{embedded} in the physical world, which they can influence and be influenced by in ways not accounted for by the leaky abstraction of their interface.
				\citet{orseau2012space} consider a fully embedded version of AIXI
				\citep{hutter2004universal,everitt2018universal} and conclude that in this setting: ``as soon as the agent and environment interact, the boundary between them may quickly blur or disappear'' \citep{orseau2012space}, but these works do not attempt to resolve the questions this raises about identifying descendants.  Thus, a more general and real-world applicable notion of ``instantiating a descendent'' is needed.

				Alternatively, one could imagine a ``know it when we see it'' approach to defining the concept.  However, such an approach might not scale well to regulating systems that could find ways of replicating and/or engineering new systems that humans would not easily recognize as cases of replication and/or engineering.  Thus, a characterization of ``instantiating descendants'' that is simultaneously rigorous and real-world applicable is missing.  The reader is invited ponder potential approaches to formalizing this problem.

				\textit{Approach 2: Heritability results} Develop an adequate definition for ``instantiating a descendant'', as well has ``heritable purpose'', and use these conceptions in one of two ways:
				\begin{itemize}
						\item[(a)] \textbf{Possibility results:} Develop AI systems with the heritable purpose to serve and protect humanity as a whole, in particular by avoiding existential risks and MPAI deployment events; or
						\item[(b)] \textbf{Impossibility results:} Develop demonstrations or arguments that Approach 2(a) is too difficult or risky and that Approach 1 is better.
				\end{itemize}
				These approaches are more difficult than Approach 1 because they involve more steps and concepts.   Nonetheless, some attempts in this direction have been made.  \citet{yudkowsky2013tiling,fallenstein2015vingean} and others consider AI systems reasoning about the heritable properties of their descendants using logic, which remains a topic of ongoing research.  One remaining challenge is to maintain the strength of descendants' reasoning in the face of self-reference issues, which is addressed to some extent---at least asymptotically---by \citet{garrabrant2016logical}.
			\end{itemize}

			It could also be valuable to empirically evaluate the propensity of agents based on current machine learning techniques to create descendants.  For instance \citet{leike2017ai} devise a toy grid-world environment for studying self-modification, where they consider the behavior of reinforcement learning algorithms.  Considering more complex environments where descendants are still easy to identify by construction would be a good next step.  Learning to predict which behaviors are likely to instantiate descendants in such settings would be also be useful.

			\Cse Progress on possibility results in Approach 2(a) would be dual purpose, in that the results would likely create the theoretical capability for other purposes aside from ``serve and protect humanity'' to be inherited and proliferated.  As well, progress on defining the notion of descendant in Approach 1 could be re-purposed for a better understanding of heritability in general, and could thereby indirectly contribute to dual purpose progress within Approach 2(a).

		\directiondef{ethic}{human-compatible ethics learning}
			It is conceivable that human-favorable behavior norms for a powerful AI system interacting with human society could be derived from some more fundamental ethical abstraction, such as loyalty or reciprocity of an agent toward other agents that have allowed its existence, which would include humans.  This research direction involves investigating that possibility.

			\Soc Many individuals experience a sense of loyalty to the people and systems that have empowered them, for example, their parents and teachers, their country of origin, the whole of human civilization, or nature.
			As a result, they choose to align their behavior somewhat with their perceptions of the preferences of those empowering systems.

			\Mot It is conceivable that  many peculiarities of human values will not be easily describable in terms of individual preferences.  There may be other implicit constraints on the behavior of individual humans that would violate the von Neumann-Morgenstern rationality axioms for individual agents, but might be valuable at the scale of group rationality.  For example, a person might reason ``I won't do X because if everyone did X it would be bad, even though if only I did X it might be slightly good.''

			Failing the development of an explicit theory for learning ``non-preferential'' human values, a fallback option might be to discover cooperative ethical principles from scratch, and then test to see if they suffice for sustainable cooperation with humans.  This would add another potential pathway to alignment, thereby reducing \riskref{urmis}.  Perhaps the ethic ``avoid acquiring too much power'' could be among the ethical principles discovered, leading to a reduction in \riskref{urprep}.  In principle, preference learning and ethics learning  could be complementary, such that partial progress on each could be combined to build more human-aligned systems.

			\Inst In addition to posing an complementary alternative to preference learning, work on \dirtitle{ethic} could yield progress on

			\begin{itemize}
			\item \Dirref{preferencelearning} and \dirref{compromisingbetween}.
			It is conceivable that a single basic principle, such as loyalty or reciprocity, would be enough to derive the extent to which an AI system should not only achieve \dirtitle{preferencelearning} with the human customer who purchases the system, but also with the engineers who designed it, and other individuals and institutions who were passively tolerant of its creation, including the public.  The system could then in theory be directed to exercise some of its own judgment to determine the relative influence various individuals and institutions had in its creation, and to use that judgment to derive appropriate compromises between conflicts in their preferences.

			\item Limited instances of \dirref{mod}.
			A system which derives its loyalties implicitly from the full history of institutions and people involved in its creation---rather than from a simple ``whom to serve'' attribute---might be more difficult to redirect to serve the purposes of a delinquent individual, thus addressing certain instances of \dirtitle{mod}.
			\end{itemize}

			\Act This direction could benefit from progress on \dirref{rigorouscoordination}, to the extent that human-compatible ethics will involve cooperation with humans.  Decentralized learning of cooperation is more likely to be applicable than centralized learning of cooperation: when an AI system learns to cooperate with a human, the human's beliefs and policies are not being controlled by the same training process as the AI system's.  That is, any group that includes humans and AI systems working together is a decentralized learning system.

			Implicit progress and insights might also be drawn from working on other research directions in this report, such as \dirrefs{transparency,preferencelearning,deference,self,compromisingbetween}.
			AI researchers will likely encounter disagreements with each other about how to operationalize ethical concepts such as loyalty or reciprocity to humanity, just as developing technical definitions of concepts like cause, responsibility, and blame have also been topics of debate among AI researchers \citep{mccoy2012blame,halpern2015cause}.  Hence operationalizing these concepts may need to go through numerous rounds of discussion and revision before researchers would converge on satisfactory definitions of what constitutes ethics learning, and what ethics are human-compatible.

			\Cse In order to selectively advance technology that would enable human/machine cooperation rather than only machine/machine cooperation,
			studies of decentralized machine/machine cooperation will need to be thoughtful about how humans would integrate into the system of cooperating agents.  Otherwise, these research directions might increase the probability of runaway economies of AI systems that cooperate well with each other at the exclusion of human involvement, increasing \risksrefs{econ,humenf}.

		\directiondef{self}{self-indication uncertainty}
			AI systems can be copied, and can therefore be implemented in numerous distinct environments including test environments, deployment environments, and corrupted environments created by hackers.
			It is possible that powerful AI systems should be required to be built with some awareness of this fact, which we call ``self-indication uncertainty''.

			\Soc Self-indication uncertainty is not a matter of necessary practical concern for most humans in their daily life.
			However, suppose a human named Alice awakes temporarily uncertain about whether she is still dreaming.
			Alice may be viewed as being uncertain about whether she is ``Real Alice'' or ``Dream Alice'', a kind of self-indication uncertainty.
			To put it another way, Alice is uncertain about whether her current perceptions and actions are taking place in the ``real world'' or the ``dream world''.

			A more familiar but perhaps more tenuous analogy is the following.
			Suppose Alex is a supporter of a certain political party is considering staying home instead of voting, because he expects his candidate to win.
			He might find himself thinking thoughts along the lines of ``If I stay home, does that mean many other supporters of my party will also stay home?  And if so, doesn't that mean we'll lose?''  Now, consider the mental subroutines within Alex that are deciding whether he should stay home, and generating the above question in his mind.
			These subroutines may be viewed as uncertain about whether they are deciding just for the one voter (Alex), or for a large number of ``copies'' of the same decision-making procedure inside the minds of many other supporters of her party.
			In other words, the vote-or-stay-home \emph{subroutine} has self-indication uncertainty about who (and in particular, how many party members) it is operating within.

			\Mot See instrumental motivations.

			\Inst Progress on modeling or training \dirtitle{self} could be useful for some instances of:
			\begin{itemize}
				\item \Dirref{corrigibility}.
				Ensuring that an AI system that is able to wonder if it is a misspecified version of its ``true self'' could aid in motivating the system to seek out corrections for those misspecifications.
				For example, consider an AI system which, after real-world deployment, maintains some degree of uncertainty about whether it is operating in a pre-deployment test environment.  Such a system might be more likely to comply with shut-down commands if it believes non-compliance in the test environment would result in non-deployment and therefore no opportunity to pursue its real objective in the real world.
				It may even be the case that some degree of \dirtitle{self} of this form is needed for an AI system to exhibit the degree of ``humility'' that humans naturally exhibit and would like to see exhibited in AI systems.
				That is to say, it remains an open question whether implicit or explicit \dirtitle{self} is a necessary condition for \dirtitle{corrigibility}.

				\item \Dirref{deference}, \dirref{hoal}, and \dirref{equilibria}.
				A computerized decision algorithm that knows it is being implemented on many different machines at once might reason, when making a decision, about the consequence of all of its copies making that same decision, rather than fallaciously assuming that only one of its copies will do it.
				This could allow an individual AI system acting within a collective or hierarchy of other AI systems to derive and follow principles that are appropriate for the entire group, playing a role in the multi-agent dynamics of \dirabbrev{hoal} and \dirtitle{equilibria}.
				Perhaps \dirtitle{deference} is such a principle.
				\item \dirref{open}.
				For each decision a game-playing agent takes, it could be made uncertain about whether the resulting action is taking place in its own reasoning (its true self) or in another player's model of the agent's reasoning (an approximate representation of its true self).
				\item \Dirref{mod}.
				An AI system which knows it might be running on a hacker's computer might be intrinsically motivated to take extra precautions before a highly-impactful action like disabling a portion of the electrical grid.

			\end{itemize}

			\Act Any particular method for representing indexical uncertainty for an agent can be readily tested for efficacy in solving Decentralized POMDPs (Dec-POMDPs), a well-known problem class for stating decentralized coordination problems between agents with a common objective \citep{bernstein2002complexity}.
			For inspiration, Dec-POMDPs can also be related to older work on decision problems with imperfect memory \citep{piccione1997interpretation}, by observing that solving a Dec-POMDP is equivalent to solving a ``memoryless POMDP'', i.e., a POMDP where the solution policy is prohibited from passing information from one time step to the next.
			\citet{piccione1997interpretation}, and later \citet{taylor2016memoryless} using more modern notation, have shown that any memoryless POMDP solution may be viewed as a Nash equilibrium, by viewing each memory loss as the creation of a new agent with the same utility function but with a certain form of uniform uncertainty about which agent it is.
			The form of uncertainty used in the equilibrium, which Bostrom calls the ``Self-indication assumption'' \citep{bostrom2007sleeping}, could be incorporated in decentralized Dec-POMDP solvers and tested for scalability.
			Future work would need to account for the possibility that copies of an AI system can be made with only minor variations, in which case perhaps a non-Boolean concept will be needed for whether to view two such systems as sharing the same identity.

			\Cse How to ideally implement self-indication uncertainty for an AI system is a subtle question, because it might enable the system to reason about whether it is operating within a test environment or a real environment.  If the system is successful in determining whether or not it is undergoing a test, it could choose to exhibit different behavior in the test than in reality, thus invalidating tests used to assess the system's safety.  Hence, there are ways in which self-indication uncertainty as a capability could undermine certain approaches to safety.

	\nodedef{Single/multi control}{smcontrol}
		\fl{3}
			Analogues of \dirref{safeshutdown}, \dirref{corrigibility}, and \dirref{deference} all exist for a multi-agent system serving a single human principal.  Achieving solutions to these problems in a single/multi control setting will likely involve challenges that did not exist in the single/single paradigm.  This section aims to identify some such problems.  For instance, what precisely does it mean to shut down an AI system if it is able to transmit a copy of its source code to another existing AI system prior to being turned off?

			Unlike in the other research sections of this report, actionable research directions are not identified in this section, only problems.

			\paragraph{Network design and administration: a present-day analogue.}
			The design and administration of
			communications and computing networks
			may be viewed as solving present-day analogues of single/multi control problems, to the extent that network components can be viewed as very rudimentary AI systems.

			For instance, maintaining communications networks requires methods for modifying and shutting down network components.
			The patent literature includes techniques for
			upgrading a programmable device in communication with a server \citep{san2002automatic},
			transmitting an emergency shutdown signal to a malfunctioning device in the network \citep{litwin2006method},
			and gracefully shutting down intermediate network nodes \cite{scudder2008technique}.
			Similarly, maintaining computing networks involves somewhat analogous methods, with patents including techniques for
			allowing a job on a node in a computing cluster to shut down that node if it is malfunctioning \citep{block2005node},
			and putting to sleep or terminating idle virtual machines to sleep to conserve cloud computing resources \citep{huang2014dynamic,rigolet2017automatic}.

			However, each of the above methods seems to depend on components following pre-defined protocols, rather than learned protocols.
			Moreover, irrespective of the method, most of the available literature on the shutdown and maintenance protocols for communications and computing networks is contained in patents---which generally only contain enough detail to enforce ownership disputes---as opposed to research literature that is intended to convey knowledge.  As such, there might be considerable room for academic progress in this area.

			\paragraph{Single/multi delegation will likely present novel control problems.}
			Just as novel administrative challenges arise for present-day computing networks that are not needed for operating a single computer, single/single control solutions are not likely to be entirely sufficient to solve single/multi control problems.

			Consider the problem of safe shutdown for a multi-agent system.  Any operationalization of the command ``safely hand off control to a human or other system'' from \dirnum{safeshutdown} could be deployed in parallel to hand off control from each agent to another ``backup'' agent assigned to it, such as a human overseer.  However, novel failure modes might then arise, including the following:

			\begin{itemize}
					\item \textbf{Unsafe composition of local shutdown protocols.}  Safe protocols for shutting down single agents in a multi-agent system might not be safe when applied to all the agents at once, e.g., if the resulting disruption to overall system performance would be unsafe.

                        To give a human social analogy: while it might be relatively safe for one doctor at a hospital to take a sick day when they're not feeling well, it would not be safe for all the doctors in the hospital to do so at the same time.
					\item \textbf{Malfuctioning of local shutdown protocols.}  If most agents in a multi-agent system successfully shut down as a result of a global shutdown command, but some agents remain active, the actions of the remaining agents might be highly unsafe outside of the context of rest of the system.  To give a human social analogy: the action of a human pilot taking off an airplane is normally a safe action to take, but would be an incredibly unsafe action if air traffic controllers around the world were on strike.  Thus, any procedure that takes air traffic controllers off the job had better take pilots off the job as well.
			\end{itemize}

			\noindent What present-day AI research directions could be undertaken that could begin to address these issues?  The task of identifying concrete next actions for single/multi control research, beyond the repeated local application of single/single control solutions, is a challenge left to the reader and future researchers.

	\section{Relevant multistakeholder objectives}\label{sec:relevantmultistakeholder}
\begin{figure}[H]
				\centering
				\includegraphics[width=\textwidth]{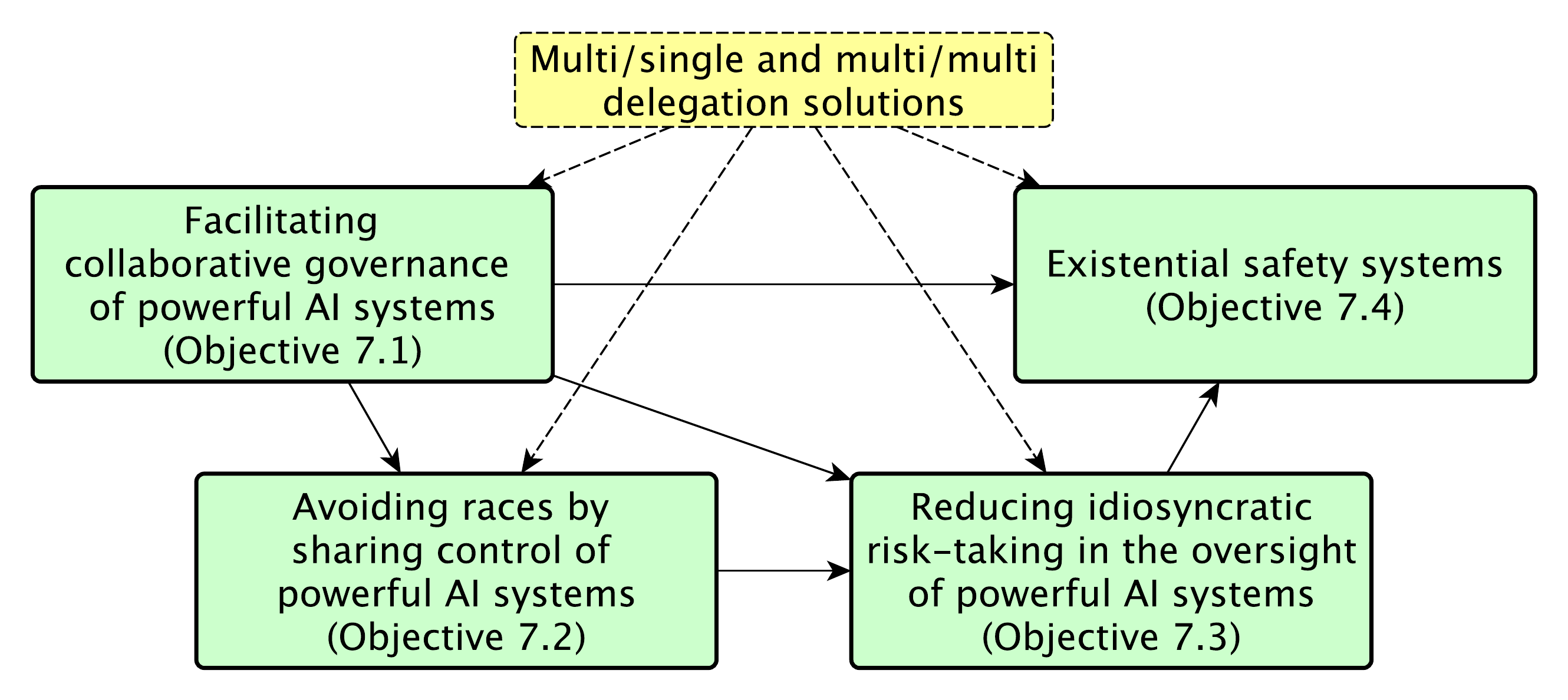}
				\caption{multi-stakeholder objectives}
				\label{fig:objectives}
			\end{figure}

		\fl{3}
		\newcommand{\heavyrevision}{{\color{red} AC: This section is currently under heavy revision.\\}}
			Before proceeding to discuss research directions on multi/single and multi/multi delegation, this section outlines some objectives that \secrefs{ms,mm} will build upon in their scenario-driven motivations. These objectives may also serve as general, high-level guidelines in the furtherance of multi/single and multi/multi delegation research. A diagram of the objectives and their most direct relationships is provided in \figref{objectives}.

			\paragraph{Note on the meaning of ``misalignment''.}  In a setting involving multiple stakeholders with diverse values, what should be considered an ``aligned'' AI system?  While there is much room for debate about what constitutes alignment from the perspective of all of humanity, by contrast there is a great deal of agreement among people that the world becoming unsurvivable to humanity would be a bad outcome.  More generally, there may be many outcomes that nearly everyone would agree are worse than the status quo, such that the concept of \emph{misalignment} might be more agreeably meaningful than \emph{alignment} in many multi-stakeholder scenarios of interest.
			In any case, for the purpose of this report, MPAI will continue to refer to AI systems whose deployment would be unsurvivable to humanity, as it was defined in \secref{mpai}.

		\objectivedef{facilitatingcollaborative}{facilitating collaborative governance}

		As time progresses and the impacts of AI technology increase, existential safety concerns and other broadly important issues will likely lead to an increased pressure for states and companies to collaborate in the governance of AI technology.

		\paragraph{What is collaborative governance?}

		For the purposes of this report, \emph{collaboration} between stakeholders in the oversight of AI technology refers to the exchange of reliable information and commitments between the stakeholders.  \emph{Collaborative governance} of AI technology refers to collaboration between stakeholders specifically in the legal governance of AI technology.  The stakeholders could include representatives of governments, companies, or other established groups.

		Making the governance of AI technology more collaborative, i.e., involving more exchange of information and commitments in the governance process, is not guaranteed to be safer or more effective, as elaborated somewhat below.

		Moreover, the technical properties of AI systems themselves can add to or detract from the options available for multiple stakeholders to collaborate in the oversight of the systems' activities.  We therefore adopt the following objective:

		\centerbox{
		\textbf{\objectiveref{facilitatingcollaborative}} is to make it easier for diverse stakeholders to collaborate in the oversight of powerful AI technologies, by the co-development of AI technology and accompanying governance techniques that will capture the benefits of collaboration in certain aspects of governance while avoiding forms of collaboration that would be unsafe or unnecessarily costly relative to independent governance.
		}

		\noindent This objective may be somewhat complex to achieve, because the potential benefits collaborative governance may also come with a variety of pitfalls that need to be avoided, as follows.

		\paragraph{Potential benefits of collaborative governance.}  Consider a scenario where some powerful new AI capability is being implemented by multiple human institutions, collaboratively or independently, to pursue one or more purposes, such as:
			\begin{itemize}

				\item efficient distribution of electricity from power plants in a safe and equitable manner;
				\item global health research requiring difficult-to-negotiate privacy policies for patients;
				\item education tools that might enable the spread of cultural values that are difficult to agree upon; or
				\item environmental monitoring or protection systems that might require difficult-to-negotiate economic policies.
			\end{itemize}

		There are a number of reasons why the developing institutions might be motivated to collaborate in the governance of this technology, including:
		\begin{itemize}
			\item[A)] to ensure fair representation of diverse views and other objectives in governing their system(s);
			\item[B)] to pool the collective knowledge and reasoning abilities of the separate institutions; or
			\item[C)] to ensure sufficient weight is given to other objectives that are of interest to everyone involved (such as existential safety), relative to objectives only of interest to one person or institution.
		\end{itemize}

		Items B and C here point to an existential safety argument for collaboration in the governance of AI systems:
		a committee of representatives from different institutions of would be less likely to accidentally (by B) or intentionally  (by C) take risks that a single institution might be willing to take.  This consideration is elaborated further in \objectiveref{reducingidiosyncratic}.

		\paragraph{Pitfalls of collaborative governance.}  In pursuing collaborative governance for AI systems, it is important to be mindful that collaborative governance does not guarantee better outcomes than independent governance.	In general, too much collaboration or the wrong kinds of collaboration between institutions can in general lead to a variety of problems:

		\renewcommand{\textit}[1]{\item \textbf{#1:}}
		\begin{itemize}
			\textit{Fragility} if the institutions become more dependent upon one another through collaboration, a failure of one institution risks failure of the other.
			\textit{Interference} the institutions' operations could become entangled in unexpected ways, leading to unexpected errors.
			\textit{Collusion} by collaborating, the institutions could gain too much power or influence relative to other institutions or the public; antitrust and competition laws exist to prevent these outcomes.
			\textit{Groupthink} membership in a group can sometimes cloud the judgement of individuals, by a process known as \emph{groupthink}
			\citep{janis1971groupthink,hart1990groupthink,janis2008groupthink,esser1998alive,janis2008groupthink,benabou2012groupthink}.  In groupthink, individual beliefs are warped to match the prevailing group consensus.  Collaboration between institutions might reduce groupthink within each institution by exposing individuals to views from outside their institution, but it could also increase groupthink if the institutions begin to view themselves as a single large group.

		\end{itemize}

		\noindent Innovations in collaborative governance for powerful AI systems should aim to account for these and other failure modes of collaborative decision-making that would be harmful to many objectives, including safety.

		\paragraph{How and when should governance be collaborative?} When, and in what ways, can collaborative governance of AI systems be more effective than independent governance by essentially separate institutions?  This is a daunting and multi-faceted question that is beyond the scope of this report to resolve.  However, we do aim instigate some technical thinking in this area, particularly as pertaining to existential safety.

		\paragraph{Sources of historical lessons.}
		Absent a satisfying theory of how and when to collaborate in the governance of powerful AI systems, studies of successes and failures in the oversight of safety-critical technologies could yield informative lessons with implications at various scales of governance.

		On the failure side, \citet{sasou1999team} have developed a broad taxonomy of team decision-making failures in the oversight of safety-critical systems, through  examining case studies in aviation, nuclear power, and the shipping industry.
		Charles Perrow's widely cited book \emph{Normal Accidents} \citep{perrow1984normal}---written partially in response to Three Mile Island nuclear accident of March 1979---predicts catastrophic failure in hazardous systems when those systems involve ``complex and tightly coupled'' interactions.  Subsequent technological disasters are also considered in the 1999 edition \citep{perrow1999normal}, such as
		the Bhopal industrial chemical leak in India in December 1984
		\citep{shrivastava1992bhopal},
		the explosion of the US space shuttle Challenger in January 1986
		\citep{vaughan1996challenger}, and
		the Chernobyl nuclear accident in Russia in April 1986 \citep{meshkati1991human}.  Perrow contrasts these events with ``normal accidents'', concluding that they involved serious managerial failures and were not inevitable consequences of the underlying technological systems.

		On the success side, positive lessons can be taken from human institutions with strong track records for the safe provision of highly valued services in hazardous industries.  This point has also been argued somewhat by \citet{dietterich2019robust}.
		There is an existing corpus of academic studies examining so-called \emph{high reliability organizations} (HROs), i.e., ``organizations that operate beneficial, highly hazardous technical systems at high capacity with very low risk, for instance, the effective management of physically (and often socially) very hazardous production processes with very low incidents of operational failure''
		\citep{laporte1995regulatory}.  Examples of organizations identified and studied closely as HROs by organizational researchers include
		\begin{itemize}
		  \item two nuclear-powered aircraft carriers \citep{rochlin1989informal,roberts1989new,roberts1990some,roberts1994decision,schulman1993negotiated},
			\item the US Federal Aviation Administration's Air Traffic Control system \citep{roberts1989new,klein1995organizational},
			\item several nuclear power plants \citep{klein1995organizational,laporte1995regulatory,bourrier1996organizing},
			\item  electricity providers \citep{roberts1989new,schulman2004high}, and
			\item a large California fire department \citep{bigley2001incident}.
		\end{itemize}
		\noindent HRO researchers have gone on to produce theories and recommendations for organizations in general to achieve high reliability
		\citep{laporte1996high,rochlin1999safe,roberts2001must,roberts2001systems,ericksen2005toward}.  Perhaps similar theories could someday be formulated quantitatively as principles for multi/single and multi/multi AI delegation in powerful AI systems.

		\paragraph{Summary.} Collaborative governance of AI systems is attractive from the perspective of issues that concern everyone, such as existential safety.  However, collaborative governance is not automatically more effective than independent governance. The objective of this subsection, \emph{\objectivetitle{facilitatingcollaborative}}, means finding collaborative AI governance techniques that are beneficial from many perspectives (including existential safety), and that avoid pitfalls of collaborative governance.  How exactly to achieve this is a complex social question that is beyond the scope of this report to answer, but is something the authors are beginning to explored somewhat at a technical level.

		\objectivedef{avoidingraces}{avoiding races by sharing control}
			If powerful AI technology is developed in a manner that makes it difficult for multiple stakeholders to share control of a single system, there is some degree of pressure competing stakeholders to race in AI development so as to secure some degree of control over the how the technology is first used. Conversely, the pressure to race can be alleviated somewhat by developing AI technology in a manner that makes it easier for multiple stakeholders to control a single system, such as by designing the system to receive inputs representing beliefs and values from multiple users. Hence, we adopt the following objective:
			\centerbox{
			\textbf{\objectiveref{avoidingraces}} is to make collaborative oversight of AI systems by companies and governments sufficiently easy and appealing as to significantly reduce pressures for AI development teams to race for first-mover advantages in the deployment of powerful AI systems, thereby reducing \riskref{races}.  The nature of the collaboration between the overseeing stakeholders could involve exchange of information, exchange of commitments, or both.
			}

			This objective may be challenging to pursue while respecting the letter and spirit of antitrust laws.  Thus, some degree of progress on  \objectiveref{facilitatingcollaborative} may be needed to ensure that control-sharing between companies cannot lead to collusion or other unfair business advantages that would harm society.

		\objectivedef{reducingidiosyncratic}{reducing idiosyncratic risk-taking}

			Consider two groups, Group 1 and Group 2, each with somewhat distinct beliefs and values, who are each involved in the governance of powerful AI capabilities that might otherwise pose a risk to global public safety.
			The two groups might be states, companies, or other common interest groups.  The AI systems under governance might be owned by one or both of the two groups, or by parties who have invited the two groups to participate in governing their systems.

			Suppose each group, in pursuit of its goals for the AI capabilities in question, would be willing to expose the global public to certain risks.  For instance, one of the groups might be willing to accept a certain level of existential risk if it means furthering a political agenda that the group believes is important.
			Since risks to the global public would negatively affect both groups, involving them both in the governance of a particular system would mean global public safety is doubly represented as a concern in the governance of that system, and might therefore be expected to have safety benefits relative to involving just one of the groups.  Hence, we adopt the following objective:

			\centerbox{
			\textbf{\objectiveref{reducingidiosyncratic}} is to co-develop AI technologies and accompanying governance techniques that enable multiple governing stakeholders to collaborate on mitigating the idiosyncratic risk-taking tendencies of each stakeholder, while still enabling the group to take broadly agreeable risks.
			}

			This objective is non-trivial to achieve.  Involving more groups in governance is not automatically helpful from a safety perspective, as discussed somewhat already in \secref{facilitatingcollaborative}.  For instance, the added complexity could render coordination more difficult for the governing body, or create a diffusion of responsibility around issues that are well known to concern everyone.

			Progress in \objectiveref{facilitatingcollaborative} can be expected to benefit this objective somewhat, insofar as consideration of risks will arise in the process of collaborative governance.
			Also, since races in AI development might cause the racing parties to take risks in order to best the competition, progress in \objectiveref{avoidingraces} benefits this objective as well.

			Beyond progress in \objectivenums{facilitatingcollaborative,avoidingraces}, there may also be ways to specifically promote the avoidance of risks to public safety, e.g., by designing AI systems that can be instructed to safely shutdown without much difficulty, and granting each member of a diverse governance committee authorization to initiate a shutdown procedure.

		\objectivedef{existentialsafety}{existential safety systems}

				In this report, an \emph{existential safety system} is any somewhat autonomous organizational unit, comprising some combination of humans and/or automated tools, which exists and operates primarily for the purpose of protecting global public safety at a scale relevant to existential risk.

				\paragraph{Examples of existential safety systems.}
				For concreteness, consider the following potential mandates for a hypothetical existential safety system:

				\renewcommand{\textit}[1]{\item \textbf{(#1)}}
				\begin{itemize}
					\textit{manufacturing oversight} The system monitors the worldwide distribution of manufacturing capabilities, for the purpose of warning human authorities if the capability to easily build a destructive technology might be developing within a particular group or region.
					\textit{technological forecasting} The system aids in the forecasting of technological developments, for the purpose of identifying if hazardous advancements are on the horizon and warn human authorities to prepare for and/or avert them.
					\textit{conflict prevention} The system aids in the monitoring of other powerful entities under the control of human authorities (such as states, corporations, or AI systems), and predicts potentially catastrophic conflicts between the entities, for the purpose of warning humans with access to legitimate means of diffusing the potential conflicts.   Peacekeeping and counterterrorism are both instances of conflict prevention.
					\textit{shutdown issuance} The system is involved in issuing shutdown commands to powerful automated systems, so that those systems can be quickly deactivated and investigated if they come to pose a substantial risk to global public safety.
				\end{itemize}
				\noindent Certain agencies of present-day human governments might already be viewed as existential safety systems.
				AI technology is not strictly speaking necessary to implement an existential safety system, but could play an invaluable role by assisting in the processing of large amounts of data, composing simulations, or automating certain judgements that are costly for humans to carry out at scale.

				\paragraph{Benefits and risks of existential safety systems.} Such systems could be extremely valuable to humanity because of the safety they can create for allowing other activities to be pursued at scale.  On the other hand, existential safety systems may be difficult to manage because of the potential they create for the accidental or intentional misuse of power.  A system with the potential to monitor and/or impact global public safety has great potential for influence, which could be quite harmful if  misused.

				\paragraph{Challenges to developing existential safety systems.}  Because of the potential for misuse of any of the monitoring or intervention capabilities that existential safety systems would employ, any viable plan for developing new existential safety systems would likely be faced with strong pressures to involve a geopolitically diverse representation in governing the system.  If those pressures are too difficult to resolve, the result could be that the safety system is never developed.  Failure to develop the safety system could in turn could imply either a considerable sacrifice of existential safety, or a considerable dampening of other valuable human activities that are deemed unsafe to pursue without an existential safety system in place to safeguard them.

				Hence, the potential development of AI existential safety systems could benefit from \objectiveref{facilitatingcollaborative}, which might diffuse political tensions regarding who would control or benefit from the systems' operations. As well, progress on \objectiveref{reducingidiosyncratic} is directly relevant to ensuring that existential safety systems would manage risks in a safe and broadly agreeable way.

\section{Multi/single delegation research}\label{sec:ms}
	\fl{2}
		\fl{3}
		\newcommand{\notrefactored}{{\color{red} AC: this section has not yet been refactored to represent changes in terminology and argument structure in \secref{relevantmultistakeholder}.  Once \secref{relevantmultistakeholder} is more stable, these sections will be rewritten at least somewhat, and possibly significantly reorganized.}\\}
		\emph{We now return to the task of outlining actionable technical research directions in this and the subsequent section.}

		Multi/single delegation refers to the problem faced by multiple stakeholders delegating to a single AI system.  This problem---or class of problems---may be key to ensuring that powerful AI systems are capable of benefiting people and institutions from a broad range of geographic, cultural, and individual backgrounds.  Existential safety is one such broadly valued benefit, and per the theme of this report, the multi/single delegation solutions here will be examined for their potential role in reducing existential risks.  However, there are likely many other broadly valuable benefits that could be derived from multi/single delegation solutions.

	\nodedef{Multi/single comprehension}{mscomprehension}
		The single/single comprehension solutions in \secref{ss} above can easily be scaled to help multiple users to understand the same AI system.  As such, there seem to be few problems in multi/single comprehension that are not subsumed by single/single comprehension.  However, at least one issue is not:

		\directiondef{privacyoperating}{privacy for operating committees}

			A committee charged with the task of operating or overseeing a powerful AI system should have the ability to discuss the system's performance in a private setting not observable to the system.  This could be a challenge if the system is sufficiently intelligent as to potentially infiltrate or infer the content of their apparently-closed-door discussions.

			\Soc When an employee undergoes a performance review at work, it helps if the people reviewing the employee can discuss the employee in private.  But, suppose the employee inadvertently finds out the identity of a coworker who granted them a negative review that was intended to be private.  Ideally, the employee can disregard the information and continue acting in a professional manner that is uninfluenced by the knowledge of the coworker's identity or privately expressed opinions.

			\Mot Consider a scenario where a socially intelligent AI system is being used to serve some broadly impactful societal purpose (e.g., for any of the purposes discussed in \secref{facilitatingcollaborative}).  Suppose also that the system is overseen by a committee which holds meetings to discuss the system's performance.  If the committee members are concerned that the details of their conversations might be leaked, they might be reluctant to raise certain issues.  For instance, suppose a committee member begins to suspect that the system is misaligned and  ought to be shut down.  That member might feel some reluctance to propose the shut-down if there is some chance the proposal would be rejected and the misaligned system would, upon becoming aware of the proposal, take actions to make shut-down more difficult. This reluctance to discuss shutdown would be reduced if the committee were able to discuss the system in a setting guaranteed to be private, or else guaranteed not to influence the system's behavior.

			\Act Techniques within the domains of cybersecurity and physical building architecture are outside the scope of this report, but must be mentioned as obvious first-line approaches to enabling private deliberation.  Indeed, research on the technical problem of maintaining privacy from an intelligent adversary is as old as the field of cryptography.  However, in applications of cryptography, one does not design the adversary.  As AI designers, we have the opportunity to decrease the degree to which AI systems would behave intentionally or unintentionally in ways that would violate our privacy.
			\begin{itemize}
				\item With improved techniques for \dirref{transparency}, it may be possible to determine by inspection if and when an AI system is deducing private information from public information.  In such cases, the system could potentially be rolled back to an earlier state with less awareness of the private information, and repaired to ignore it.

				\item There is an existing literature on \emph{non-discrimination} in statistics and machine learning, which has developed techniques for avoiding discrimination based on certain variables. For instance,

				\citet{dwork2012fairness} present an algorithm for maximizing utility subject to a fairness constraint.
				\citet{hardt2016equality} present a method of adjusting a learned predictor (after its training is complete) so as to eliminate a  certain rigorously defined form of discrimination.
				\citet{kilbertus2017avoiding} further develop these concepts using causal models of discrimination based on certain types of information, and put forward algorithms which more finely isolate and eliminate the impacts of discriminatory causal pathways.

				Now, returning to the committee in the scenario-driven motivation: suppose we view the committee's private deliberation process as part of the causal history generating the committee's final endorsed communications to the AI system.
				As a safeguard in case the private deliberations would later be accidentally revealed to the AI system (e.g., via a computer security leak), we could impose a constraint upon the AI system that its behavior should not discriminate upon the private information if it is observed.   That is to say, the private deliberations should be ignored even if seen.

                    Mathematically, this problem is quite similar to the non-discrimination criteria pursued by \citet{dwork2012fairness, hardt2016equality} and \citet{kilbertus2017avoiding}.    Hence, building upon their work to apply it in more complex settings could be a useful way forward.  Ideally, one would not want to depend on only this safeguard to maintain privacy, and of course physical and cybersecurity measures to maintain privacy should be applied by default.

					\item If committee members are interested in privacy for themselves as individuals, perhaps principles from differential privacy \citep{dwork2011differential} could be generalized to allow new members to join the committee without too much being revealed about the influence they are having over its decisions.

			\end{itemize}

			\Cse If AI systems are designed to respect our privacy even when our private information is accidentally leaked, we might fail to notice when the leaks are happening (since the AI systems do not act on the information), and we might then develop a false sense of security that the leaks are not happening at all.  Later, if a malicious actor or malfunction disables the ``respecting privacy'' feature of an AI system, there would be lot of leaked private information available for the system to take action on.  As such, the actionable directions above should be taken as complementary, not supplementary, to standard physical and cybersecurity measures to maintain privacy.

	\nodedef{Multi/single instruction}{msinstruction}
		\directiondef{modelingcommittees}{modeling human committee deliberation}
			A system whose purpose is to serve a committee may need to model not only the preferences of the committee members but also the dynamics of the committee's deliberation process.  This objective is in some sense opposite in spirit to \dirref{privacyoperating}.  Finding a way to serve each of these objectives in some way is an interesting and potentially important meta-problem.

			\Soc A good CEO thinks not only about the individual wishes of their shareholders, but also about the relative weight of the shareholders' investments, and how their opinions and preferences will update at a meeting of the shareholders as they discuss and deliberate together.

			\Mot Consider the task of building any broadly impactful AI system that will serve or be governed by a committee.  Some approaches to human/machine instruction involve the AI system maintaining a model of the human's intentions.  To the extent that such modeling may be necessary, in the case of an AI system serving a committee, the intentions of the constituent committee members may need to be modeled to some degree.  Even if the committee elects a single delegate to faithfully convey their intentions to the AI system, inferring the intentions of the delegate may require modeling the committee structure that selects and/or directs the delegate, including the relative levels of authority of various committee members if they are not equal.

			\begin{itemize}

				\item For \objectiveref{facilitatingcollaborative}, it will save the committee time if the system is able to flesh out a lot of unspecified details in the committee's instructions, which might require imagining what the committee would decide upon if a much longer meeting were held to hammer out the details.

				\item For \objectiveref{avoidingraces}, the committee's instructions would need to be followed well enough to encourage continued collaboration of the stakeholders in operating and maintaining the shared system, as opposed to splintering their efforts in order to build or manage separate competing systems. For this, it might help for the system to be able to model the outcome of deliberations where one of the committee members (or the institution they represent) is considering separating from and competing with the remainder of group in some way.
				\item For \objectiverefs{reducingidiosyncratic,existentialsafety}, it would make sense to ensure that the system follows the committee's instructions with a level of caution that the committee's summary judgement would consider reasonable if the system were examined more closely.  For this, it might help to be able to model the committee's deliberative processes for accepting and rejecting risks, which might be a simpler problem than modeling arbitrary deliberation.

			\end{itemize}

			It is possible that modeling the committee's deliberations might be unnecessary for serving the committee, just as modeling a single human's deliberations is sometimes unnecessary for serving the human.  However, in domains where modeling of human intentions is necessary to serve humans, modeling of deliberation would also seem important because of the impact of deliberation upon intention.

			\Act One approach to modeling committee deliberation would be to use multi-agent system models.  That is to say,
			\begin{itemize}
				\item[1.] Assume the committee members behave similarly to an existing AI model for multi-agent interaction;
				\item[2.] Fit the AI model parameters to the observed behavior of the committee, and
				\item[3.] Use the fitted model to predict and reason about the committee's future behavior and/or opinions.
			\end{itemize}

			In step (1), for any particular committee there is the question of whether it should be modeled as comprising agents with the same goal or slightly different goals:

			\begin{itemize}
				\item \emph{Shared goal.}  In cases where the goals of the human committee members are highly aligned, it might make sense to model their interactions using some of the AI systems examined in \dirref{rigorouscoordination}.

				\item \emph{Different goals.} To allow for the possibility of multiple goals among the committee members, one could instead model their deliberation as a multi-agent negotiation process.  How should the humans in this exchange be modeled?

                    There is little existing work implementing formal models of spoken negotiation between more than two agents at once, but perhaps some inspiration could be taken from two-agent negotiation models, such as those studied by \citet{rosenfeld2014negochat} or \cite{lewis2017deal}.

			\end{itemize}

			Whether modeling a committee with a shared goal or divergent goals, one would also need to be judicious and perhaps innovative in step (1) to use or develop a model that accounts for known facts about human cognition, such as those explored in \dirref{humancognitive}.

			\Cse Because this objective is directly contrary to \dirref{privacyoperating} where the goal is to prevent the AI system from acting on the committee's deliberations, it should be approached with some caution.  Just as modeling single humans could make it easier to manipulate those humans, modeling committee deliberations may also make it easier to intentionally or unintentionally manipulate the committee's judgments.  Hence, a judicious awareness of how to apply this technology would be needed to ensure it is applied beneficially.

		\directiondef{moderatinghuman}{moderating human belief disagreements}
		This direction is concerned with the development of a service for assisting in the resolution of belief disagreements between conflicting stakeholders or committee members involved in the oversight of impactful systems or processes, including AI systems.

			\Soc A workshop event organizer is faced with the challenge of finding a time and place for a workshop that will not only be pleasing to the participants, but also will meet the participants' advanced approval enough for them to choose to attend.
			This involves the organizer not only accounting for the conflicting preferences of the attendees, but also perhaps for disagreements and misconceptions among their beliefs.  For example, suppose participants A, B, and C would only be willing to attend the workshop if it were held at a certain beautiful and secluded venue, while participants X, Y, and Z would all enjoy that venue but falsely believe that it would be difficult for them to reach by car.  To get everyone to attend, the organizer might need to dispel some misconceptions about the venue in their advertisement for the workshop (e.g., ``Just a 20-minute taxi ride from PQR airport'').  For some issues, the organizer might need to facilitate disagreements among the workshop invitees without having direct access to the truth.  For example, suppose invitee A is planning to attend a private event called the Q Conference and has complained to the workshop organizer that the proposed workshop date conflicts with the Q Conference, but invitee B has said that the Q Conference date will not conflict with the workshop.  Then, the workshop organizer may need to ask A and B to check with the Q Conference organizers until the disagreement is settled and an agreeable workshop date can be chosen.

			\Mot
			\begin{itemize}
				\item For \objectiveref{facilitatingcollaborative},  consider the development of a powerful AI technology to serve a broadly useful societal purpose, that would be governed or overseen by a set of individuals entrusted by society to pursue those purposes responsibly.  If a disagreement moderation service is able to tease out and get to the bottom of sources of disagreement among the overseers, this could help to ensure that truth prevails in the overseers' collective wisdom.

				\item For \objectiveref{avoidingraces}, consider two competing AI development teams who each believe their own approach to AI development is safer and more likely to succeed.  If a trusted third-party disagreement resolution system is able to help the teams to settle their disagreement and effectively agree in advance about who would be likely to win in a development race, the two teams might be able to agreeably combine their efforts in a way that grants slightly more influence over the joint venture to the would-be race winner(s).

				\item For \objectiverefs{reducingidiosyncratic,existentialsafety}, consider a powerful AI system being governed or operated by a commitee of overseers.  A disagreement moderation service might be able to identify when a conflict of interest or other idiosyncratic bias might be affecting the risk assessments of one of the overseers.  Then, the disagreement moderation service could encourage that person to further reflect upon their potential source of bias, or recommend recusing that overseer from the decision.
			\end{itemize}

			\Act There are a number of theoretical results that suggest two agents can reach agreement through communication after making different observations in the world.  Aumann's Agreement Theorem \citep{aumann1976agreeing} shows if that two Bayesian agents with a common prior also have common knowledge of one another's posteriors and the fact that they are both Bayesian, then those posteriors must be equal.
            This raises the question of how common knowledge of posteriors could be established.  Encouragingly, \citet{aaronson2005complexity} showed an effective procedure for exchanging messages such that the agents will converge on agreement within an $\varepsilon$ margin of error after exchanging $1/\varepsilon^2$ bits of information.  \citet{hellman2013almost} generalizes Aumann's theorem in a different direction, by relaxing the assumption of common priors and deriving a bound on posterior disagreement as a function of prior disagreement.  Perhaps an effectively computable version of Hellman's result could also be derived.

            There is also existing work in the social sciences on moderating disagreements.  For instance, \citet{luskin2002considered} and \citet{fishkin2005experimenting} have examined examine human-assisted deliberative procedures for resolving disagreements among laypeople.  Perhaps some such techniques could be automated by building on techniques and interfaces developed in \dirref{aiad}.

			\Cse Disagreement moderation services that push too hard toward reaching agreement and not enough toward truth-seeking could cause institutions to enter deluded states of ``groupthink'' \citep{janis1971groupthink}.  More generally, to the extent that progress on disagreement moderation might involve studying the dynamics of how human beliefs evolve, some of those results could also enable technologies that would be able to manipulate human beliefs in dishonest or otherwise undesirable ways.  As such, work on human disagreement moderation is somewhat ``dual purpose'', and should be therefore be shared and applied judiciously.

		\directiondef{compromisingbetween}{resolving planning disagreements}
			This direction is concerned with assisting in the formulation of plans that multiple stakeholders can agree to execute together, especially in cases where belief or value disagreements might exist between the stakeholders that cannot be brought closer to agreement by moderation techniques such as \dirref{moderatinghuman}.  In such cases, perhaps only a mutually agreeable plan can be found, in lieu of an agreement about underlying beliefs or values.

			\Soc Consider a group of company directors in the process of forming or running a company.  If these directors encounter what appears to be an impassable disagreement about the company's priorities, a period of dysfunction could ensue, or perhaps the company will split or shut down.  However, if the CEO of the company is able to devise or identify a plan that addresses all of the directors' concerns somewhat, perhaps everyone will buy into the plan, and smooth operations for the company can continue.  This ability to devise agreeable plans is a key capability for a CEO.  In a more diffuse sense, this planning ability can also be seen happening even before a company is formed, through the history of legal professionals creating and revising standardized bylaws for forming companies.  Standardized bylaws establish broadly agreeable norms for how companies should operate---including details on how the directors can leave or depart the company if irreconcilable disagreements arise later---such that company directors can readily agree in advance to the bylaws as a plan for governing the company in case of disputes.

			\Mot In all of the multi-stakeholder objectives in \secref{relevantmultistakeholder}, some of the human stakeholders governing the AI system may have conflicting preferences about what the system should do, or differing beliefs that cannot be resolved through further discussion.  If an AI-based component of the system is able to assist them in arriving at a policy that is nonetheless appealing to all of the stakeholders involved, this improves the probability of stakeholders choosing to collaborate in its further development, deployment, and/or operation.

			\Act There are a number of subproblems here that one could begin to address:
			\begin{itemize}
			\item \emph{Accommodating preference disagreements.}  Perhaps this could be achieved by weighting the AI system's model of humans' reward functions \citep{harsanyi1980cardinal}.

			\item \emph{Accommodating belief disagreements.}  When disagreements about facts cannot be resolved through discourse, in order for an AI system to serve multiple stakeholders in a manner that is efficient and agreeable to each of them, the system might need to explicitly model the differences in beliefs between the stakeholders.  \citet{critch2017servant} provide conditions on the structure of such plans that are necessary and sufficient for a plan to be subjectively Pareto optimal to the stakeholders before the plan is executed.

			\item \emph{Rewarding stakeholder engagement.}  In soliciting statements of disagreement between stakeholders, it would help if an AI system could make use of communications from the stakeholders in such a way that the stakeholders do not individually regret sharing information with the system, the way honest participants in a Vickrey-Clarke-Groves auction do not regret placing their bids \citep{groves1973incentives}.  For instance, particular incentive structures might be needed to alleviate or compensate for fears among stakeholders that they might upset one another by revealing their disagreements.  On the other hand, stakeholders might also need reassurance that they will not be unfairly punished  for revealing their idiosyncratic preferences and/or biases, or exploited for revealing  confusions or misconceptions in their beliefs.
			This could be accomplished by allowing stakeholders to share information privately with only the AI system, and not the other stakeholders.  For the stakeholders to prefer this option, they might require a high level of trust that the AI system will make appropriate use of their private information without revealing it.  Perhaps this could be achieved through the adaptation of differential privacy methods \citep{dwork2011differential}.
		\end{itemize}

		\Cse This research direction has potential side effects that are similar to those considered in \dirref{moderatinghuman}, namely, that the resulting techniques could be used to manipulate us humans in ways we would not endorse.  Another potential side effect might be that if the automated formation of broadly agreeable plans obviated the need for humans to settle belief disagreements in order to get things done, then accuracy of human beliefs could gradually deteriorate over time, from a lack of incentive for humans to seek out truth in the process of settling disagreements.

	\nodedef{Multi/single control}{mscontrol}

			This section is concerned with control techniques that could be used for any single AI system serving a committee or diverse group of human stakeholders. As usual thoughout this report, \emph{control} refers to stop-gap measures for when the humans' comprehension and/or instruction techniques are not working.

			Just as in single/single delegation, the overseers of a powerful AI system should retain the ability to shut down or otherwise override the system in at least some circumstances, as a separate fallback procedure if the communication abstractions that normally allow humans to comprehend and instruct the system begin to fail.  A variety of committee structures could be considered for authorizing override commands.  For example, consider shutdown commands.  For some systems, perhaps a shutdown command should require full consensus among all of its overseeing stakeholders to be authorized.  For other systems, perhaps it makes sense for every stakeholder to have unilateral authority to command a shutdown.
		\directiondef{sharexec}{shareable execution control}

			One way to help ensure that multiple stakeholders continue to endorse a system's operation would be to grant each stakeholder control over some aspect of the system's execution.

			\Soc
			When a company tasks a hiring committee with deciding whether to hire a particular candidate, if one member of the committee is sufficiently strongly opposed to hiring the candidate, typically the candidate is rejected.  This is because teams function better when everyone is sufficiently happy with the work environment that they do not want to leave, and if one committee member is very strongly opposed to a candidate, their opposition might be representative of some broader problem the company as a whole would face in employing the candidate.  This is true to the extent that a good hiring committee is one selected to be representative of the company's needs as a whole, with regards to hiring the candidate.

			\Mot This direction is relevant to \objectiverefs{avoidingraces,reducingidiosyncratic}.  Specifically, if the Alpha Institute is sufficiently concerned about the risks that the Beta Institute might take with a new AI technology, the Alpha Institute might be willing to grant the Beta Institute some level of direct control over the Alpha Institute's usage of the technology in exchange for the Beta Institute granting the Alpha Institute similar control over the Beta Institute's usage.

			\Act Perhaps the simplest example of a computer system with shareable execution control is one that requires passwords from multiple users to be entered before it will execute.  More general is the concept of secure multi-party computation; see \citet{du2001secure} for a review.

			However, in some cases, depending on an entirely cryptographic control-sharing mechanism might not be satisfying due to general concerns about cybersecurity risks, and there may be a desire to physically separate system components and share them out between stakeholders.  For instance,  \citet{martic2018scaling} have put forward a method for achieving a separation of trained AI system components that could be applied in this way, and hypothesize that it might be applicable to any setting where training the AI system is very expensive.

			Related is the concept of ``federated control'' for computer systems.  For instance, \citet{kumar2017federatedcontrol} have begun to examine methods of global optimization directed by locally controlled units, although not in a manner that grants the individual units the potential to unilaterally control the entire collective, e.g., via a shutdown command.  For very large numbers of stakeholders to control specifically the learning process of an AI system, some ideas from so-called ``federated learning'' might be applied \citep{konevcny2016federated,smith2017federated,mcmahan2016communication}.  However, these approaches also do not grant any special unilateral controls to the individual participants in the process.

			For any of these methods to be valuable in practice, one would need to ensure that the individual stakeholders sharing control of the system do not shut down their system components so often as to render the system useless and hence not worth sharing.
			For instance, this could happen if there is widespread doubt or disagreement about whether the system is operating correctly.

			Progress on \dirrefs{transparency,moderatinghuman,compromisingbetween} might be helpful in addressing such scenarios.

			 \Cse The ability to threaten the shutdown of a powerful AI system that is deeply integrated with the well-being and functioning of human society is a privilege that could easily be abused if a malicious actor gained access to the shutdown mechanism(s).  As such, access to such control mechanisms, if they exist at all, should only be granted to highly secure and trustworthy institutions, and the access itself should likely be revokable in the case of suspected abuse or security compromises.  Conversely, this concern also presents a general argument against the creation of AI systems that would cause widepread harm to humans in the case of a surprise shutdown event.

\section{Multi/multi delegation research}\label{sec:mm}

	\fl{2}
		\fl{3}
			This section addresses technical problems and solutions arising for multiple human stakeholders delegating to multiple AI systems. Multi/multi delegation encompasses novel problems not addressed by single/single, single/multi, and multi/single delegation, many of which will be important to ensuring powerful AI systems will bring about robustly beneficial outcomes for all human persons.

			Some of these problems may also be relevant to existential safety.  For instance, in multi/multi delegation, some of the human/machine interactions might cross what would otherwise be natural stakeholder boundaries within the composite multi/multi interaction, such as personal property lines or state boundaries.  As such, solutions may require more than mere compositions of human/human and machine/machine interaction methods, to avoid risks that could arise from coordination failures or conflicts.  In terms argued by \citet{rahwan2018society}, novel tools will likely be needed to program, debug, and maintain an ``algorithmic social contract'' between humans and mediated by AI systems.

	\nodedef{Multi/multi comprehension}{mmcomprehension}
			Multi/multi delegation raises novel problems in human/AI comprehension.  For instance, what happens when Stakeholder A wishes to comprehend an AI system that is being developed, owned, or used primarily by Stakeholder B?  How can A respect B's privacy in this process?  And, if given only limited opportunities to observe B's system, how can A use those opportunities judiciously to answer only their most pressing and important questions about B?

		\directiondef{prepotencefree}{capacity oversight criteria}
			This research direction is concerned with the identification of quantitative and qualitative criteria describing what capacities might be either necessary or sufficient for any given research group to develop a prepotent AI system.  Such criteria could be used to define registration and auditing requirements for AI development groups, creating opportunities for outside stakeholders to comprehend and assess the safety and ethics of otherwise proprietary AI systems.  Outside oversight is thus an aspect of multi/multi comprehension: it allows stakeholders other than the developers and owners of a given AI system to understand how the system works and is being used.

			Many would argue that the potential for capabilities far less than prepotence should be sufficient to trigger outside oversight of powerful proprietary AI systems.  Others might argue that too much oversight can stifle innovation and deprive society of invaluable scientific advancements.  Without taking a side on this age old debate of regulation versus innovation---which is liable to be settled differently in different jurisdictions---it might still be easy to agree that the capacity to develop and deploy a prepotent AI system is definitely sufficient to warrant outside oversight.  Therefore, success in this research direction could potentially yield agreeable worldwide limits on what is acceptable for AI development groups to do without outside oversight.

			\Hist The eventual need for oversight standards for AI development may be similar to the present-day NIH guidelines for research involving recombinant or synthetic nucleic acid molecules \citep{nih2013guidelines}, or the NSABB's recommendations for the evaluation and oversight of proposed gain-of-function research \citep{nsabb2016recommendations}.

			\Soc It is common for business regulations to apply to a company only when that company acquires a threshold amount of a certain resource.  For instance, various regulations for farming in the United States are triggered by threshold levels of production, land area, service connections, or fuel storage \citep{epa2019laws}.  These rules ensure that regulatory effort is commensurate with the scale of a company's potential for impact.  Analogous principles could be used to oversee the usage of any large amounts of data, communication bandwidth, or processing power thought to be sufficient for accidentally or intentionally developing a prepotent AI system.

			\Mot As human society's potential to develop powerful AI systems increases, at some point we must collectively draw some agreed upon lines between computing activities that
			\begin{enumerate}[1)]
				\item should be considered obviously safe,
				\item should require self-applied safety precautions,
				\item should require third-party regulatory oversight for potentially dangerous system developments, or
				\item should not be permitted under any conditions (e.g., the development of a misaligned, prepotent system).
			\end{enumerate}

			\noindent Where should these lines be drawn?  As with any standards, a balance will need to be established between the necessary and the unnecessary.  The present research direction would aim to strike this balance using empirical and mathematical research on what exactly is necessary, and what exactly is sufficient, to avoid the development of prepotent systems, as well as systems that might risk non-existential but nonetheless catastrophic destablizations of human society.  Clarifying our shared understanding of (1)-(4) above is directly relevant to \objectiveref{facilitatingcollaborative}, and hence also to \objectiverefs{avoidingraces,reducingidiosyncratic,existentialsafety}.

			\Act How can one determine what capacities are necessary or sufficient to build prepotent systems, without actually taking the risk of building a large number of prepotent systems to experiment with?  To answer this question safely, a combination of as theoretical and empirical approaches will likely be needed, enabling both quantitative and qualitative assessments.

			Empirical work in this area could begin by quantifying how computational resources like processing speed, memory, and parallelism can be translated---under various algorithmic paradigms---into the ability to out-perform humans or other algorithms in game play.  Measurements of this nature are already commonplace in AI development for competitive games.  As well, in the training of generative adversarial networks \citep{goodfellow2014generative}, there is somewhat of an art to preventing either the discriminator from outperforming the generator too early during training, by limiting the number of intermediate computational steps afforded to discriminator.

			Such research could conceivably lead to general insights regarding balances between learning processes.  For instance, if one system is able to learn much faster than another, when does this result in an equilibrium that strongly favors the fast learner?
			If sufficiently general, answers to such questions could be applied to algorithmic models of human cognition along the lines of  \citet{griffiths2015rational}, so as to make and test predictions about resource levels that might be necessary or sufficient for a system to learn too quickly for human society to adapt to the system.

			Further in the future, empirical findings should eventually be formalized into a theory that allows for the reliable prediction of when one system will be competitively dominant over another, without needing to run the systems in a competition to observe the results. Perhaps these ideas could be practically useful well before any risk of prepotence exists.  For instance, suppose one wishes to ensure a reasonably equitable distribution of  technological resources between two distinct human populations or groups.  This might be operationalized as requiring that neither group should become ``relatively prepotent'' with respect to the other.  Perhaps this requirement could even be formalized as an agreement or treaty to prevent the development of ``relatively prepotent'' AI technologies.  A mathematical theory adequate to address this question might also help to estimate what resources would be necessary or sufficient for an entirely non-human system to achieve competitive dominance over humanity as a whole, i.e., prepotence.

			\Cse There are a number of potential negative side effects of research in this area:
			\begin{itemize}
				\item Experimenting with ``relative prepotence''---i.e., the competitive dominance of AI systems over humans or other AI systems in multi-agent scenarios---could select for the creation of AI systems with generalizable tendencies leading to absolute prepotence.

				\item Publishing results on capacities that are either necessary or sufficient for prepotence could encourage malicious actors to obtain those capabilities.  This suggests some level of discretion in distributing such findings.

				\item Consider the way published speed limits on highways lead to everyone driving at or very near to the speed limit.  Publishing recommended computing capacity limits for development teams might similarly encourage many individuals and/or institutions to obtain computing resources that that are just just short of triggering registration or auditing criteria.  This suggests setting registration and auditing criteria with the expectation that many actors will operate just short of triggering the criteria.
			\end{itemize}

			\nodedef{Multi/multi instruction}{mminstruction}
				When a single AI system receives an instruction form a single human stakeholders in a multi/multi delegation scenario, those instructions will need to be taken in a manner that does not interfere too much with the other humans and AI systems in the interaction.  This presents many novel challenges for human/AI instruction research, of which the following is just a single illustrative example.

				\directiondef{socialcontract}{social contract learning}
				This research direction is concerned with enabling AI systems to respect the ``terms'' of a social contract with multiple stakeholders, including existing institutions such as states, businesses, and human civilization as a whole.

				\Hist There is a point of view in moral and political philosophy known as \emph{social contract theory} \citep{rousseau1766contrat,rousseau2002social}.  In this view, ``persons' moral and/or political obligations are dependent upon a contract or agreement among them to form the society in which they live'' \citep{friend2004social}.  The relevance of a social contract to shaping the impact of science and innovation was argued by \citet{gibbons1999science}.

				\Soc Suppose Alice works for Alphacorp and Bob works for Betacorp.  Neither Alice nor Bob has read the relevant sections of state and federal legal code governing their companies.  Nonetheless, some things just feel wrong to do.  For instance, suppose Alice and Bob go on a date, and Alice knowingly presents Bob with an opportunity to sell Betacorp widgets to Alphacorp at an inflated price that Bob knows is exorbitant for Alphacorp.  Common sense might say that Alice is acting in ``bad faith'' with respect to her Alphacorp duties.  But what is ``bad faith'' exactly?  Even if Bob doesn't quite know the definition, he might be uncomfortable with the deal.  He might even turn down the deal, not out of loyalty to Betacorp's shareholders--who would in fact stand to benefit from the sale---but out of respect for the ethical norm that Alice should be more professional in her representation of Alphacorp.  While this norm might technically be enforceable by state or federal law enforcement's protection of Alphacorp's right to terminate Alice if she acts in bad faith to her company duties, Bob's respect for the norm is more difficult to explain in purely legal terms.  It seems Bob has learned to respect a certain kind of social order in business dealings that he is not willing to associate with violating.

				\Mot Ideally, powerful AI technology should avoid disrupting human society at scales that would pose significant risks to humanity's continued existence.

				Thus, an existential catastrophe may be viewed as an extreme form of disruption to social order, which might be entirely preventable if less extreme risks of disruption are also avoided.  In particular, maintaining certain forms of social order might be necessary to avoid \riskref{hazardoussocial}, and might be integral to pursuing  \objectiverefs{avoidingraces,reducingidiosyncratic,existentialsafety}.

				\Act The self-driving car industry presents a natural opportunity to observe when and how learning algorithms can respect the implicit terms of a social contract \citep{leben2017rawlsian,rahwan2018society,contissa2017ethical}.  For instance, when two self-driving cars interact, there are at least four agents involved: the two cars, and their two passengers.  Each car needs to take actions that will respect the other vehicle while protecting their own passenger sufficiently well to retain their loyalty as a customer of the car manufacturer and/or ride provider.  With larger numbers of cars, car manufacturers will also need to ensure their cars avoid collectively causing coordination failures in the form of traffic jams.  Viewed  at this larger scale, any given self-driving car will implicitly be serving numerous human and institutional stakeholders, in way that needs to strike a 'deal' between these many stakeholders for the self-driving car industry to unfold and continue operating successfully.

				There is already a strong interest in identifying end-to-end training methods for self-driving cars \citep{bojarski2016end}, as well as interest in the ethical problems the industry could face \citep{goodall2016can}.  Imitation learning via reward learning is already being explored for this application \citep{laskey2017dart}.

				It seems plausible that a better understanding of the social aspects of driving may be crucial to progress in this area, including aspects of driver-to-driver communication via movement \citep{brown2017trouble}, and how to plan through a series of such signaling behaviors \citep{fisac2019hierarchical}.  Safety and ethics solutions for driverless vehicles that are sufficiently respectful of human-driven vehicles, and that will alleviate rather than precipitate

				large-scale coordination problems like traffic jams, may lead to many insights and principles for the safe and gradual introduction of autonomous agents into society.

				\Cse As with any safety-critical technology, there is always the risk of premature deployment.  For example, if self-driving car algorithms are deployed at scale before their interaction effects are well understood, car accidents and/or large-scale traffic coordination problems could result.  On the other hand, if the self-driving car industry is sufficiently careful to avoid such failures, there might still be subsequent risks that safety or ethical solutions for self-driving cars could be prematurely deployed in other areas where those solutions would not result in adequate safety or ethics.

\nodedef{Multi/multi control}{mmcontrol}
		\directiondef{mod}{reimplementation security}
			This research direction is concerned with preventing individual stakeholders from modifying or otherwise reimplementing individual AI systems in a multi/multi delegation scenario, in cases where such modifications would jeopardize the safety or ethics of their overall interaction.

			\Soc Suppose Bob has been entrusted with the capability to make large payments from his employer's bank account.  One hopes that an outsider could not easily induce Bob to abuse that capability simply by serving Bob a recreational drug that would distort his sense of safety or ethics.

			That is to say, one hopes that Bob will not be vulnerable to attempts to 'modify' him in ways that would compromise his judgement.  For this and other reasons, some institutions conduct regular drug testing to ensure the judgment of their members is unlikely to be compromised by drugs.

			\Mot  In general, many measures may be needed to lower the risk of unauthorized modifications to publicly available AI technologies.  For instance:

			\begin{itemize}
				\item[(1)] If any AI system could be modified and/or scaled up to versions that would threaten public safety, then before sharing the system with the public, its code should probably be obfuscated to diminish the risk of unauthorized individuals modifying or scaling it up.  The degree of effort and security should be commensurate with the degree of risk.
				\item[(2)] If large numbers of research and engineering developers are employed in the task of developing or maintaining a near-prepotent AI system, protocols may be desired for allowing the many developers to carry out experiments and make changes to the system without having read access to its full source code.
				\item[(3)] Access to hardware sufficient to reverse-engineer the software components of near-prepotent AI systems should be closely monitored and in many cases restricted; see also \dirref{prepotencefree}.
			\end{itemize}

			Without appropriate security measures to prevent unsafe reimplimentations of powerful AI systems, careless AI developers could precipitate \risksrefs{uncdev,urprep,urmis,invdep}, and malicious or indifferent developers could precipitate \risksref{voldep}.  On a societal scale, ensuring powerful AI systems cannot be easily modified to disregard safety or ethical guidelines is a manner of pursuing \objectiveref{reducingidiosyncratic}, and might be a necessary for \objectiveref{existentialsafety}.

			\Act Problems (1) and (2) above might benefit from program obfuscation techniques \citep{anckaert2007program, bitansky2011program}, which allow potential adversaries to interact with a program without being able to easily understand its inner workings.
			Determining obfuscation techniques that work well with present-day machine learning systems, without slowing down their operation significantly, would be a good start.

			To address (3), cloud computing resources could be safeguarded by machine learning techniques for intrusion detection \citep{buczak2015survey}.
			Large deployments of offline computing resources might also be detectable in some cases
			by repurposing smart supply-chain monitoring systems currently used for demand forecasting \citep{carbonneau2008application}.

			\Cse It would be quite a problem if a powerful, incorrigible AI system used a combination of reimplementation security techniques to prevent all humans from correcting its code.  Or, imagine an AI-based malware system that somehow makes it extremely technically or socially difficult to restore its host hardware to a clean state.  Indeed, anywhere that repairs to computer systems might be needed, reimplementation security techniques could conceivably be abused to make the repairs more difficult.

		\directiondef{equilibria}{human-compatible equilibria}
			This research direction is concerned with developing a more realistic understanding of game-theoretic equilibria and population equilibria where some of the agents involved are humans, and where the human agents are guaranteed not to be destroyed or replaced by the dynamics of the interaction.

			\Soc The following scenario describes a \emph{disequilibrium}.
			Suppose Alice runs a small business, and to attract more clients, she opens a small blogs for sharing news and insights relevant to her work.

			She soon learns that many other business owners in her industry are outsourcing their blog-writing to advertising companies that specialize in \emph{search engine (ranking) optimization} (SEO).
			SEO companies make a systematic study of search engines like Google, and learn how to optimize webpage content to rank more highly in search engine results \citep{beel2009academic}.
			So, Alice contacts an SEO firm to begin outsourcing some of her blog design.
			At first she only outsources decisions regarding the layout of the blog.
			However, when she falls unacceptably behind her competitors in Google's search rankings, she decides to outsource her choice of headlines to the SEO firm as well.
			Eventually, Alice entirely replaces herself in her role as a blog writer, with an SEO firm writing entire blog posts on her behalf, by imitating the style and content of posts from more successful companies.  In this story, Alice was not at equilibrium with Google in her role as a blog-writer, and was eventually replaced by the SEO firm.

			\Mot There are a number of reasons why there might be no human-compatible program equilibrium in a given game:
					\begin{itemize}
						\item (speed) The human may simply be too slow relative to a software system that would replace them.
						\item (decision quality) The human might make worse decisions than a software counterpart.
						\item (transparency / verifiability) A human is not able to make the contents of their mind readable to others in the way a computer can produce a record of its internal processes.  This could lead to less trust in the human relative to trust that could be placed in an AI system, and therefore weaker performance from the human in games that require trust.

					\end{itemize}

 			Any of these issues could lead to \risksref{econ}, and further to \riskref{humenf}.  Therefore, a need exists to identify ``human-compatible equilibria'': economic and social roles wherein there would be little or no incentive to replace a human being with an AI system.  A simple and trivial example would be a game where the counterparty checks ``Are you human?'' and grants you a reward if and only if you pass the check.  Is this the only sort of game where a human, practically speaking, would be irreplaceable?

			\Act To begin thinking about this dynamic in a simple case, consider a two-player game wherein each player designs and submits a computer program, after which the programs interact in some way that yields a pair of payoffs for the players.
			The programs submitted are said to be in a \emph{program equilibrium} \citep{tennenholtz2004program} if each player has no incentive, from her own perspective, to replace her program with a new version upon seeing the opponent's program.
			This concept is importantly different from the concept of a Nash equilibrium: whereas Nash disequilibrium involves an incentive to change strategies, program disequilibrium involves an incentive to replace an agent in its entirety.

			Using this framework, one can meaningfully ask whether a human being $H$ and an AI system $Q$ are in a program equilibrium, by modeling the human's decision-making process as a probabilistic program $P$, along the lines described by \citet{stuhlmuller2015modeling} or \citet{griffiths2015rational}.  Informally, then, one might define a \emph{human-compatible equilibrium} for a given game as a triple $(H,P,Q)$, where $(P,Q)$ constitute a program equilibrium, and $P$ played against $Q$ is an excellent predictor of $H$ played against $Q$.  In such a case, one would have some assurance of a stable relationship between $H$ and $Q$.  The level of assurance would of course depend on our willingness (and $Q$'s willingness) to rely on $P$ as a theoretical model of $H$.

			What sorts of programs $P$ could make sense to use here?  Or, what instructions could one offer a human to make the human more likely to achieve a human-compatible equilibrium with a an AI system?  One might worry that any program $P$ that achieves an equilibrium with $Q$ would have to be prohibitively different from a human being.

			However, it is known that systems with differing goals, but who are highly transparent to one another (e.g., able to read one another's source codes) are capable of cooperative equilibria arising from the ability to simulate or write proofs about one another's future actions before they are taken.
			This has already been shown possible by \citet{critch2019parametric} for agents who use theorem-provers to decide whether to cooperate with one another in a Prisoner's Dilemma, using a generalization of L\"{o}b's theorem \citep{lob1955solution} to circumvent stack-overflow issues that would otherwise arise from agents reasoning about one another's reasoning.
			As a parallel effort, Chapter 11 of \citet{agentmodels} explores a few scenarios with probabilistic programs that have the ability to sample simplified instances of one another; however, a stack overflow occurs if the agents can make unrestricted function calls to each other.
			This problem could be circumvented by probabilistic programs that exploit the structure of L\"{o}b's theorem in their procedure for deciding whether to cooperate.

			Hence, implementing a ``stack overflow resistant human-compatible program equilibrium'' is a natural and actionable next step.
			Probabilistic program models of humans taken from cognitive science could be used as stand-ins for human agents in early experiments, and perhaps later used by real-world AI systems to assess the cooperativity of humans around them.
			This could serve to ensure that human beings are not excluded from a highly-collaborative machine economy that might otherwise exclude us because of the difficulty of mathematically proving our trustworthiness.

			\Cse Progress toward modeling human-compatible equilibria might yield progress toward modeling general equilibria in games and populations.  Such concepts could potentially be misused, accidentally or intentionally, to develop networks or populations of AI systems that interact very stably with one another, but poorly with humans, or in a manner incompatibly with human morals or ethics.

\ \\

\noindent
This concludes the final research direction examined in this report.

\section{Further reading}\label{sec:further}

	\fl{3}

		For further reading on existential risk from artificial intelligence, see:
			\begin{itemize}
				\item \cite{good1966speculations}. Speculations concerning the first ultraintelligent machine.  \emph{Advances in computers 6}, 31--88.
				\item \cite{yudkowsky2008artificial}.  Artificial intelligence as a positive and negative factor in global risk.  \emph{Global catastrophic risks 1} (303), 184.

				\item \cite{bostrom2012superintelligent}.  The superintelligent will: Motivation and instrumental rationality in advanced artificial agents.  \emph{Minds and Machines 22(2)}, 71--85.
			\end{itemize}
		For reading on existential risk in general, see:
			\begin{itemize}
				\item \cite{matheny2007reducing}.  Reducing the risk of human extinction. \emph{Risk analysis 27(5)}, 1335--1344.
				\item \cite{bostrom2013existential}. Existential risk prevention as a global priority.  \emph{Global Policy 4(1)} 15--21.

			\end{itemize}

		\subsection{Related research agendas}\label{sec:related}
		Several related technical research agendas having themes in common with this report are described below, ordered by year.  Familiarity with these related agendas \emph{is not} a prerequisite to reading this report, but they will make for valuable follow-up reading because of their varied perspectives on the risks and benefits of AI technology.  As well, since \secref{ss,sm,ms,mm} of this report may be viewed as coarsely describing a long-term research agenda aiming to understand and improve interactions between humans and AI systems (as described in \secref{flowthrough}), these related agendas can be compared and contrasted with the implicit long-term agenda present in this report, as follows.

		\newcommand{\agenda}[2]{\vspace{1em}\par\noindent\textbf{#2} \citep{#1}}

		\newcommand{\nomulti}[1]{The research directions in #1 do not directly address alignment or delegation for AI systems serving multiple stakeholders\xspace}

		\newcommand{\nomultinohuman}[1]{\nomulti{#1}, and do not address the modeling of human cognition\xspace}

		\newcommand{\noxrisk}[1]{There is no direct discussion of existential risk in #1\xspace}

		\newcommand{\nobigrisks}[1]{There is no discussion of existential or global catastrophic risk in #1\xspace}

		\agenda{soares2014aligning}{Aligning Superintelligence with Human Interests (ASHI)} lays out research directions intended to address three problems:  ``How can we create an agent that will reliably pursue the goals it is given? How can we formally specify beneficial goals? And how can we ensure that this agent will assist and cooperate with its programmers as they improve its design, given that mistakes in the initial version are inevitable?''  ASHI also introduced the concept of ``alignment'' for AI systems, a key concept in this report.

				Aside from idiosyncratic differences in focus and approach, this report aims to expand and improve upon the narrative of ASHI in several regards.
				\begin{itemize}

				\item \nomultinohuman{ASHI}.

				\item This report avoids expository dependence on any ``superintelligence'' concept (see \secref{prepotence}), such as that developed by \citet{bostrom2014superintelligence}.   Instead, this report focuses on the minimal properties of an AI system that could lead to an irreversible loss of control for humanity, namely, prepotence.

				\item This report also aims to avoid the appearance of dependency on an economic ``agent'' concept, by building fewer arguments that depend on attributing ``agency'', ``beliefs'', or ``desires'' to AI systems in general (even if these concepts make sense for some systems).  Instead, we categorize AI systems (prepotent AI and MPAI) according to the impact the systems will have, or could have, upon society.
				\end{itemize}

		\noxrisk{ASHI}, although it is written with concerns similar to this report, specifically, that the deployment of powerful AI systems could have ``an enormous impact upon humanity'' and ``cause catastrophic damage''.  It also cites artificial intelligence as a positive and negative factor in global risk \citep{yudkowsky2008artificial}.

		\agenda{russell2015research}{Research Priorities for Robust and Beneficial Artificial Intelligence (RPRBAI)} lays out a number of research areas for ensuring that AI remains robust and beneficial to human society.
		Many research \directions in this report may be viewed as approaches to the broader priorities outlined in RPRBAI.
		For example, and \dirref{formalverification} addresses the RPRBAI ``Verification'' heading, and \dirref{preferencelearning} addresses the ``Validity'' heading.  The section on ``Law and Ethics Research'' can be viewed as addressing multistakeholder delegation problems.

		\noxrisk{RPRBAI}, although there is some consideration given to the risks of losing control of AI systems in the future, which could correspond roughly to the concept of \emph{prepotence} explored in this report.   By contrast, this report takes a much less balanced view of the risks and rewards of AI development, and adopts existential safety as its explicit objective. 	The distinctiveness of this objective from provable beneficence has already been elaborated somewhat in \secref{omitteddebates}.

		\agenda{amodei2016concrete}{Concrete Problems in AI Safety (CPAS)} examines open problems arising from the potential for accidents in machine learning systems.  Accidents are defined as ``unintended and harmful behavior that may emerge from poor design of real-world AI systems''.  CPAS set forth five practical research areas relating to accident risk, ``categorized according to whether the problem originates from having the wrong objective function (avoiding side effects and avoiding reward hacking), an objective function that is too expensive to evaluate frequently (scalable supervision), or undesirable behavior during the learning process (safe exploration and distributional shift).''

		There are a few key differences between the research directions of this report and those covered in CPAS:
		\begin{itemize}
		\item \nomultinohuman{CPAS}.

		\item CPAS seems mostly focussed on mitigating accident risk, whereas this report also considers the intentional deployment of destructive AI technologies, as well as hazardous social conditions that might precipitate risky AI deployments, as key guiding concerns throughout its research directions.
		\end{itemize}

		\noxrisk{CPAS}.  While the authors acknowledge ``concerns about the longer-term implications of powerful AI'', they also rightly argue that ``one need not invoke these extreme scenarios to productively discuss accidents''.  After all, the term ``AI safety'' should encompass research on any safety issue arising from the use of AI systems, whether the application or its impact is small or large in scope.

		\agenda{taylor2016alignment}{Alignment for Advanced Machine Learning Systems (AAMLS)} examines eight research areas attempting to address the question, ``As learning systems become increasingly intelligent and autonomous, what design principles can best ensure that their behavior is aligned with the interests of the operators?''
		As such, AAMLS is similar in focus to \secref{ss} of this report, particularly \dirref{preferencelearning}, and the concept of  ``alignment'' used in AAMLS corresponds roughly to the concept of ``preference alignment'' used here.

		\nomultinohuman{AAMLS}.  \noxrisk{AAMLS}, although \emph{Superintelligence} \citep{bostrom2014superintelligence} is cited in its introduction, indicating concern for global-scale risks and benefits as a key motivation.

        \agenda{leike2018scalable}{Scalable Agent Alignment via Reward Modeling: a research direction (SAARM)} defines the ``agent alignment problem'' as asking ``how do we create agents that behave in accordance with the user's intentions?''.  The authors argue that ``alignment becomes more important as ML performance increases, and any solution that fails to scale together with our agents can only serve as a stopgap''.
				SAARM proposes \emph{reward modelling} as a candidate solution to the agent alignment problem, i.e., learning a reward function from human feedback and optimizing it using (e.g. deep) reinforcement learning, which corresponds closely to \dirref{preferencelearning} as described in this report.

				There are at least a few important differences to draw between SAARM and this report:
				\begin{itemize}
					\item \nomulti{SAARM}.

					\item SAARM is situated in the paradigm of reinforcement learning, whereas this report avoids assumptions about which types of AI systems could be important sources of existential risks in the future.

					\item SAARM also highlights the importance of being able to trust the alignment of AI systems, and discusses methods which could help build such trust.

					The issue of how much a users feel they can trust an AI system is not addressed directly in this report (although comprehension and control techniques can be used to legitimately build trust).

				\end{itemize}

				\nobigrisks{SAARM}.

	\section{Acknowledgements}
		\fl{3}
		In forming the ideas presented in this report, the lead author is grateful for helpful conversations on the topic of existential risk from artificial intelligence, each spanning at least three contiguous hours, with each of
			Abram Demski,
			Anna Salamon,
			Eliezer Yudkowsky,
			Jaime Fernandez Fisac,
			Jessica Taylor,
			Michael Dennis,
			Nate Soares,
			Nick Bostrom,
			Owain Evans,
			Owen Cotton-Barratt,
			Patrick LaVictoire,
			Paul Christiano,
			Rohin Shah,
			Sam Eisenstat,
			Scott Garrabrant,
			Stuart Armstrong,
			Stuart Russell,
			and
			Toby Ord.
			Contiguous intervals of dedicated conversation time on these topics have been indispensable in developing the state of understanding represented in this document.  Helpful editorial feedback was also received from
			Allan Dafoe,
			Daniel Filan,
			Jaan Tallinn,
			Jess Riedel,
			Lawrence Chan,
			Richard Ngo,
			Roger Grosse,
			and
			Rosie Campbell.
			We are also grateful to Martin Fukui for assistance in assembling hyperlinks for inclusion in the bibliography.

\bibliography{main,main-old,orgsafety}

\auxdef{dirtot}{\numberstring{dircounter}}
\end{document}